\documentclass[twocolumn]{aastex61} 
\usepackage{graphicx}
\usepackage{natbib}
\usepackage{subfloat}
\def\arcdeg{\hbox{$^\circ$}}

\shorttitle{REDDOT}
\shortauthors{Winters et al.}

\begin{document}

\title{The Solar Neighborhood XLV. The Stellar Multiplicity Rate of
  M Dwarfs Within 25 pc}

\correspondingauthor{Jennifer G. Winters}
\email{jennifer.winters@cfa.harvard.edu}


\author{Jennifer G. Winters} 
\altaffiliation{Visiting Astronomer, Cerro Tololo Inter-American
  Observatory. CTIO is operated by AURA, Inc. under contract to the
  National Science Foundation.}
\affil{Harvard-Smithsonian Center for Astrophysics, 60 Garden Street,
  Cambridge, MA 02138, USA}

\author{Todd J. Henry}
\altaffiliation{Visiting Astronomer, Cerro Tololo Inter-American
  Observatory. CTIO is operated by AURA, Inc. under contract to the
  National Science Foundation.}
\affil{RECONS Institute, Chambersburg, Pennsylvania, 17201}

\author{Wei-Chun Jao}
\altaffiliation{Visiting Astronomer, Cerro Tololo Inter-American
  Observatory. CTIO is operated by AURA, Inc. under contract to the
  National Science Foundation.}
\affil{Department of Physics and Astronomy, Georgia State University,
  Atlanta, GA 30302-4106}

\author{John P. Subasavage}
\altaffiliation{Visiting Astronomer, Cerro Tololo Inter-American
  Observatory. CTIO is operated by AURA, Inc. under contract to the
  National Science Foundation.}
\affil{The Aerospace Corporation, 2310 E. El Segundo Boulevard, El
  Segundo, CA 90245}

\author{Joseph P. Chatelain}
\altaffiliation{Visiting Astronomer, Cerro Tololo Inter-American
  Observatory. CTIO is operated by AURA, Inc. under contract to the
  National Science Foundation.}
\affil{Las Cumbres Observatory, 6740 Cortona Dr., Suite 102, Goleta, CA 93117}

\author{Ken Slatten}
\affil{RECONS Institute, Chambersburg, Pennsylvania, 17201}

\author{Adric R. Riedel}
\altaffiliation{Visiting Astronomer, Cerro Tololo Inter-American
  Observatory. CTIO is operated by AURA, Inc. under contract to the
  National Science Foundation.}
\affil{Space Telescope Science Institute, Baltimore, MD 21218}

\author{Michele L. Silverstein}
\altaffiliation{Visiting Astronomer, Cerro Tololo Inter-American
  Observatory. CTIO is operated by AURA, Inc. under contract to the
  National Science Foundation.}
\affil{Department of Physics and Astronomy, Georgia State University,
  Atlanta, GA 30302-4106}

\author{Matthew J. Payne} 
\affil{Harvard-Smithsonian Center for Astrophysics, 60 Garden Street,
  Cambridge, MA 02138, USA}

\begin{abstract}
\label{abstract}

We present results of the largest, most comprehensive study ever done
of the stellar multiplicity of the most common stars in the Galaxy,
the red dwarfs.  We have conducted an all-sky volume-limited
survey for stellar companions to 1120 M dwarf primaries known to
lie within 25 pc of the Sun via trigonometric parallaxes.  In addition
to a comprehensive literature search, stars were explored in new
surveys for companions at separations of 2\arcsec~to 300\arcsec.  A
reconnaissance of wide companions to separations of 300\arcsec~was
done via blinking archival images. $I-$band images were used to search
our sample for companions at separations of 2\arcsec~to 180\arcsec.
Various astrometric and photometric methods were used to probe the
inner 2\arcsec~to reveal close companions.  We report the discovery of
20 new companions and identify 56 candidate multiple systems.

We find a stellar multiplicity rate of 26.8$\pm$1.4\% and a stellar
companion rate of 32.4$\pm$1.4\% for M dwarfs. There is a broad peak
in the separation distribution of the companions at 4 -- 20 AU, with a
weak trend of smaller projected linear separations for lower mass
primaries. A hint that M dwarf multiplicity may be a function of
tangential velocity is found, with faster moving, presumably older,
stars found to be multiple somewhat less often.  We calculate that
stellar companions make up at least 17\% of mass attributed to M
dwarfs in the solar neighborhood, with roughly 11\% of M dwarf mass
hidden as unresolved companions. Finally, when considering all M dwarf
primaries and companions, we find that the mass distribution for M
dwarfs increases to the end of the stellar main sequence.



\end{abstract} 


\section{Introduction}
\label{sec:intro}

Much like people, stars arrange themselves in various configurations
-- singles, doubles, multiples, clusters, and great aggregations known
as galaxies. Each of these collections is different, depending on the
proximity of the members and the shared history and composition of the
stars involved. Stellar multiples and their properties (e.g.,
separations, mass ratios, etc.) provide fundamental clues about the
nature of star formation, the distribution of baryonic mass in the
Universe, and the evolution of stellar systems over time. How stars
are parceled into singles, doubles, and higher order multiples also
provides clues about the angular momentum distribution in stellar
systems and can constrain whether or not planets may be found in these
systems \citep{Holman(1999),Raghavan(2010),Wang(2014),Winn(2015),
  Kraus(2016)}. Of all the populations in our Galaxy, the nearest
stars provide the fundamental framework upon which stellar
astrophysics is based because they contain the most easily studied
representatives of their kinds. Because M dwarfs, often called ``red
dwarfs'', dominate the nearby stellar population, accounting for
roughly 75\% of all stars \citep{Henry(2006)}, they are a critical
sample to study in order to understand stellar multiplicity.

Companion searches have been done for M dwarfs during the past few
decades, but until recently, most of the surveys have had
inhomogeneous samples made up of on the order of 100 targets.  Table
\ref{tab:previousmmult} lists these previous efforts, with the survey
presented in this work listed at the bottom for comparison. With
samples of only a few hundred stars, our statistical understanding of
the distribution of companions is quite weak, in particular when
considering the many different types of M dwarfs, which span a factor
of eight in mass \citep{Benedict(2016)}.  In the largest survey of M
dwarfs to date, \citet{Dhital(2010)} studied mid-K to mid-M dwarfs
from the Sloan Digital Sky Survey that were not nearby and found
primarily wide multiple systems, including white dwarf components in
their analysis. In the next largest studies, only a fraction of the M
dwarfs studied by \citet{Janson(2012),Janson(2014a)} had trigonometric
distances available, leading to a sample that was not
volume-limited. \citet{Ward-Duong(2015)} had a volume-limited sample
with trigonometric parallaxes from {\it Hipparcos}
(\citealt{Perryman(1997)}, updated in \citealt{vanLeeuwen(2007)}), but
the faintness limit of {\it Hipparcos} ($V$ $\sim$ 12) prevented the
inclusion of later M dwarf spectral types.\footnote{These final three
  studies were underway simultaneously with the study presented here.}


\begin{deluxetable*}{lclccl}
\centering
\setlength{\tabcolsep}{0.03in}
\tablewidth{0pt}
\tabletypesize{\scriptsize}
\tablecaption{Previous M Dwarf Multiplicity Studies - Techniques \label{tab:previousmmult}}
\tablehead{\colhead{Reference}           &
	   \colhead{\# of Stars}         &
 	   \colhead{Technique}           &
	   \colhead{Search Region}       &
           \colhead{MR\tablenotemark{*}} &
           \colhead{Notes}               }
                                                                                           
\startdata                                                                                 
\citet{Skrutskie(1989)}     &   55           &   Infrared Imaging      &  2 --- 14\arcsec    &  \nodata        & multiplicity not reported  \\  
\citet{Henry(1990)}         &   27           &   Infrared Speckle      &  0.2 --- 5\arcsec   & 34 $\pm$ 9      &                            \\  
\citet{Henry(1991)}         &   74           &   Infrared Speckle      &  0.2 --- 5\arcsec   & 20 $\pm$ 5      &                            \\  
\citet{Fischer(1992)}       & 28-62          &   Various               &  various            & 42 $\pm$ 9      & varied sample              \\  
\citet{Simons(1996)}        &   63           &   Infrared Imaging      &  10 --- 240\arcsec  & 40              &                            \\  
\citet{Delfosse(1999c)}     &  127           &   Radial Velocities     &  $<$1.0\arcsec      & \nodata         & multiplicity not reported  \\  
\citet{Law(2006b)}          &   32           &   Lucky Imaging         &  0.1 --- 1.5\arcsec &7 $^{+7}_{-3}$    & M5 - M8                    \\ 
\citet{Endl(2006)}          &   90           &   Radial Velocities     &  $<$1.0\arcsec      &  \nodata        & Jovian search              \\  
\citet{Law(2008)}           &   77           &   Lucky Imaging         &  0.1 --- 1.5\arcsec &13.6 $^{+6.5}_{-4}$& late-type M's             \\
\citet{Bergfors(2010)}      &  124           &   Lucky Imaging         &  0.2 --- 5\arcsec   & 32 $\pm$ 6      & young M0 - M6              \\
\citet{Dhital(2010)}        & 1342           &   Sloan Archive Search  &  7 --- 180\arcsec   &  \nodata        & wide binary search         \\
\citet{Law(2010)}           &   36           &   Adaptive Optics       &  0.1 --- 1.5\arcsec &  \nodata        & wide binary search         \\
\citet{Dieterich(2012)}     &  126           &   HST-NICMOS            &  0.2 --- 7.5\arcsec &  \nodata        & brown dwarf search         \\
\citet{Janson(2012)}        &  701           &   Lucky Imaging         &  0.08 --- 6\arcsec  &  27 $\pm$ 3     & young M0 - M5              \\
\citet{Janson(2014a)}       &  286           &   Lucky Imaging         &  0.1 --- 5\arcsec   &  21-27          & $>$ M5                     \\ 
\citet{Ward-Duong(2015)}    &  245           &   Infrared AO           &  10 --- 10,000 AU   &  23.5 $\pm$ 3.2 & K7 - M6                    \\ 
\hline                                                                                       
This survey                 & 1120           &   Various               & 0 --- 300\arcsec    &  26.8 $\pm$ 1.4 & all trig. distances        \\ 
\enddata                                                                                     

\tablenotetext{*}{Multiplicity Rate}

\end{deluxetable*}

Considering the significant percentage of all stars that M dwarfs
comprise, a study with a large sample (i.e., more than 1000 systems)
is vital in order to arrive at a conclusive understanding of red dwarf
multiplicity, as well as to perform statistical analyses of the
overall results, and on subsamples based on primary mass, metallicity,
etc. For example, using a binomial distribution for error analysis, an
expected multiplicity rate of 30\% on samples of 10, 100, and 1000
stars, respectively, yields errors of 14.5\%, 4.6\%, and 1.4\%,
illustrating the importance of studying a large, well-defined sample
of M dwarfs, preferably with at least 1000 stars.

  
 
Here we describe a volume-limited search for stellar companions to
1120 nearby M dwarf primary stars. For these M dwarf
primaries\footnote{We refer to any collection of stars and their
  companion brown dwarfs and/or exoplanets as a system, including
  single M dwarfs not currently known to have any companions.} with
trigonometric parallaxes placing them within 25 pc, an all-sky
multiplicity search for stellar companions at separations of 2\arcsec
~to 300\arcsec ~was undertaken. A reconnaissance for companions with
separations of 5--300\arcsec~was done via the blinking of digitally
scanned archival SuperCOSMOS $BRI$ images, discussed in detail in
$\S$\ref{subsec:blinking_search}.  At separations of 2\arcsec ~to
10\arcsec, the environs of these systems were probed for companions
via $I-$band images obtained at telescopes located in both the
northern and southern hemispheres, as outlined in
$\S$\ref{subsec:imaging_search}.  The Cerro Tololo Inter-American
Observatory / Small and Moderate Aperture Research Telescope System
(CTIO/SMARTS) 0.9m and 1.0m telescopes were utilized in the southern
hemisphere, and the Lowell 42in and United States Naval Observatory
(USNO) 40in telescopes were used in the northern hemisphere (see \S
\ref{subsec:imaging_search} for specifics on each telescope).  In
addition, indirect methods based on photometry were used to infer the
presence of nearly equal magnitude companions at separations less than
$\sim$2\arcsec ~(\S \ref{subsec:sub_arc}). Various subsets of the
sample were searched for companions at sub-arcsecond separations using
long-term astrometry at the CTIO/SMARTS 0.9m (\S \ref{subsubsec:pbs})
and {\it Hipparcos} reduction flags (\S \ref{subsubsec:hip}). Finally,
an extensive literature search was conducted (\S
\ref{subsec:lit}). Because spectral type M is effectively the end of
the stellar main sequence, the stellar companions revealed in this
search are, by definition, M dwarfs, as well. We do not include brown
dwarf companions to M dwarfs in the statistical results for this
study, although they are identified.

In the interest of clarity, we first define a few terms. {\it
  Component} refers to any physical member of a multiple
system. The {\it primary} is either a single star or the most massive
(or brightest in $V$) component in the system, and {\it companion} is
used throughout to refer to a physical member of a multiple
system that is less massive (or fainter, again in $V$) than the
primary star. Finally, we use the terms `red dwarf' and `M dwarf'
interchangeably throughout.


\section{Definition of the Sample}
\label{sec:sampledef}

\subsection{Astrometry}
\label{subsec:astrometry}

The RECONS 25 Parsec Database is a listing of all stars, brown
  dwarfs, and planets thought to be located within 25 pc, with
  distances determined only via accurate trigonometric
  parallaxes. Included in the database is a wealth of information on
  each system: coordinates, proper motions, the weighted mean of the
  parallaxes available for each system, $UBVRIJHK$ photometry,
  spectral types in many cases, and alternate names. Additionally
  noted are the details of multiple systems: the number of components
  known to be members of the system, the separations and position
  angles for those components, the year and method of detection, and
  the delta-magnitude measurement and filter in which the relative
  photometry data were obtained. Its design has been a massive
  undertaking that has spanned at least eight years, with expectations
  of its release to the community in 2019.

The 1120 systems in the survey sample have published trigonometric
parallaxes, $\pi_{trig}$, of at least 40 mas with errors of 10 mas or
less that have been extracted from the RECONS 25 Parsec
  Database.  As shown in Table \ref{tab:pisource}, there are three
primary sources of trigonometric parallax data for M dwarfs currently
available.  The {\it General Catalogue of Trigonometric Stellar
  Parallaxes, Fourth Edition} \citep{vanAltena(1995)}, often called
the {Yale Parallax Catalogue} (hereafter YPC), is a valuable
compendium of ground-based parallaxes published prior to 1995 and
includes just under half of the nearby M dwarf parallaxes for our
  sample, primarily from parallax programs at the Allegheny,
Mt.~Stromlo, McCormick, Sproul, US Naval, Van Vleck, Yale, and Yerkes
Observatories. The {\it Hipparcos} mission (initial release by
\citet{Perryman(1997)}, and revised results used here by
\citet{vanLeeuwen(2007)}; hereafter HIP) updated 231 of those
parallaxes, and contributed 229 new systems for bright ($V$
$\lesssim$~12.5) nearby M dwarfs. Overall, 743 systems have parallaxes
from the YPC and HIP catalogs.



The next largest collection of parallaxes measured for nearby M dwarfs
is from the RECONS\footnote{REsearch Consortium On Nearby Stars, {\it
    www.recons.org}} team, contributing 308 red dwarf systems to the
25 pc census via new measurements \citep{Jao(2005), Jao(2011),
  Jao(2014), Costa(2005), Costa(2006), Henry(2006), Subasavage(2009),
  Riedel(2010), Riedel(2011), Riedel(2014), vonBraun(2011),
  Mamajek(2013), Dieterich(2014), Winters(2017), Bartlett(2017),
  Jao(2017),Henry(2018),Riedel(2018)}, published in {\it The Solar
  Neighborhood} series of papers (hereafter TSN) in {\it The
  Astronomical Journal}.\footnote{A few unpublished measurements used
  in this study are scheduled for a forthcoming publication in this
  series.}  Finally, other groups have contributed parallaxes for an
additional 69 nearby M dwarfs.  As shown in Table
\ref{tab:pisource}, RECONS' work in the southern hemisphere creates a
balanced all-sky sample of M dwarfs with known distances for the first
time, as the southern hemisphere has historically been under-sampled.
An important aspect of the sample surveyed here is that because all
1120 systems have accurate parallaxes, biases inherent to
photometrically-selected samples are ameliorated.




\begin{deluxetable}{lcc}
\centering
\setlength{\tabcolsep}{0.03in}
\tablewidth{0pt}
\tabletypesize{\small}
\tablecaption{Parallax Sources for Multiplicity Search \label{tab:pisource}}
\tablehead{\colhead{Reference}             &
	   \colhead{\# of Targets}         &
 	   \colhead{\# of Targets}         \\
	   \colhead{   }                   &
	   \colhead{North of $\delta$ = 0} &
	   \colhead{South of $\delta$ = 0} }

\startdata
YPC                        &  389  & 125  \\ 
HIP                        &   83  & 146  \\ 
RECONS - published         &   31  & 272  \\ 
RECONS - unpublished       &    2  &   3  \\ 
Literature (1995-2012)     &   51  &  18  \\ 
\hline
TOTAL                      &  556  & 564  \\
\enddata

\end{deluxetable}


A combination of color and absolute magnitude limits was used to
select a sample of {\it bona fide} M dwarfs.  Stars within 25 pc were
evaluated to define the meaning of ``M dwarf'' by plotting spectral
types from RECONS \citep{Riedel(2014)}, \citet{Gray(2003)},
\citet{Reid(1995)}, and \citet{Hawley(1996)} versus $(V-K)$ and $M_V$.
Because spectral types can be imprecise, there was overlap between the
K and M types, so boundaries were chosen to split the types at
carefully defined $(V-K)$ and $M_V$ values.  A similar method was
followed for the M-L dwarf transition using results primarily from
\citet{Dahn(2002)}.  These procedures resulted in ranges of 8.8 $\leq
M_V \leq$ 20.0 and 3.7 $\leq (V-K) \leq$ 9.5 for stars we consider to
be M dwarfs.  For faint stars with no reliable $V$ available, an
initial constraint of $(I-K) \leq$ 4.5 was used to create the sample
until $V$ could be measured.  These observational parameters
correspond to masses of 0.63 $< M/M_{\odot} <$ 0.075, based on
the mass-luminosity relation presented in \citet{Benedict(2016)}. We
note that no M dwarfs known to be companions to more massive stars are
included in this sample. Systems that contained a white dwarf
component were excluded from the sample, as the white dwarf was
previously the brighter and more massive primary.



\begin{figure}
  \begin{center}
  \includegraphics[scale=0.35,angle=90]{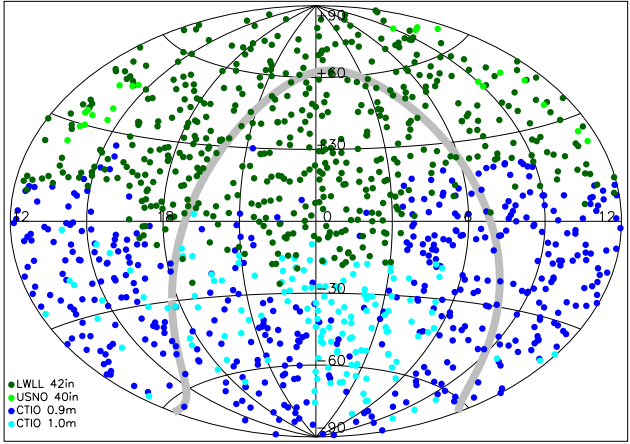}
  \caption{The distribution on the sky of all 1120 M dwarf primaries
    examined for multiplicity. Different colors indicate the different
    telescopes that were utilized for the CCD imaging search: royal
    blue for the CTIO/SMARTS 0.9m, dark green for the Lowell 42in,
    cyan for the CTIO/SMARTS 1.0m, and bright green for the USNO
    40in. The Galactic plane is outlined in gray. Illustrated is the
    uniformity of the sample in both hemispheres, due in large part to
    RECONS' parallax work in the southern
    hemisphere. \label{fig:all_sky_aitoff}}
  \end{center}
  \end{figure}

Imposing these distance, absolute magnitude, and color criteria yields
a sample of 1120 red dwarf primaries as of January 1, 2014, when the
companion search sample list was frozen, with some new parallaxes
  measured by RECONS being added as they became available. The
astrometry data for these 1120 systems are listed in Table
\ref{tab:astrdata}.  Included are the names of the M dwarf primary,
coordinates (J2000.0), proper motion magnitudes and position angles
with references, the weighted means of the published trigonometric
parallaxes and the errors, and the number of parallaxes included in
the weighted mean and references.  We note that for multiple systems,
the proper motion of the primary component has been assumed to be the
same for all members of the system. All proper motions are from
SuperCOSMOS, except where noted.  Proper motions with the reference
`RECONS (in prep)' ~indicate SuperCOSMOS proper motions that will be
published in the forthcoming RECONS 25 Parsec Database (Jao et al., in
prep), as these values have not been presented previously. In the
cases of multiple systems for which parallax measurements exist for
companions, as well as for the primaries, these measurements have been
combined in the weighted means. The five parallaxes noted as `in prep'
will be presented in upcoming papers in the TSN series. Figure
\ref{fig:all_sky_aitoff} shows the distribution on the sky of the
entire sample investigated for multiplicity.  Note the balance in the
distribution of stars surveyed, with nearly equal numbers of M dwarfs
in the northern and southern skies.

\subsection{Sample Selection Biases}
\label{subsec:biases}
We describe here how the sample selection process could bias the
result of our survey.

We note that our sample is volume-limited, not volume-complete. If we
assume the 188 M dwarf systems in our sample that lie within 10 pc
comprise a volume-complete sample and extrapolate to 25 pc assuming a
uniform stellar density, we expect 2938 M dwarf systems to lie within
25 pc.

We cross-matched our sample of M dwarf primaries to the recently
available parallaxes from the {\it Gaia} Data Release 2 (DR2)
\citep{Gaia(2016a),GaiaDR2(2018)} and found that 90\% (1008 primaries)
had {\it Gaia} parallaxes that placed them within 25 pc. Four percent
fell outside of 25 pc with a {\it Gaia} DR2 parallax. The remaining
6\% (69 primaries) were not found to have a {\it Gaia} DR2 parallax,
but 47 (4\%) are known to be in multiple systems with
separations between the components on the order of or less than
1\arcsec. Nine of these 47 multiple systems are within the ten parsec
horizon. A few of the remaining 22 that are not currently known to be
multiple are definitively nearby, but have high proper motion (e.g.,
GJ~406) or are bright (e.g., GJ~411). We do not make any corrections
to our sample based on this comparison because it is evident that a
sample of stars surveyed for stellar multiplicity based on the {\it
  Gaia} DR2 would neglect binaries. We look forward, however, to the
{\it Gaia} DR3 which will include valuable multiplicity information.

\begin{figure}

  \includegraphics[scale=0.35,angle=90]{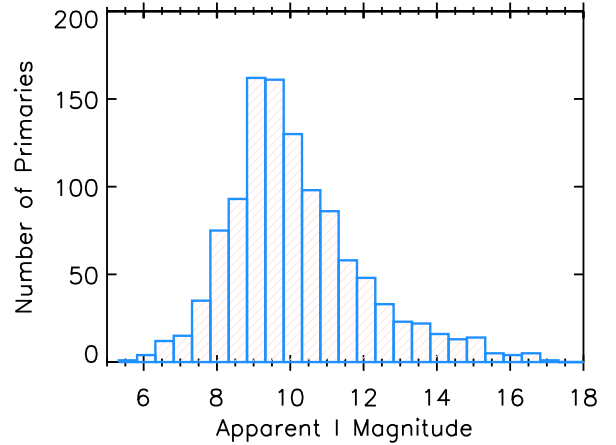}

\vspace{0.25cm}
  \caption{The distribution of $I-$band magnitudes of our sample of
    1120 M dwarfs known to lie within 25 pc, illustrating that most
    (85\%) of the target stars are brighter than $I$ =
    12. \label{fig:imags}}
  \end{figure}

Figure \ref{fig:imags} shows the distribution of the apparent $I$
magnitudes of the red dwarfs surveyed, with a peak at $I$ =
8.5--9.5. Because brighter objects are generally targeted for parallax
measurements before fainter objects for which measurements are more
difficult, 85\% of the sample is made up of bright stars ($I <$
12.00), introducing an implicit Malmquist bias. As unresolved multiple
systems are usually over-luminous, this survey's outcomes are biased
toward a larger multiplicity rate.

We have also required the error on the published trigonometric
parallax to be $\leq$10 mas in order to limit the sample to members
that are reliably within 25 pc. Therefore, it is possible that
binaries were missed, as perturbations on the parallax due to an
unseen companion can increase the parallax error. Forty-five M dwarf
systems with YPC or HIP parallaxes were eliminated from the sample due
to their large parallax errors. We cross-checked these 45 targets
against the {\it Gaia} DR2 with a search radius of 3\arcmin, to
mitigate the positional offset of these typically high proper motion
stars. Twenty-nine were returned with parallaxes by {\it Gaia}, 19 of
which remained within our chosen 25 pc distance horizon. Four of these
19 had close companions detected by {\it Gaia}. If we assumed that the
16 non-detections were all multiple systems and all within 25 pc, the
sample size would increase to 1155, and the multiplicity rate would
increase by 0.9\%. We do not include any correction due to this
bias. We note that the parallaxes measured as a result of RECONS'
astrometry program, roughly one-third of the sample, would not factor
into this negative bias, as all of these data were examined and stars
with astrometric perturbations due to unseen companions flagged.

Additionally, there is mass missing within 25 pc in the form of M
dwarf primaries \citep{Winters(2015)}. However, because the
multiplicity rates decrease as a function of primary mass (see \S
\ref{subsubsec:mr_by_mass} and Figure \ref{fig:run_mass}), the
percentages of `missing' multiple systems in each mass bin are
effectively equal. Based on the 10 pc sample, as above, we expect 969
M dwarfs more massive than 0.30 M$_{\odot}$ within 25 pc, but have 506
in our sample. The MR of 28.2\% for the estimated 463 missing systems
results in 131 (14\%) missing multiples in this primary mass
subset. We expect 1109 M dwarfs with primaries 0.15 $-$ 0.30
M$_{\odot}$ within 25 pc, but have 402 in our sample. The MR of 21.4\%
for the estimated 707 missing systems results in 151 (14\%) missing
multiples in this primary mass subset. Finally, we expect 859 M dwarfs
with primaries 0.075 $-$ 0.15 M$_{\odot}$ within 25 pc, but have 212
in our sample. The MR of 16.0\% for the estimated 647 missing systems
results in 104 (12\%) missing multiples in this primary mass
subset. Therefore, we do not include a correction for this bias.

\begin{deluxetable}{lccccccccc}
\centering
\setlength{\tabcolsep}{0.03in}
\tablewidth{0pt}
\tabletypesize{\tiny}
\tablecaption{Astrometry Data \label{tab:astrdata}}
\tablehead{\colhead{Name}                &
 	   \colhead{R.A.}                &
 	   \colhead{Decl.}               &
 	   \colhead{$\mu$}               &
 	   \colhead{P.A.}                &
 	   \colhead{Ref}                 &
 	   \colhead{$\pi$}               &
 	   \colhead{$\sigma_{\pi}$}       &
 	   \colhead{\# $\pi$}            &
 	   \colhead{Ref}                 \\
           \colhead{    }                &
 	   \colhead{(hh:mm:ss)}     &
 	   \colhead{(dd:mm:ss)}     &
 	   \colhead{(\arcsec yr$^{-1}$ )} &
 	   \colhead{(deg)}               &
 	   \colhead{   }                 &
 	   \colhead{(mas)}               &
 	   \colhead{(mas)}               &
 	   \colhead{        }            &
 	   \colhead{   }                 \\
           \colhead{(1) }                &
 	   \colhead{(2) }                &
 	   \colhead{(3)  }               &
 	   \colhead{(4)  }               &
 	   \colhead{(5) }                &
 	   \colhead{(6)}                 &
 	   \colhead{(7)  }               &
 	   \colhead{(8)  }               &
 	   \colhead{(9)  }               &
 	   \colhead{(10)}                }
\startdata
 GJ~1001ABC                   &00 04 36.45 &$-$40 44 02.7 &   1.636 & 159.7 & 71 &   77.90                                 &   2.04 & 2   & 15,68            \\ 
 GJ~1                         &00 05 24.43 &$-$37 21 26.7 &   6.106 & 112.5 & 28 &  230.32                                 &   0.90 & 2   & 68,69            \\ 
 LHS~1019                     &00 06 19.19 &$-$65 50 25.9 &   0.564 & 158.7 & 72 &   59.85                                 &   2.64 & 1   & 69               \\ 
 GJ~1002                      &00 06 43.19 &$-$07 32 17.0 &   2.041 & 204.0 & 39 &  213.00                                 &   3.60 & 1   & 68               \\ 
 GJ~1003                      &00 07 26.71 &$+$29 14 32.7 &   1.890 & 127.0 & 38 &   53.50                                 &   2.50 & 1   & 68               \\
 LHS~1022                     &00 07 59.11 &$+$08 00 19.4 &   0.546 & 222.0 & 38 &   44.00                                 &   6.30 & 1   & 68               \\
 L~217-28                     &00 08 17.37 &$-$57 05 52.9 &   0.370 & 264.0 & 40 &   75.17                                 &   2.11 & 1   & 73               \\ 
 HIP~687                      &00 08 27.29 &$+$17 25 27.3 &   0.110 & 233.8 & 28 &   45.98                                 &   1.93 & 1   & 69               \\
 G~131-26AB                   &00 08 53.92 &$+$20 50 25.4 &   0.251 & 194.4 & 53 &   54.13                                 &   1.35 & 1   & 53               \\
 GJ~7                         &00 09 04.34 &$-$27 07 19.5 &   0.715 & 079.7 & 72 &   43.61                                 &   2.56 & 2   & 68,69            \\ 
 LEHPM~1-255                  &00 09 45.06 &$-$42 01 39.6 &   0.271 & 096.7 & 72 &   53.26                                 &   1.51 & 1   & 73               \\ 
\enddata

\tablecomments{The first 10 lines of this Table are shown to
  illustrate its form and content.}

\tablenotetext{a}{The weighted mean parallax includes the parallax of
  both the primary and the secondary components.}

\tablerefs{
(1) \citet{Andrei(2011)};
(2) \citet{Anglada(2012)}; 
(3) \citet{Bartlett(2017)};
(4) \citet{Benedict(1999)}; 
(5) \citet{Benedict(2000b)}; 
(6) \citet{Benedict(2001)}; 
(7) \citet{Benedict(2002c)}; 
(8) \citet{Biller(2007)}; 
(9) \citet{Costa(2005)}; 
(10) \citet{Costa(2006)}; 
(11) \citet{Dahn(2002)}; 
(12) \citet{Deacon(2001)}; 
(13) \citet{Deacon(2005a)}; 
(14) \citet{Deacon(2005b)}; 
(15) \citet{Dieterich(2014)}; 
(16) \citet{Dupuy(2012)}; 
(17) \citet{Fabricius(2000)}; 
(18) \citet{Faherty(2012)}; 
(19) \citet{Falin(1999)}; 
(20) \citet{Gatewood(1993)}; 
(21) \citet{Gatewood(2003)}; 
(22) \citet{Gatewood(2008)}; 
(23) \citet{Gatewood(2009)}; 
(24) \citet{Henry(1997)}; 
(25) \citet{Henry(2006)}; 
(26) \citet{Henry(2018)}; 
(27) \citet{Hershey(1998)}; 
(28) \citet{Hog(2000)}; 
(29) \citet{Ianna(1996)}; 
(30) \citet{Jao(2005)}; 
(31) \citet{Jao(2011)};
(32) \citet{Jao(2017)};
(33) \citet{Khovritchev(2013)}; 
(34) \citet{Lepine(2005a)}; 
(35) \citet{Lepine(2009)};
(36) \citet{Lurie(2014)}; 
(37) \citet{Luyten(1979a)}; 
(38) \citet{Luyten(1979b)}; 
(39) \citet{Luyten(1980a)}; 
(40) \citet{Luyten(1980b)}; 
(41) \citet{Martin(1998a)}; 
(42) \citet{Martinache(2007)}; 
(43) \citet{Martinache(2009)}; 
(44) \citet{Monet(2003)}; 
(45) \citet{Pokorny(2004)}; 
(46) \citet{Pourbaix(2003)}; 
(47) \citet{Pravdo(2006)}; 
(48) \citet{Pravdo(2009)}; 
(49) RECONS (in prep);
(50) \citet{Reid(2003a)}; 
(51) \citet{Riedel(2010)}; 
(52) \citet{Riedel(2011)}; 
(53) \citet{Riedel(2014)}; 
(54) \citet{Riedel(2018)}; 
(55) \citet{Schilbach(2009)}; 
(56) \citet{Schmidt(2007)}; 
(57) \citet{Shakht(1997)}; 
(58) \citet{Shkolnik(2012)}; 
(59) \citet{Smart(2007)}; 
(60) \citet{Smart(2010b)}; 
(61) \citet{Soderhjelm(1999)}; 
(62) \citet{Subasavage(2005a)}; 
(63) \citet{Subasavage(2005b)}; 
(64) \citet{Teegarden(2003)}; 
(65) \citet{Teixeira(2009)}; 
(66) \citet{Tinney(1995)}; 
(67) \citet{Tinney(1996)}; 
(68) \citet{vanAltena(1995)}; 
(69) \citet{vanLeeuwen(2007)}; 
(70) \citet{vonBraun(2011)}; 
(71) \citet{Weis(1999)}; 
(72) \citet{Winters(2015)};
(73) \citet{Winters(2017)}.}
\end{deluxetable}



\subsection{Optical and Infrared Photometry}
\label{subsec:photometry}


Existing $VRI$ photometry for many of the M dwarfs in the sample was
culled from the literature, much of which has been presented
previously for the southern M dwarfs in
\citet{Winters(2011),Winters(2015),Winters(2017)}; however, a number
of M dwarfs in the sample had no published reliable optical photometry
available. As part of the effort to characterize the M dwarfs in the
survey, new absolute photometry in the Johnson-Kron-Cousins
$V_JR_{KC}I_{KC}$\footnote{These subscripts will be dropped
  henceforth. The central wavelengths for the $V_J$, R$_{KC}$, and
  I$_{KC}$ filters at the 0.9m are 5438\AA, 6425\AA, and 8075\AA,
  respectively; filters at other telescopes are similar.}  filters was
acquired for 81, 3, and 49 stars at the CTIO/SMARTS 0.9m, CTIO/SMARTS
1.0m, and Lowell 42in telescopes, respectively, and is presented here
for the first time.  Identical observational methods were used at all
three sites.  As in previous RECONS efforts, standard star fields from
\citet{Graham(1982)}, \citet{Bessel(1990)}, and/or
\citet{Landolt(1992), Landolt(2007), Landolt(2013)} were observed
multiple times each night to derive transformation equations and
extinction curves.  In order to match those used by Landolt, apertures
14$\arcsec$~in diameter were used to determine the stellar fluxes,
except in cases where close contaminating sources needed to be
deblended.  In these cases, smaller apertures were used and aperture
corrections were applied.  Further details about the data reduction
procedures, transformation equations, etc., can be found in
\citet{Jao(2005)}, \citet{Winters(2011)}, and \citet{Winters(2015)}.


In addition to the 0.9m, 1.0m, and 42in observations, three stars were
observed at the United States Naval Observatory (USNO) Flagstaff
Station 40in telescope.  Basic calibration frames, bias and sky flats
in each filter are taken either every night (bias) or over multiple
nights in a run (sky flats) and are applied to the raw science data.
Standard star fields from \citet{Landolt(2009),Landolt(2013)} were
observed at multiple airmasses between $\sim$1.0 and $\sim$2.1 each
per night to calculate extinction curves.  All instrumental
magnitudes, both for standards and science targets, are extracted by
fitting spatially-dependent point spread functions (PSFs) for each
frame using Source Extractor ({\it SExtractor} \citep{Bertin(1996)}
and {\it PSFEx} \citep{Bertin(2011)}, with an aperture diameter of
14\arcsec.  Extensive comparisons of this technique to basic aperture
photometry have produced consistent results in uncrowded fields.


Optical and infrared photometry for the 1448 components of the 1120 M
dwarf systems is presented in Table \ref{tab:photdata}, where
available. $JHK_s$ magnitudes were extracted from 2MASS
\citep{Skrutskie(2006)} and confirmed by eye to correspond to the star
in question during the blinking survey. Included are the names of the
M dwarfs (column 1), the number of known components in the systems
(2), J2000.0 coordinates (3, 4), $VRI$ magnitudes (5, 6, 7), the
number of observations and/or references (8), the 2MASS $JHK_s$
magnitudes (9, 10, 11), and the photometric distance estimate.  Next
are listed the $\Delta$$V$ magnitudes between stellar companions and
primaries (12), the deblended $V$ magnitudes $V_{db}$ (13), and
estimated masses for each component (14). Components of multiple
systems are noted with a capital letter (A,B,C,D,E) after the name in
the first column.  If the names of the components are different, the
letters identifying the primary and the secondary are placed within
parentheses, e.g., LHS1104(A) and LHS1105(B).  If the star is a
companion in a multiple system, `0' is given in column (2). `J' for
joint photometry is listed with each blended magnitude. Brown dwarf
companions are noted by a `BD' next to the `0' in column 2, and often
do not have complete photometry, if any.




\begin{deluxetable*}{lccccccccccccccc}
\centering
\setlength{\tabcolsep}{0.03in}
\tablewidth{0pt}
\tabletypesize{\tiny}
\tablecaption{Photometry Data \label{tab:photdata}}
\tablehead{\colhead{Name}                &
	   \colhead{\# Obj}              &
 	   \colhead{R.A.}                &
 	   \colhead{Decl.}               &
 	   \colhead{$V_J$}               &
 	   \colhead{$R_{KC}$}            &
 	   \colhead{$I_{KC}$}            &
 	   \colhead{\# nts/ref}          &
 	   \colhead{$J$}                 &
 	   \colhead{$H$}                 &
 	   \colhead{$K_s$}               &
 	   \colhead{$\pi_{ccd}$}          &
 	   \colhead{$\sigma_{\pi}$}       &
 	   \colhead{$\Delta$$V$}         &
 	   \colhead{$V_{db}$}            &
 	   \colhead{Mass}                \\
           \colhead{     }               &
	   \colhead{     }               &
 	   \colhead{(dd:mm:ss)}     &
 	   \colhead{(hh:mm:ss)}     &
 	   \colhead{(mag)}               &
 	   \colhead{(mag)}               &
 	   \colhead{(mag)}               &
 	   \colhead{     }               &
 	   \colhead{(mag)}               &
 	   \colhead{(mag)}               &
 	   \colhead{(mag)}          &
 	   \colhead{(pc)}                &
 	   \colhead{(pc)}           &
 	   \colhead{(mag)}               &
 	   \colhead{(mag)}          &
 	   \colhead{(M$_{\odot}$)}       \\
           \colhead{(1)}                 &
	   \colhead{(2)}                 &
 	   \colhead{(3)}                 &
 	   \colhead{(4)}                 &
 	   \colhead{(5)}                 &
 	   \colhead{(6)}                 &
 	   \colhead{(7)}                 &
 	   \colhead{(8)}                 &
 	   \colhead{(9)}                 &
 	   \colhead{(10)}                &
 	   \colhead{(11)}                &
 	   \colhead{(12)}                &
 	   \colhead{(13)}                &
 	   \colhead{(14)}                &
 	   \colhead{(15)}                &
 	   \colhead{(16)}                }
                                                                                           
\startdata
 GJ1001B                     &0BD &00 04 34.87 &$-$40 44 06.5  &  22.77J  & 19.04J  & 16.67J  &   /10\tablenotemark{d}                   &  13.11J                      &  12.06J                    &  11.40J                   & \nodata  & \nodata  &   \nodata                    & \nodata &  \nodata   \\           
 GJ1001C                     &0BD &00 04 34.87 &$-$40 44 06.5  &  \nodata & \nodata & \nodata &  \nodata                                 &  \nodata                     &  \nodata                   &  \nodata                  & \nodata  & \nodata  &   \nodata                    & \nodata &  \nodata   \\
 GJ1001A                     &3   &00 04 36.45 &$-$40 44 02.7  &  12.83   & 11.62   & 10.08   &   /40                                    &   8.60                       &   8.04                     &   7.74                    &  12.5    &    1.9   &   \nodata                    &  12.83  &  0.234   \\ 
 GJ0001                      &1   &00 05 24.43 &$-$37 21 26.7  &   8.54   &  7.57   &  6.41   &   /4                                     &   5.33                       &   4.83\tablenotemark{a}    &   4.52                    &   5.6    &    0.9   &   \nodata                    &   8.54  &  0.411   \\ 
 LHS1019                     &1   &00 06 19.19 &$-$65 50 25.9  &  12.17   & 11.11   &  9.78   &   /21                                    &   8.48                       &   7.84                     &   7.63                    &  16.6    &    2.6   &   \nodata                    &  12.17  &  0.335   \\ 
 GJ1002                      &1   &00 06 43.19 &$-$07 32 17.0  &  13.84   & 12.21   & 10.21   &   /40                                    &   8.32                       &   7.79                     &   7.44                    &   5.4    &    1.0   &   \nodata                    &  13.84  &  0.116   \\ 
 GJ1003                      &1   &00 07 26.71 &$+$29 14 32.7  &  14.16   & 13.01   & 11.54   &   /37                                    &  10.22                       &   9.74                     &   9.46                    &  36.0    &    7.0   &   \nodata                    &  14.16  &  0.203   \\
 LHS1022                     &1   &00 07 59.11 &$+$08 00 19.4  &  13.09   & 12.02   & 10.65   &   /37                                    &   9.39                       &   8.91                     &   8.65                    &  28.9    &    5.2   &   \nodata                    &  13.09  &  0.311   \\
 L217-028                    &1   &00 08 17.37 &$-$57 05 52.9  &  12.13   & 11.00   &  9.57   &   /40                                    &   8.21                       &   7.63                     &   7.40                    &  13.2    &    2.0   &   \nodata                    &  12.13  &  0.293   \\ 
 HIP000687                   &1   &00 08 27.29 &$+$17 25 27.3  &  10.80   &  9.88   &  8.93   &   /35                                    &   7.81                       &   7.17                     &   6.98                    &  18.5    &    3.2   &   \nodata                    &  10.80  &  0.582   \\
\enddata

\tablecomments{The first 10 lines of this Table are shown to
  illustrate its form and content.}

\tablecomments{A `J' next to a photometry value indicates that the
  magnitude is blended due to one or more close companions. A `]' next
to the photometric distance estimate indicates that the joint
photometry of the multiple system was used to calculate the distance
estimate, which is thus likely underestimated. A 'u' following the
photometry reference indicates that we present an update to previously
presented RECONS photometry.}

\tablecomments{$^{a}$ $2MASS$ magnitude error greater than 0.05 mags;
  $^{b}$ An assumption was made regarding the $\Delta$mag; $^{c}$ A
  conversion to $\Delta$$V$ was done from a reported magnitude
  difference in another filter; $^{d}$ Photometry in SOAR filters and
  not converted to Johnson-Kron-Cousins system; $^{e}$
  \citet{Barbieri(1996)}; $^{f}$ \citet{Benedict(2000b)}; $^{g}$
  \citet{Benedict(2016)}; $^{h}$ \citet{Henry(1999)}; $^{i}$
  \citet{Henry(1999),Tamazian(2006)}; $^{j}$ \citet{Segransan(2000)};
  $^{k}$ \citet{Delfosse(1999c)}; $^{l}$ \citet{Diaz(2007)}; $^{m}$
  \citet{Duquennoy(1988b)}; $^{n}$ \citet{Herbig(1965)}; $^{o}$
  Photometry for `AC', instead of for the `B' component, was
  mistakenly reported in \citet{Davison(2015)}.}

\tablerefs{(1) this work; 
(2) \citet{Bartlett(2017)};
(3) \citet{Benedict(2016)};
(4) \citet{Bessel(1990)}; 
(5) \citet{Bessell(1991)}; 
(6) \citet{Costa(2005)}; 
(7) \citet{Costa(2006)}; 
(8) \citet{Dahn(2002)};
(9) \citet{Davison(2015)};
(10) \citet{Dieterich(2014)}; 
(11) \citet{Harrington(1980)}; 
(12) \citet{Harrington(1993)}; 
(13) \citet{Henry(2006)}; 
(14) \citet{Henry(2018)};
(15) \citet{Hog(2000)}; 
(16) \citet{Hosey(2015)};
(17) \citet{Jao(2005)};
(18) \citet{Jao(2011)}; 
(19) \citet{Jao(2017)}; 
(20) \citet{Koen(2002)}; 
(21) \citet{Koen(2010)}; 
(22) \citet{Lepine(2009)}; 
(23) \citet{Lurie(2014)};
(24) \citet{Reid(2002)}; 
(25) \citet{Riedel(2010)}; 
(26) \citet{Riedel(2011)};
(27) \citet{Riedel(2014)};
(28) \citet{Riedel(2018)}; 
(29) \citet{Weis(1984)}; 
(30) \citet{Weis(1986)}; 
(31) \citet{Weis(1987)}; 
(32) \citet{Weis(1988)}; 
(33) \citet{Weis(1991a)}; 
(34) \citet{Weis(1991b)}; 
(35) \citet{Weis(1993)}; 
(36) \citet{Weis(1994)}; 
(37) \citet{Weis(1996)}; 
(38) \citet{Weis(1999)}; 
(39) \citet{Winters(2011)}; 
(40) \citet{Winters(2015)};
(41) \citet{Winters(2017)}.
}
\end{deluxetable*}

For new photometry reported here, superscripts are added to the
references indicating which telescope(s) was used to acquire the $VRI$
photometry: `09' for the CTIO/SMARTS 0.9m, `10' for the CTIO/SMARTS
1.0m, `40' for the USNO 40in, and `42' for the Lowell 42in. If the
$\Delta$$V$ is larger than 3, the magnitude of the primary is treated
as unaffected by the companion(s).  All masses are estimated from the
absolute $V$ magnitude, which has been calculated from the deblended
$V$ magnitude for each star in column (13), the parallax in Table
\ref{tab:astrdata}, and the empirical mass-luminosity relations of
\citet{Benedict(2016)}.  If any type of assumption or conversion was
made regarding the $\Delta$$V$ (as discussed in \S
\ref{subsubsec:deblend}), it is noted.


As outlined in \citet{Winters(2011)}, photometric errors at the 0.9m
are typically 0.03 mag in $V$ and 0.02 mag in $R$ and $I$.  To verify
the Lowell 42in data\footnote{No rigorous comparisons are yet possible
  for our sample of red dwarfs for the CTIO/SMARTS 1.0m and USNO 40in,
  given only three stars observed at each.}, Table
\ref{tab:phot_compare} presents photometry for four stars observed at
the Lowell 42in and at the CTIO/SMARTS 0.9m, as well as six stars with
$VRI$ from the literature.  Results from the 42in and 0.9m match to
0.06 mag, except for the $R$ magnitude of GJ 1167, which can be
attributed to a possible flare event observed at the time of
observation at the 42in, as the $V$ and $I$ magnitudes are consistent.
This object is, in fact, included in a flare star catalog of UV
Cet-type variables \citep{Gershberg(1999A)}.  An additional six stars
were observed by Weis\footnote{All photometry from Weis has been
  converted to the Johnson-Kron-Cousins (JKC) system using the
  relation in \citet{Bessell(1987)}.}, and the photometry matches to
within 0.08 mag for all six objects, and typically to 0.03 mag.  Given
our typical 1$\sigma$ errors of at most 0.03 mag for $VRI$, we find
that the Lowell 42in data have differences of 2$\sigma$ or less in 28
of the 30 cases shown in Table \ref{tab:phot_compare}.





\begin{deluxetable}{lcccccl}
\centering
\setlength{\tabcolsep}{0.03in}
\tablewidth{0pt}
\tabletypesize{\scriptsize}
\tablecaption{Overlapping Photometry Data \label{tab:phot_compare}}
\tablehead{\colhead{Name}                &
	   \colhead{($V-K$)}             &
 	   \colhead{$V_J$}               &
	   \colhead{$R_{KC}$}            &
	   \colhead{$I_{KC}$}             &
	   \colhead{\# obs}              &
           \colhead{tel/ref}             \\
	   \colhead{   }                 &
	   \colhead{(mag)}               &
           \colhead{(mag)}               &
           \colhead{(mag)}               &
	   \colhead{(mag)}               &
	   \colhead{   }                 &
           \colhead{     }               }

\startdata
2MA~J0738$+$2400      & 4.86    & 12.98 & 11.81   & 10.35    &   1    & 42in \\
                      &         & 12.98 & 11.83   & 10.35    &   2    & 0.9m \\ 
G~43-2                & 4.76    & 13.23 & 12.08   & 10.67    &   1    & 42in \\
                      &         & 13.24 & 12.07   & 10.66    &   2    & 0.9m  \\ 
2MA~J1113$+$1025      & 5.34    & 14.55 & 13.27   & 11.63    &   1    & 42in \\
                      &         & 14.50 & 13.21   & 11.59    &   2    & 0.9m  \\ 
GJ~1167               & 5.59    & 14.16 & 12.67   & 11.10    &   1    & 42in \\
                      &         & 14.20 & 12.82   & 11.11    &   1    & 0.9m  \\ 
\hline
LTT~17095A            & 4.22    & 11.12 & 10.12   &  9.00    &   1    & 42in \\
                      &         & 11.11 & 10.11   &  8.94    &  ...   & 1    \\
GJ~15B                & 5.12    & 11.07 &  9.82   &  8.34    &   2    & 42in \\
                      &         & 11.06 &  9.83   &  8.26    &  ...   & 3    \\
GJ~507AC              & 3.96    &  9.52 &  8.56   &  7.55    &   1    & 42in \\
                      &         &  9.52 &  8.58   &  7.55    &  ...   & 3    \\
GJ~507B               & 4.64    & 12.15 & 11.06   &  9.66    &   1    & 42in \\
                      &         & 12.12 & 11.03   &  9.65    &  ...   & 3    \\
GJ~617A               & 3.64    &  8.59 &  7.68   &  6.85    &   1    & 42in \\
                      &         &  8.60 &  7.72   &  6.86    &  ...   & 3    \\
GJ~617B               & 4.67    & 10.74 &  9.67   &  8.29    &   1    & 42in \\
                      &         & 10.71 &  9.63   &  8.25    &  ...   & 2    \\
\enddata

\tablerefs{(1) \citet{Weis(1993)}; (2) \citet{Weis(1994)}; (3)
  \citet{Weis(1996)}.}

\end{deluxetable}

\section{The Searches and Detected Companions}
\label{sec:search}

Several searches were carried out on the 1120 nearby M dwarfs in an
effort to make this the most comprehensive investigation of
multiplicity ever undertaken for stars that dominate the solar
neighborhood. Information about the surveys is collected in Tables
6--12, including a statistical overview of the individual surveys in
Table \ref{tab:methods_stats}. Note that the number of detections
includes confirmations of previously reported multiples in the
literature. Specifics about the Blink Survey are listed in Table
\ref{tab:supercosmos_info}. Telescopes used for the CCD Imaging Survey
in Table \ref{tab:telescopes}, while detection limit information for
the CCD Imaging Survey is presented in Tables \ref{tab:det_lim_stars}
and \ref{tab:det_lim_sum}.  Results for confirmed multiples are
collected in Table \ref{tab:multinfo}, whereas candidate, but as yet
unconfirmed, companions are listed in Table \ref{tab:suspects}.

We report the results of each search here; overall results are
presented in \S \ref{sec:results_all}.


\begin{deluxetable}{lcccc}
\centering
\setlength{\tabcolsep}{0.03in}
\tablewidth{0pt}
\tabletypesize{\scriptsize}
\tablecaption{Companion Search Technique Statistics \label{tab:methods_stats}}
\tablehead{\colhead{Technique}           &
           \colhead{Separation}          &
	   \colhead{Searched}            &
 	   \colhead{Searched}            &
	   \colhead{Detected}            \\
           \colhead{}                    &
	   \colhead{(\arcsec)}           &
	   \colhead{(\#)}                &
	   \colhead{(\%)}                &
           \colhead{(\#)}                }

\startdata
Image Blinking         &  5--300         & 1110  &   99   &  64     \\
                       &                 &       &        &         \\
CCD Imaging            &  2--10          & 1120  &  100   &  44     \\
                       &                 &       &        &         \\
RECONS Perturbations   &  $<$ 2          &  324  &   29   &  39     \\ 
HR Diagram Elevation   &  $<$ 2          & 1120  &  100   &  11     \\
Distance Mismatches    &  $<$ 2          & 1112  &   99   &  37     \\
{\it Hipparcos} Flags  &  $<$ 2          &  460  &   41   &  31     \\
Literature/WDS Search  &   all           & 1120  &  100   & 290     \\
                       &                 &       &        &         \\
Individual companions  &  TOTAL          & 1120  &  100   &  310    \\
\enddata

\end{deluxetable}


\begin{deluxetable}{lcccc}
\centering
\setlength{\tabcolsep}{0.03in}
\tablewidth{0pt}
\tabletypesize{\scriptsize}
\tablecaption{Blink Survey Information \label{tab:supercosmos_info}}
\tablehead{\colhead{Filter}              &
	   \colhead{Epoch Span}          &
 	   \colhead{DEC Range}           &
	   \colhead{Mag. Limit}          &
           \colhead{$\Delta$$\lambda$}   \\
	   \colhead{   }                 &
	   \colhead{(yr)}                &
	   \colhead{(deg)}               &
           \colhead{(mag)}               &
           \colhead{(\AA)}               }

\startdata
$B_J$ (IIIaJ)        & 1974 - 1994 &  all-sky                        & $\sim$20.5 &  3950 - 5400  \\
$R_{59F}$ (IIIaF)    & 1984 - 2001 &  all-sky                        & $\sim$21.5 &  5900 - 6900  \\
$I_{IVN}$ (IVN)      & 1978 - 2002 &  all-sky                        & $\sim$19.5 &  6700 - 9000  \\
$E_{POSS-I}$ (103aE) & 1950 - 1957 &  $-$20.5 $<$ $\delta$ $<$ $+$05 & $\sim$19.5 &  6200 - 6700  \\
$I_{KC}$             & 2010 - 2014 &  all sky                        & $\sim$17.5 &  7150 - 9000  \\
\enddata

\end{deluxetable}

\subsection{Wide-Field Blinking Survey: Companions at 5--300\arcsec}
\label{subsec:blinking_search}

Because most nearby stars have large proper motions, images of the
stars taken at different epochs were blinked for common proper motion
(CPM) companions with separations of 5--300\arcsec. A wide companion
would have a similar proper motion to its primary and would thus
appear to move in the same direction at the same speed across the sky.
Archival SuperCOMOS $B_JR_{59F}I_{IVN}$\footnote{These subscripts will
  be dropped henceforth.} photographic plate images 10\arcmin~$\times$
10\arcmin~in size were blinked using the Aladin interface of the
Centre de Donne{\`e}s astronomiques de Strasbourg (CDS) to detect
companions at separations greater than $\sim$5\arcsec. These plates
were taken from 1974--2002 and provide up to 28 years of temporal
coverage, with typical epoch spreads of at least 10 years.
Information for the images blinked is given in Table
\ref{tab:supercosmos_info}, taken from \citet{Morgan(1995)},
\citet{Subasavage(2007)}, and the UK Schmidt webpage.\footnote{{\it
    http://www.roe.ac.uk/ifa/wfau/ukstu/telescope.html}} Candidates
were confirmed to be real by collecting $VRI$ photometry and
estimating photometric distances using the suite of relations in
\citet{Henry(2004)}; if the distances of the primary and candidate
matched to within the errors on the distances, the candidate was
deemed to be a physical companion. In addition to recovering 63
  known CPM companions, one new CPM companion (2MA0936-2610C) was
discovered during this blinking search, details of which are given in
\S \ref{subsec:newcomp}. No comprehensive search for companions at
angular separations larger than 300\arcsec ~was conducted.



\subsubsection{Blink Survey Detection Limits}
\label{subsubsec:det_lim_blink}

The CPM search had two elements that needed to be evaluated in order
to confidently identify objects moving with the primary star in
question: companion brightness and the size of each system's proper
motion. 



A companion would have to be detectable on at least two of the three
photographic plates in order to notice its proper motion, so any
companion would need to be brighter than the magnitude limits given in
Table \ref{tab:supercosmos_info} in at least two images.  Because the
search is for {\it stellar} companions, it is only necessary to be
able to detect a companion as faint as the faintest star in the
sample, effectively spectral type M9.5 V at 25 pc.  The two faintest
stars in the sample are DEN 0909-0658, with $VRI$ = 21.55, 19.46,
17.18 and RG0050-2722 with $VRI$ = 21.54, 19.09, 16.65.  The $B$
magnitudes for these stars are both fainter than the mag$\sim$20.5
limit of the $B$ plate, and thus neither star was detected in the $B$
image; however, their $R$ and $I$ magnitudes are both brighter than
the limits of those plates and the stars were identified in both the
$R$ and $I$ images. Ten other objects are too faint to be seen on the
$B$ plate, but as is the case with DEN0909-0658 and RG0050-2722, all
are bright enough for detection in the $R$ and $I$ images.


 
  
  
  \begin{figure}[ht!]
  \centering
  {\includegraphics[scale=0.33,angle=90]{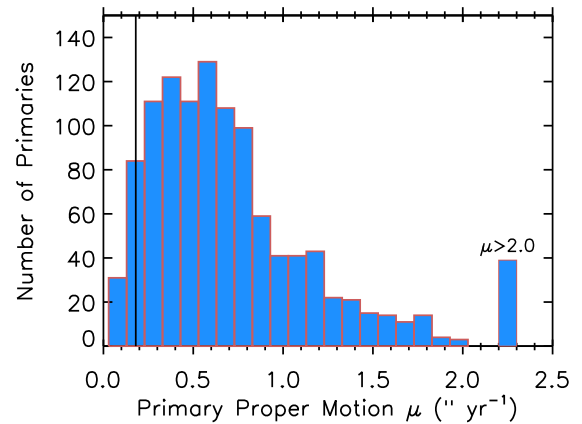}}

\vspace{0.25cm}

  \caption{Histogram of the proper motion of the primary (or single)
    component in each system, with the vertical line indicating $\mu$
    $=$ 0\farcs18 yr$^{-1}$, the canonical lower proper motion limit
    of Luyten's surveys. The majority (95\%) have proper motions,
    $\mu$, larger than 0\farcs18 yr$^{-1}$.  \label{fig:mu_hist}}
  \end{figure}

The epoch spread between the plates also needed to be large enough to
detect the primary star moving in order to then notice a companion
moving in tandem with it.  As shown in the histogram of proper motions
in Figure \ref{fig:mu_hist}, most of the survey stars move faster than
0\farcs18 yr$^{-1}$, the historical cutoff of Luyten's proper motion
surveys.  Hence, even a 10-year baseline provides 1\farcs8 of motion,
our adopted minimum proper motion detection limit, easily
discerned when blinking plates.  However, 58 of the stars in the
survey ($\sim$5\% of the sample) have $\mu$ $<$ 0\farcs18 yr$^{-1}$,
with the slowest star having $\mu$ $=$ 0\farcs03 yr$^{-1}$; for this
star, to detect a motion of 1\farcs8 the epoch spread would need to be
60 years. For 18 stars with Decl $-$20 $<$ $\delta$ $<$
$+$5$^{\circ}$, the older POSS-I plate (taken during 1950--1957) was
used for the slow-moving primaries. This extended the epoch spread by
8--24 years, enabling companions for these 18 stars to be detected,
leaving 40 slow-moving stars to search.

The proper motions of 151 additional primaries were not initially able
to be detected confidently because the epoch spread of the SuperCOSMOS
plates was less than 5 years. These 151 stars, in addition to the 40
stars with low $\mu$ mentioned above that were not able to be blinked
using the POSS plates, were compared to our newly-acquired $I-$band
images taken during the CCD Imaging Survey, extending the epoch spread
by almost twenty years in some cases. Wherever possible, the
SuperCOSMOS $I-$band image was blinked with our CCD $I-$band image,
but sometimes a larger epoch spread was possible with either the $B-$
or $R-$band plate images. In these cases, the plate that provided the
largest epoch spread was used. In order to upload these images to {\it
  Aladin} to blink with the archival SuperCOSMOS images, World
Coordinate System (WCS) coordinates were added to the header of each
image so that the two images could be aligned properly. This was done
using {\it SExtractor} for the CTIO/SMARTS 0.9m and the USNO 40in
images and the tools at Astrometry.net for the Lowell 42in and the
CTIO/SMARTS 1.0m images.


After using the various techniques outlined above to extend the image
epoch spreads, 1110 of 1120 stars were sucessfully searched in the
Blink Survey for companions. In ten cases, either the primary star's
proper motion was still undetectable, the available CCD images were
taken under poor sky conditions and many faint sources were not
visible, or the frame rotations converged poorly.  A primary result
from this Blink Survey is that in the separation regime from
10--300\arcsec, where the search is effectively complete, we find a
multiplicity rate of 4.7\% (as discussed in \S
\ref{subsec:corrections}).  Thus, we estimate that only 0.5 CPM
stellar companions (10 $\ast$ 4.7\%) with separations
10--300\arcsec~were missed due to not searching ten stars during the
Blinking Survey.

\subsection{CCD Imaging Survey: Companions at 2--10\arcsec}
\label{subsec:imaging_search}



To search for companions with separations 2--10\arcsec, astrometry
data were obtained at four different telescopes: in the northern
hemisphere, the Hall 42in telescope at Lowell Observatory and the USNO
40in telescope, both in Flagstaff, AZ, and in the southern hemisphere,
the CTIO/SMARTS 0.9m and 1.0m telescopes, both at Cerro Tololo
Inter-American Observatory in Chile. Each M dwarf primary was observed
in the $I_{KC}$ filter with integrations of 3, 30, and 300 seconds in
order to reveal stellar companions at separations 2--10\arcsec.  This
observational strategy was adopted to reveal any close equal-magnitude
companions with the short 3-second exposures, while the long
300-second exposures would reveal faint companions with masses at the
end of the main sequence.  The 30-second exposures were taken to
bridge the intermediate phase space. Calibration frames taken at the
beginning of each night were used for typical bias subtraction and
dome flat-fielding using standard $IRAF$ procedures. 




\begin{deluxetable}{lcccc}
\centering
\setlength{\tabcolsep}{0.03in}
\tablewidth{0pt}
\tabletypesize{\scriptsize}
\tablecaption{Telescopes Utilized for CCD Imaging Search and $VRI$ Photometry \label{tab:telescopes}}
\tablehead{\colhead{Telescope}           &
	   \colhead{FOV}                 &
 	   \colhead{Pixel Scale}         &
	   \colhead{\# Nights}           &
           \colhead{\# Objects}         }
\startdata
Lowell 42in      & 22.3\arcmin ~$\times$ 22.3\arcmin  & 0\farcs327 px$^{-1}$ & 21  &  508  \\ 
USNO 40in        & 22.9\arcmin ~$\times$ 22.9\arcmin  & 0\farcs670 px$^{-1}$ & 1   &   22  \\
CTIO/SMARTS 0.9m & 13.6\arcmin ~$\times$ 13.6\arcmin  & 0\farcs401 px$^{-1}$ & 16  &  442  \\ 
CTIO/SMARTS 1.0m & 19.6\arcmin ~$\times$ 19.6\arcmin  & 0\farcs289 px$^{-1}$ &  8  &  148  \\ 
\enddata

\end{deluxetable}

\begin{figure}
  \centering
  {\includegraphics[scale=0.35,angle=90]{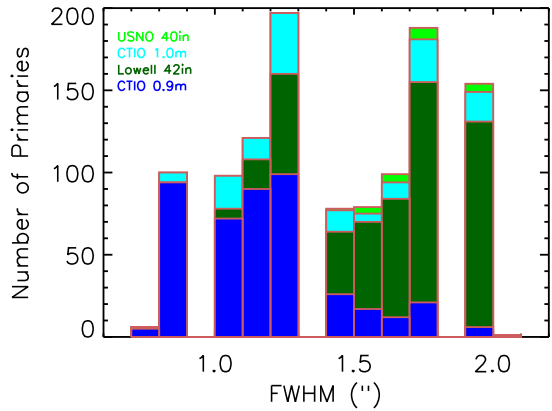}}
  \caption{The seeing FWHM measured for target star frames used in
    the $I-$band CCD imaging search.  The four different telescopes
    used are represented as royal blue for the CTIO/SMARTS 0.9m,
    dark green for the Lowell 42in, cyan for the CTIO/SMARTS 1.0m,
    and bright green for the USNO 40in.  Note the generally
    superior seeing conditions for targets observed at the
    0.9m. \label{fig:imag_info}}
  \end{figure}

Technical details for the cameras and specifics about the
observational setups and numbers of nights and stars observed at each
telescope are given in Table \ref{tab:telescopes}.  The telescopes
used for the imaging campaign all have primary mirrors roughly 1m in
size and have CCD cameras that provide similar pixel scales.  Data
from all telescopes were acquired without binning pixels.  The
histogram in Figure \ref{fig:imag_info} illustrates the seeing
measured for the best images of each star surveyed at the four
different telescopes.  Seeing conditions better than 2\arcsec~were
attained for all but one star, GJ 507, with some stars being observed
multiple times.  While the 0.9m has a slightly larger pixel scale than
the 1.0m and the 42in, as shown in Figure \ref{fig:imag_info}, the
seeing was typically better at that site, allowing for better
resolution.  Only 22 primaries (fewer than 2\% of the survey) were
observed at the USNO 40in, so we do not consider the coarser pixel
scale to have significantly affected the survey.  Overall, the data
from the four telescopes used were of similar quality and the results
could be combined without modification.

 
  
A few additional details of the observations are worthy of note:

$\bullet$ A total of 442 stars were observed at the CTIO/SMARTS 0.9m
telescope, where consistently good seeing, telescope operation, and
weather conditions make observations at this site superior to those at
the other telescopes used, as illustrated in Figure
\ref{fig:imag_info}.

$\bullet$ While being re-aluminized in December 2012, the primary
mirror at the Lowell 42in was dropped and damaged. The mask that was
installed over the damaged mirror as a temporary fix resulted in a PSF
flare before a better mask was installed that slightly improved the
PSF.  Of the 508 stars observed for astrometry at Lowell, 457 were
observed before the mishap and 51 after.


$\bullet$ $I-$band images at both the Lowell 42in and CTIO/SMARTS 1.0m
suffer from fringing, the major cause of which is night sky emission
from atmospheric OH molecules.  This effect sometimes occurs with
back-illuminated CCDs at optical wavelengths longer than roughly 700
nm where the light is reflected several times between the internal
front and back surfaces of the CCD, creating constructive and
destructive interference patterns, or fringing \citep{Howell(2000),
  Howell(2012)}.  In order to remove these fringes, $I-$band frames
from multiple nights with a minimum of saturated stars in the frame
were selected, boxcar smoothed, and then average-combined into a
fringe map.  This fringe map was then subtracted from all $I-$band
images using a modified IDL code originally crafted by
\citet{Snodgrass(2013)}.

 

Four new companions were discovered during this portion of the
survey. Details on these new companions are given in \S
\ref{subsec:newcomp}. In each case, archival SuperCOSMOS plates were
blinked to eliminate the possibility that new companions were
background objects. We detected 32 companions with separations
  2--10\arcsec, as well as 12 companions with $\rho <$ 2\arcsec,
  including the four noted above.


\subsubsection{CCD Imaging Survey Detection Limits}
\label{subsubsec:det_lim_image}

The $M_I$ range of the M dwarf sequence is roughly 8 magnitudes ($M_I
=$ 6.95 --- 14.80 mag, specifically, for our sample). Therefore, an
analysis of the detection limits of the CCD imaging campaign was done
for objects with a range of $I$ magnitudes at $\rho$ $=$
1--5\arcsec~and at $\Delta$mags $=$ 0 -- 8 in one-magnitude increments
for different seeing conditions at the two main telescopes where the
bulk (85\%) of the stars were imaged: the CTIO/SMARTS 0.9m and the
Lowell 42in.  While the companion search in the CCD frames extended to
10\arcsec, sources were detected even more easily at separations 5 --
10\arcsec ~than at 5\arcsec, so it was not deemed necessary to perform
the analysis for the larger separations.

Because the apparent $I-$band magnitudes for the stars in the sample
range from 5.32--17.18 (as shown in Figure \ref{fig:imags}), objects
with $I-$band magnitudes of approximately 8, 12, and 16 were selected
for investigation.  Only 88 primaries (7.8\% of the sample) have $I <$
8, so it was not felt necessary to create a separate set of
simulations for these brighter stars.  The stars used for the
detection limit analysis are listed in Table \ref{tab:det_lim_stars}
with their $I$ magnitudes, the FWHM at which they were observed and at
which telescope, and any relevant notes.



\begin{deluxetable}{lcccl}
\centering
\setlength{\tabcolsep}{0.03in}
\tablewidth{0pt}
\tabletypesize{\scriptsize}
\tablecaption{Stars Used for Imaging Search Detection Limit Study \label{tab:det_lim_stars}}
\tablehead{\colhead{Name}                &
	   \colhead{$I$}                 &
 	   \colhead{FWHM}                &
	   \colhead{Tel}                 &
           \colhead{Note}               \\
	   \colhead{   }                 &
	   \colhead{(mag)}               &
	   \colhead{(arcsec)}            &
           \colhead{     }               &
           \colhead{     }               }

\startdata
GJ~285                      &  8.24            & 0.8            & 0.9m    &          \\ 
LP~848-50AB                 & 12.47J           & 0.8            & 0.9m    & $\rho$$_{AB}$ $<$2\arcsec         \\ 
SIP~1632-0631               & 15.56            & 0.8            & 0.9m    &          \\ 
L~32-9A                     &  8.04            & 1.0            & 0.9m    & $\rho$$_{AB}$ $=$ 22\farcs40     \\
SCR~0754-3809               & 11.98            & 1.0            & 0.9m    &          \\
BRI~1222-1221               & 15.59            & 1.0            & 0.9m    &          \\
GJ~709                      &  8.41            & 1.0            & 42in    &          \\
GJ~1231                     & 12.08            & 1.0            & 42in    &          \\
Reference Star              & 16 (scaled)      & 1.0            & 42in    &          \\
GJ~2060AB                   &  7.83J           & 1.5            & 0.9m    & $\rho$$_{AB}$ $=$ 0\farcs485         \\
2MA~2053-0133               & 12.46            & 1.5            & 0.9m    &          \\
Reference Star              & 16 (scaled)      & 1.5            & 0.9m    &          \\
GJ~109                      &  8.10            & 1.5            & 42in    &          \\
LHS~1378                    & 12.09            & 1.5            & 42in    &          \\
2MA~0352+0210               & 16.12            & 1.5            & 42in    &          \\
Reference Star              &  8 (scaled)      & 1.8            & 0.9m    &          \\
SCR~2307-8452               & 12.00            & 1.8            & 0.9m    &          \\
Reference Star              & 16 (scaled)      & 1.8            & 0.9m    &          \\
GJ~134                      &  8.21            & 1.8            & 42in    &          \\
LHS~1375                    & 12.01            & 1.8            & 42in    &          \\
SIP~0320-0446AB             & 16.37            & 1.8            & 42in    & $\rho$$_{AB}$ $<$0\farcs33         \\
GJ~720A                     &  8.02            & 2.0            & 42in    & $\rho$$_{AB}$ $=$ 112\farcs10      \\
LHS~3005                    & 11.99            & 2.0            & 42in    &          \\
2MA~1731+2721               & 15.50            & 2.0            & 42in    &          \\
\enddata

\tablecomments{The `J' on the $I-$band magnitudes of LP~848-50AB and
  GJ~2060AB indicates that the photometry includes light from the
  companion. The other sub-arcsecond binary, SIP~0320-0446AB, has a
  brown dwarf companion that does not contribute significant light to
  the photometry of its primary star.}

\end{deluxetable}

Each of the selected test stars was analyzed in seeing conditions of
1\farcs0, 1\farcs5, and 1\farcs8, but because the seeing at CTIO is
typically better than that at Anderson Mesa, we were able to push to
0\farcs8~for the 0.9m, and had to extend to 2\farcs0~for the Lowell
42in.  These test stars were verified to have no known {\it
  detectable} companions within the 1--5\arcsec~separations explored
in this part of the project.  We note that one of the targets examined
for the best resolution test, LP 848-050AB, has an astrometric
perturbation due to an unseen companion at an unknown separation, but
that in data with a FWHM of 0.8\arcsec, the two objects were still not
resolved.  As the detection limit determination probes separations
1--5\arcsec, using this star does not affect the detection limit
analysis.  The other binaries used all had either larger or smaller
separations than the 1--5\arcsec~regions explored, effectively making
them point sources.

The $IDL$ {\sc shift} task was used to shift and add the science star
as a proxy for an embedded companion, scaled by a factor of 2.512 for
each magnitude difference.  In cases where the science star was
saturated in the frame, a reference star was selected from the shorter
exposure taken in similar seeing in which the science star was not
saturated.  Its relative magnitude difference was calculated so that
it could be scaled to the desired brightness in the longer exposure,
and then it was embedded for the analysis.  In all cases, the
background sky counts were subtracted before any scaling was done.

 
  \begin{figure}
  \centering
  
  \includegraphics[scale=0.40,angle=0]{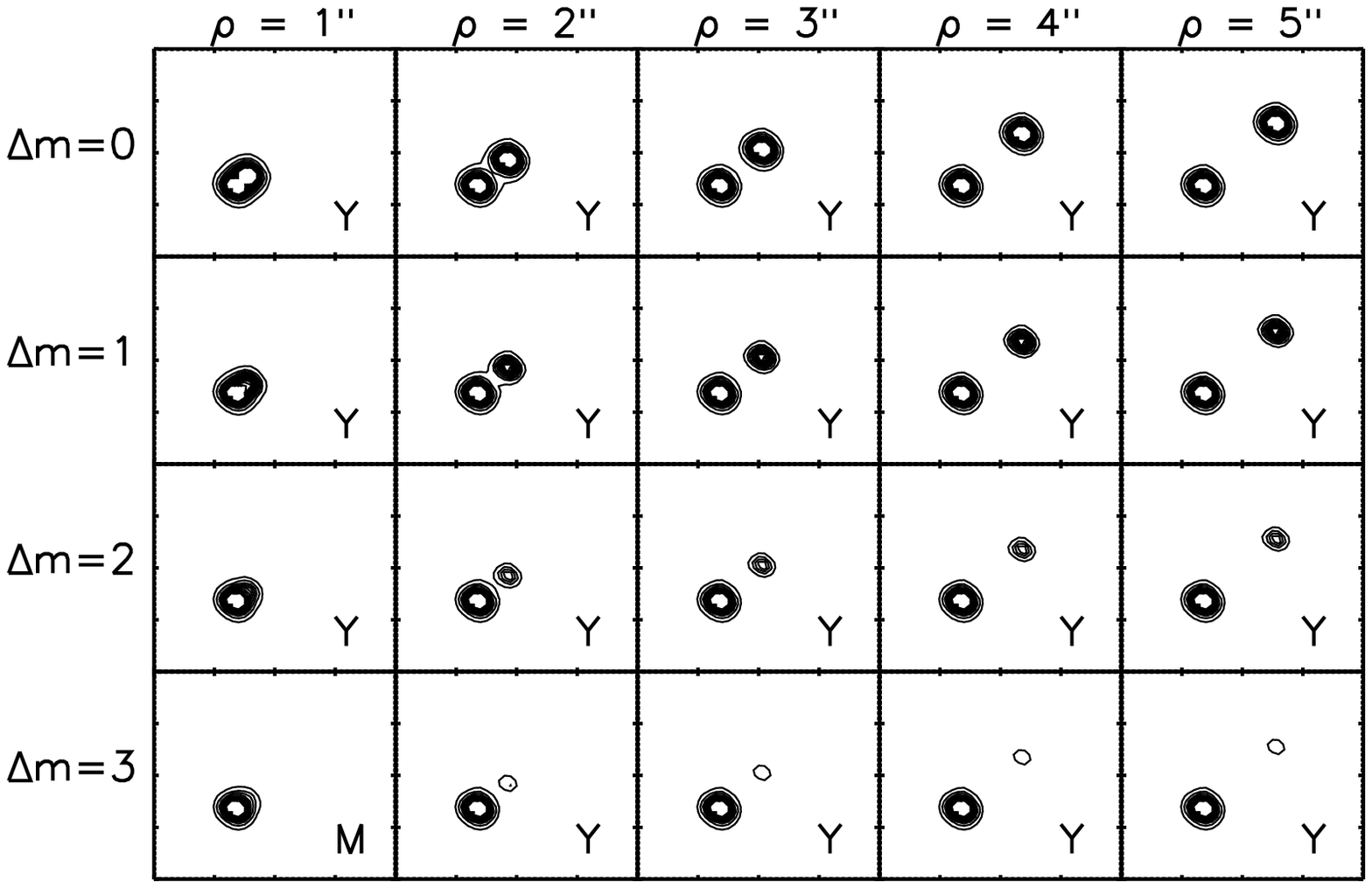}
  
  \vspace{-1.15cm}
  \includegraphics[scale=0.40,angle=0]{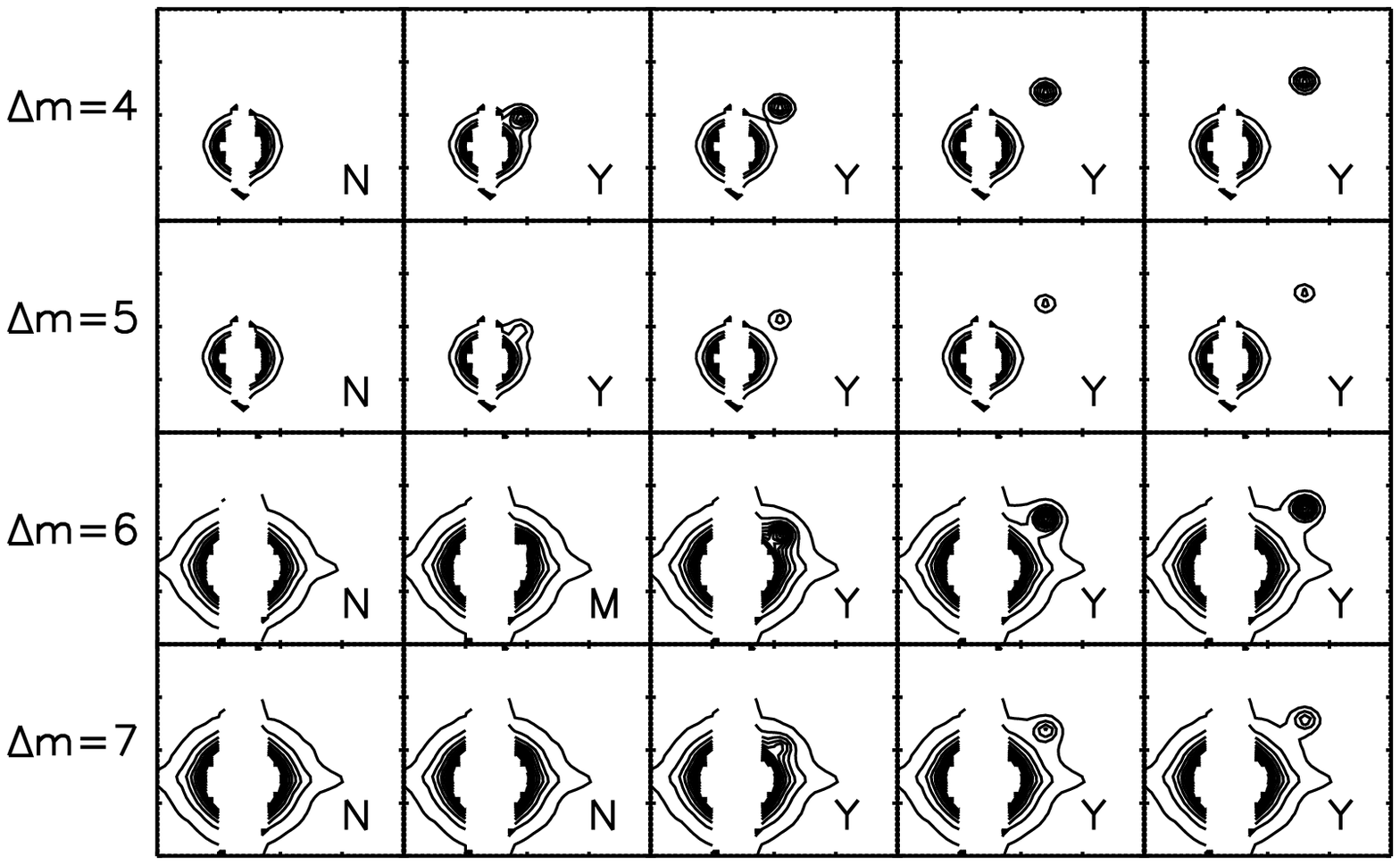}
  
  \vspace{-1.15cm}
  \includegraphics[scale=0.40,angle=0]{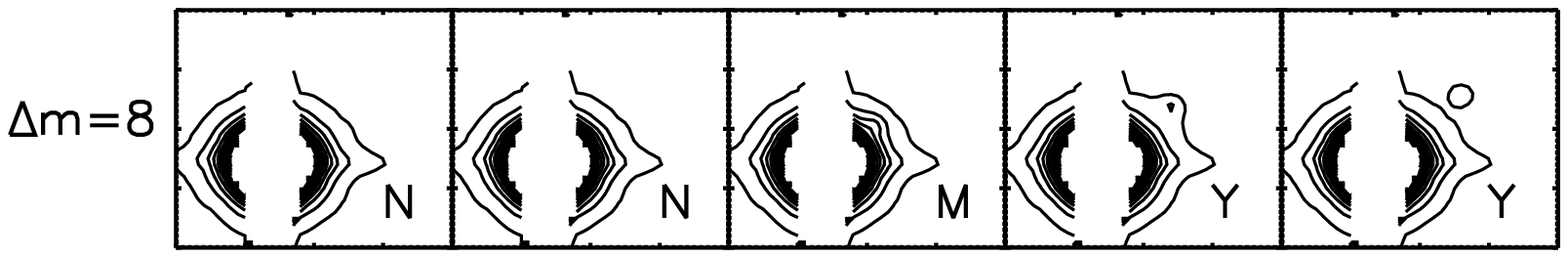}
  
  \vspace{-3.0cm}
  \caption{Detection Limits for the CTIO/SMARTS 0.9m: Contour plots
    for L~32-9A, with $I =$ 8.04 in 1\farcs0 seeing conditions for an
    embedded companion at $\rho$ $=$ 1--5\arcsec~with $\Delta$mags $=$
    0--8.  The Y, N, and M labels indicate $yes$, $no$, or $maybe$ for
    whether or not the embedded companion is detectable. The 3-second
    exposure was used for $\Delta$mags $=$ 0--3, the 30-second
    exposure was used for $\Delta$mags $=$ 4--5, and the 300-second
    exposure was used for $\Delta$mags $=$ 6--8.  Thirty-five
    simulated companions were detected, seven were undetectable, and
    three were possibly detected. \label{fig:contour1_ctio}}
  \end{figure}

 
  \begin{figure}
  \centering
  
  \includegraphics[scale=0.40,angle=0]{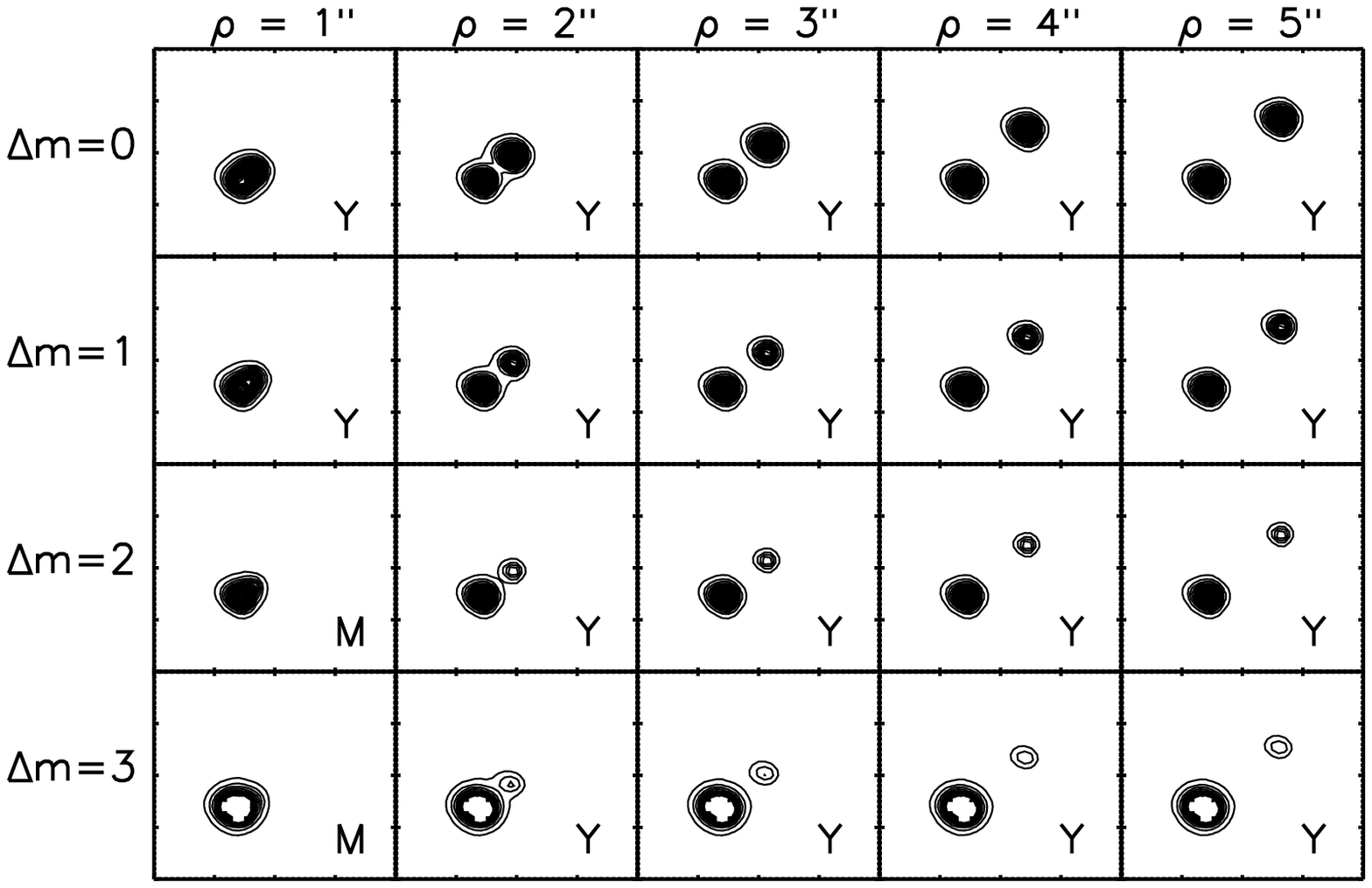}
  
  \vspace{-1.15cm}
  \includegraphics[scale=0.40,angle=0]{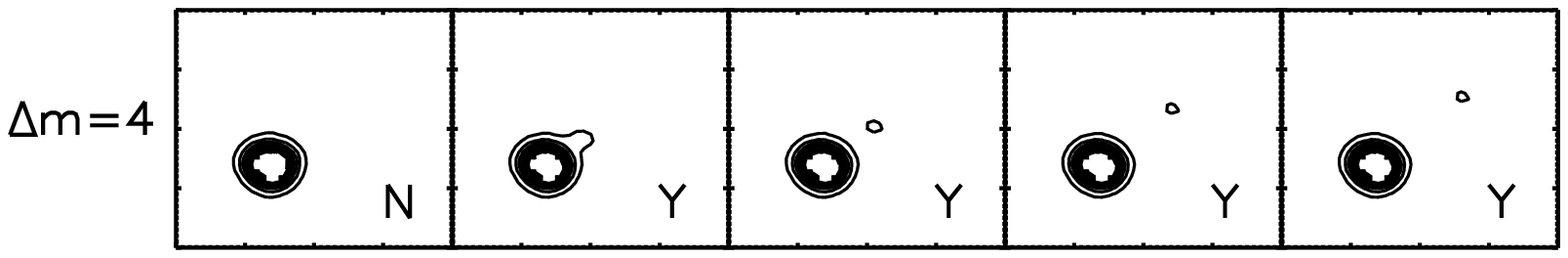}
  
  \vspace{-3.0cm}
 
  \caption{Detection Limits for the CTIO/SMARTS 0.9m: Contour plots
    for SCR~0754-3809, with $I =$ 11.98 mag at 1\farcs0 seeing
    conditions for an embedded companion at $\rho$ $=$ 1$-$5\arcsec
    ~with $\Delta$mags $=$ 0--4. The Y, N, and M labels indicate
    $yes$, $no$, or $maybe$ for whether or not the embedded companion
    is detectable. The 30-second exposure was used for $\Delta$mags
    $=$ 0, 1, 2, and the 300-second exposure was used for $\Delta$mags
    $=$ 3, 4. Twenty-two simulated companions were detected, one was
    undetectable, and two were possibly
    detected. \label{fig:contour2_ctio}}
  
  \end{figure}
 
 
  \begin{figure}
  \centering
  \includegraphics[scale=0.40,angle=0]{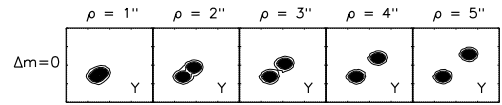}
  
  \vspace{-3.0cm}
  \caption{Detection Limits for the CTIO/SMARTS 0.9m: Contour plots
    for BRI~1222-1221, with $I =$ 15.59 mag at 1\farcs0 seeing
    conditions for an embedded companion at $\rho$ $=$ 1$-$5\arcsec
    ~with $\Delta$mags $=$ 0. The Y, N, and M labels indicate $yes$,
    $no$, or $maybe$ for whether or not the embedded companion is
    detectable. The 300-second exposure was used.  All 5 simulated
    companions were detected. \label{fig:contour3_ctio}}
  \end{figure}

Stars with $I$ = 8 were searched for companion sources in $IRAF$ via
radial and contour plots using the 3-second exposure to probe
$\Delta$$I$ $=$ 0, 1, 2, 3, the 30-second exposure for $\Delta$$I$ $=$
4, 5, and the 300-second exposure for $\Delta$$I$ $=$ 6, 7, 8.
Similarly, twelfth magnitude stars were probed at $\Delta$$I$ $=$ 0,
1, 2, 3 using the 30-second exposure and at $\Delta$$I$ $=$ 4 with
the 300-second frame.  Finally, the 300-second exposure was used to
explore the regions around the sixteenth magnitude objects for
evidence of a stellar companion at $\Delta$$I$ $=$ 0.


In total, 600 contour plots were made using $IDL$ and inspected by
eye.  A subset of 75 example plots for stars with $I$ = 8.04, 11.98,
and 15.59 observed in seeing conditions of 1\farcs0 at the 0.9m are
shown in Figures \ref{fig:contour1_ctio} - \ref{fig:contour3_ctio}.
The `Y', `N', and `M' labels in each plot indicate $yes$, $no$, or
$maybe$ for whether or not the injected synthetic companion was
detectable by eye at the separation, magnitude, and seeing conditions
explored.  As can be seen, the target star with $I$ = 8.04 is highly
saturated in the frames used for $\Delta$$I$ greater than 4.  Overall,
the companion can be detected in 62 of the 75 simulations, not
detected in eight cases, and possibly detected in five more cases.
The conditions in which the companion remains undetected in some cases
are at small $\rho$ and at $\Delta$$I$ $>$ 4, typically around bright
stars. Note that these images do not stand alone --- contour plots for
target stars are also compared to plots for other stars in the frames,
allowing an additional check to determine whether or not the star in
question is multiple.


\begin{deluxetable}{lcccccc}
\centering
\setlength{\tabcolsep}{0.03in}
\tablewidth{0pt}
\tabletypesize{\tiny}
\tablecaption{Imaging Search Detection Limit Summary \label{tab:det_lim_sum}}
\tablehead{\colhead{Seeing}              &
	   \colhead{Yes}                 &
 	   \colhead{ No }                &
	   \colhead{Maybe}               &
	   \colhead{Yes}                 &
 	   \colhead{ No }                &
	   \colhead{Maybe}               \\
	   \colhead{Conditions}          &
	   \colhead{(\#)}                &
	   \colhead{(\#)}                &
           \colhead{(\#)}                &
	   \colhead{(\#)}                &
	   \colhead{(\#)}                &
           \colhead{(\#)}                }

\startdata
                     &            & 0.9m     &         &             & 42in    &         \\
\hline                                          
FWHM $=$ 0\farcs8    & 64         & 8        &  3      & \nodata     & \nodata & \nodata \\
\hline                                             
$I$ $=$ 8 mag        & 36         & 7        & 2       & \nodata     & \nodata & \nodata \\
$I$ $=$ 12 mag       & 23         & 1        & 1       & \nodata     & \nodata & \nodata  \\
$I$ $=$ 16 mag       &  5         & \nodata  & \nodata & \nodata     & \nodata & \nodata  \\
\hline                                                
FWHM $=$ 1\farcs0    & 62         & 8        &  5      & 60          & 12      &  3       \\ 
\hline                                             
$I$ $=$ 8 mag        & 35         & 7        &  3      & 34          & 8       &  3        \\
$I$ $=$ 12 mag       & 22         & 1        &  2      & 21          & 4       & \nodata  \\
$I$ $=$ 16 mag       &  5         & \nodata  & \nodata &  5          & \nodata & \nodata  \\
\hline                                                 
FWHM $=$ 1\farcs5    & 58         & 12       &  5      & 55          & 12      &  8       \\ 
\hline                                             
$I$ $=$ 8 mag        & 33         & 9        &  3      & 33          & 6       &  6       \\
$I$ $=$ 12 mag       & 20         & 3        &  2      & 17          & 6       &  2       \\
$I$ $=$ 16 mag       &  5         & \nodata  & \nodata &  5          & \nodata & \nodata  \\
\hline                                                 
FWHM $=$ 1\farcs8    & 50         & 18       &  7      & 52          & 14      &  9       \\
\hline                                             
$I$ $=$ 8 mag        & 28         & 13       &  4      & 29          & 10      &  6       \\
$I$ $=$ 12 mag       & 18         &  5       &  2      & 19          & 4       &  2       \\
$I$ $=$ 16 mag       &  4         & \nodata  &  1      &  4          & \nodata &  1       \\
\hline                                            
FWHM $=$ 2\farcs0    & \nodata    & \nodata  & \nodata & 46          & 17      & 12        \\
\hline                                             
$I$ $=$ 8 mag        & \nodata    & \nodata  & \nodata & 24          & 12      &  9         \\
$I$ $=$ 12 mag       & \nodata    & \nodata  & \nodata & 18          & 5       &  2         \\
$I$ $=$ 16 mag       & \nodata    & \nodata  & \nodata &  4          & \nodata &  1         \\
\hline                                                                                      
TOTAL                & 234        & 46       & 20      & 213         & 55      & 32         \\ 
\enddata

\end{deluxetable}

The full range of $\Delta$$I$ for the M dwarf sequence is roughly
eight magnitudes, so $\Delta$$I >$ 8 represents detections of early L
dwarf and brown dwarf companions.  There were no companions detected
with $\Delta$$I$ $>$ 8 around the brighter stars in the simulations,
indicating that this survey was {\it not} sensitive to these types of
faint companions at separations 1--5\arcsec~around the brightest M
dwarfs in the sample, although they would be detected around many of
the fainter stars (none were found).


Table \ref{tab:det_lim_sum} presents a summary of the results of the
detections of the embedded companions.  Overall, the simulated
companions were detected 75\% of the time for all brightness ratios on
both telescopes, were not detected 17\% of the time, and were possibly
detected in 9\% of the simulations.  For the simulations of bright
stars with $I =$ 8, 70\% of the embedded companions were detected.
For stars with $I =$ 12, companions were detected in 79\% of the time,
and for the faint stars with $I =$ 16, companions were detected in
93\% of the cases tested. At $\rho$ $=$ 1\arcsec, the embedded
companions were detected in 28\% of cases, not detected in 52\% of
cases, and possibly detected in 20\% of cases. Thus, we do not claim
high sensitivity at separations this small. In total, for $\rho$
$\geq$ 2\arcsec, we successfully detected the simulated companions
86\% of the time, did not detect them 8\% of the time, and possibly
detected them 6\% of the time.


We note that this study was not sensitive to companions with large
$\Delta$mags at separations $\sim$1 --- 2\arcsec ~from their
primaries. While the long exposure $I-$band images obtained during the
direct imaging campaign would likely reveal fainter companions at
$\rho$ $\sim$2 --- 5\arcsec, the saturation of some of the observed
brighter stars creates a CCD bleed along columns in the direction in
which the CCDs read out. Faint companions located within $\sim$1---
2\arcsec ~of their primaries, but at a position angle near 0$^{\circ}$
or 180$^{\circ}$ would be overwhelmed by the CCD bleed of the
saturated star and not be detected. We do not include any correction
due to this bias, as it mostly applies to companions at separations
$<$ 2\arcsec ~from their primaries, below our stated detection limit
sensitivity.

 
  \begin{figure}
  \centering


  {\includegraphics[scale=0.35,angle=90]{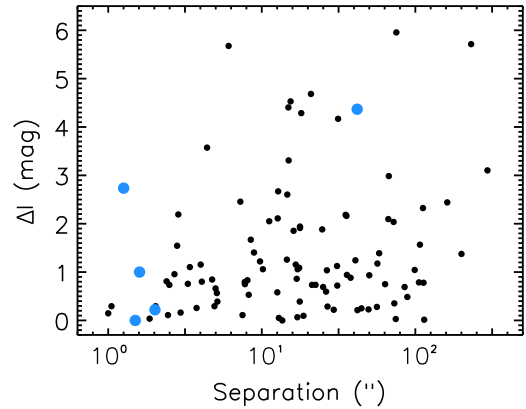}}

  \caption{Log-linear plot of $\Delta$$I$ vs.~angular separation to
    illustrate the observational limits of our Blinking and CCD
    Imaging Surveys. Solid black points indicate known companions that
    were confirmed, while the new companions discovered during our
    searches are shown as larger blue points. \label{fig:rho_deli}}
  \end{figure}


\subsubsection{Detection Limits Summary}
\label{subsubsec:det_lim_sum}

Figure \ref{fig:rho_deli} illustrates detected companions in the
Blinking and CCD Imaging Surveys, providing a comparison for the
detection limits derived here.  We note that the largest $\Delta$$I$
detected was roughly 6.0 mag (GJ 752B), while the largest
angular separation detected was 295\arcsec ~(GJ 49B).

\begin{figure}
  \centering

  {\includegraphics[scale=0.35,angle=270]{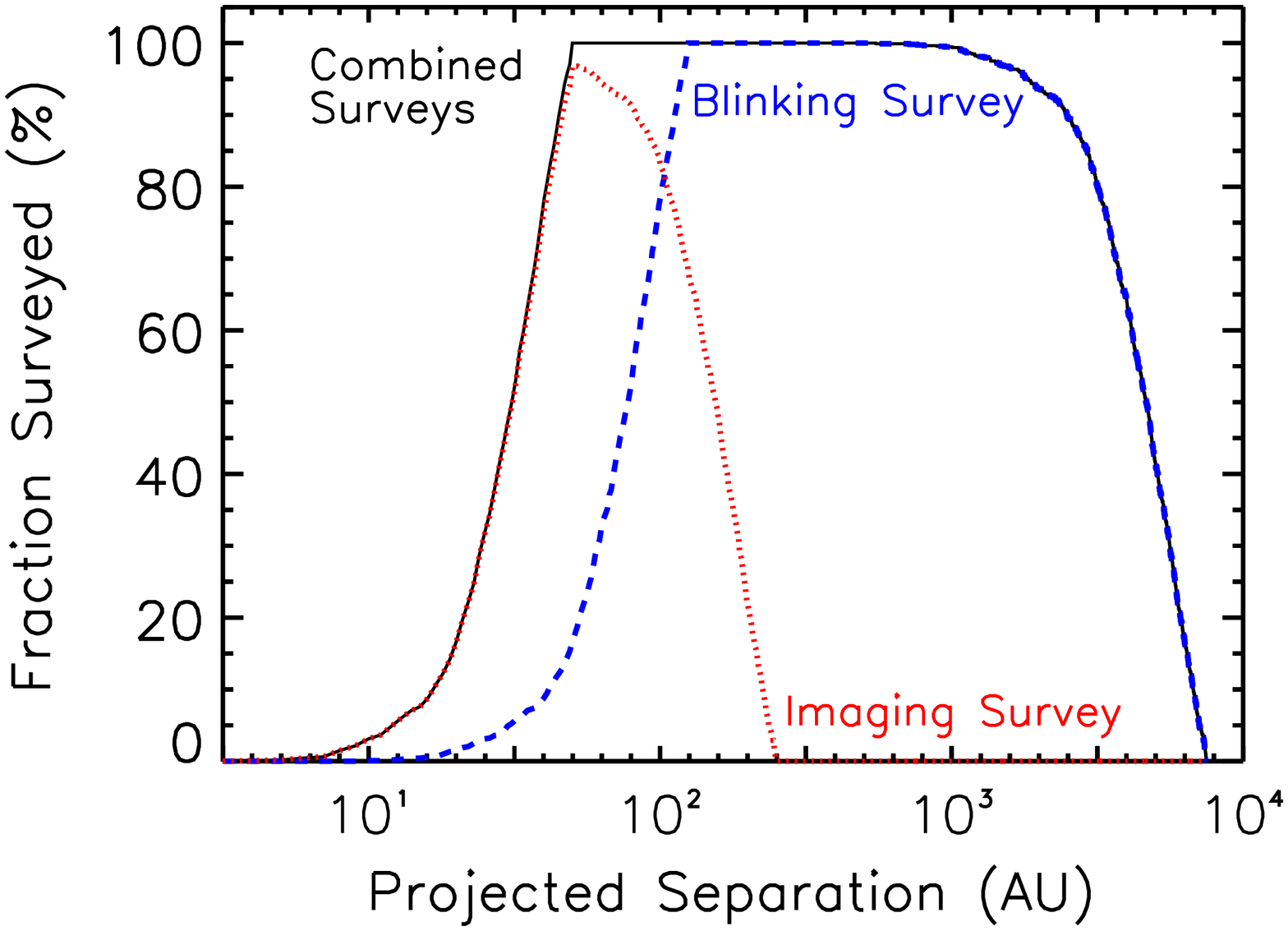}}

  \caption{The fraction of our sample surveyed as a function of
    log-projected linear separation. The curve for the imaging
    campaign is shown as a red dotted line, while the blinking
    campaign coverage is shown as a blue dashed line. The solid black
    line indicates the combined coverage of the two campaigns. We show
    that our surveys are complete for stellar companions at projected
    linear separations of roughly 50 - 1000 AU, 75\% complete at
    separations 40 - 3000 AU, and 50\% complete at separations 30 -
    4000 AU.
     \label{fig:complete}}
  \end{figure}

Figure \ref{fig:complete} indicates the coverage curves for our two
main surveys as a function of projected linear separation. Using
  the angular separation limits of each survey (2 --- 10\arcsec ~ for
  the imaging survey and 5 --- 300\arcsec ~for the blinking survey)
  and the trigonometric distances of each object to determine the
  upper and lower projected linear separation limit for each M dwarf
  primary in our sample, we show that either the imaging or blinking
  survey would have detected stellar companions at projected distances
  of 50 - 1000 AU for 100\% of our sample.





 
\begin{deluxetable*}{lclcccccccccc}
\centering
\setlength{\tabcolsep}{0.03in}
\tablewidth{0pt}
\tabletypesize{\scriptsize}
\tablecaption{Multiplicity Information for Sample \label{tab:multinfo}}
\tablehead{\colhead{Name}                &
	   \colhead{\# Obj}              &
 	   \colhead{Map}                 &
	   \colhead{RA}                  &
           \colhead{DEC}                 &
	   \colhead{$\rho$}              &
	   \colhead{$\theta$}            &
           \colhead{Year}                &
	   \colhead{Technique}           &
           \colhead{Ref}                 &
           \colhead{$\Delta$mag}         &
           \colhead{Filter}              &
           \colhead{Ref}                 \\
	   \colhead{   }                 &
	   \colhead{   }                 &
	   \colhead{   }                 &
           \colhead{(hh:mm:ss)}          &
           \colhead{(dd:mm:ss)}           &
	   \colhead{(\arcsec)}           &
           \colhead{(deg)}               &
           \colhead{     }               &
           \colhead{     }               &
           \colhead{     }               &
           \colhead{(mag)}               &
           \colhead{    }                &
           \colhead{    }                }

\startdata                                       
 GJ~1001             &  0      &  BC    & 00 04 34.87 & $-$40 44 06.5 &      0.087   & 048  &  2003  &  HSTACS &   40   &   0.01 &  222    &   40     \\ 
 GJ~1001             &  3      &  A-BC  & 00 04 36.45 & $-$40 44 02.7 &     18.2     & 259  &  2003  &  visdet &   40   &   9.91 & $V_J$   &    1     \\
 G~131-26            &  2      &  AB    & 00 08 53.92 & $+$20 50 25.4 &      0.111   & 170  &  2001  &  AO det &   13   &   0.46 &   H     &   13     \\
 GJ~11               &  2      &  AB    & 00 13 15.81 & $+$69 19 37.2 &      0.859   & 089  &  2012  &  lkydet &   62   &   0.69 &   i'    &   62     \\
 LTT~17095           &  2      &  AB    & 00 13 38.74 & $+$80 39 56.8 &     12.78    & 126  &  2001  &  visdet &  103   &   3.63 & $V_J$   &    1     \\
 GJ~1005             &  2      &  AB    & 00 15 28.06 & $-$16 08 01.8 &      0.329   & 234  &  2002  &  HSTNIC &   30   &   2.42 & $V_J$   &    9     \\
 2MA~0015-1636       &  2      &  AB    & 00 15 58.07 & $-$16 36 57.8 &      0.105   & 090  &  2011  &  AO det &   18   &   0.06 &   H     &   18     \\
 L~290-72            &  2      &  AB    & 00 16 01.99 & $-$48 15 39.3 &   $<$1       & ...  &  2007  &    SB1  &  117   &   ...  &  ...    &  ...     \\ 
 GJ~1006             &  2      &  AB    & 00 16 14.62 & $+$19 51 37.6 &     25.09    & 059  &  1999  &  visdet &  103   &   0.94 & $V_J$   &  111     \\
 GJ~15               &  2      &  AB    & 00 18 22.88 & $+$44 01 22.7 &     35.15    & 064  &  1999  &  visdet &  103   &   2.97 & $V_J$   &    1     \\
\enddata

\tablecomments{The first 10 lines of this Table are shown to
  illustrate its form and content.}

\tablecomments{The codes for the techniques and instruments used to
  detect and resolve systems are: AO det --- adaptive optics; astdet
  --- detection via astrometric perturbation, companion often not
  detected directly; astorb --- orbit from astrometric measurements;
  HSTACS --- {\it Hubble Space Telescope's} Advanced Camera for
  Surveys; HSTFGS --- {\it Hubble Space Telescope's} Fine Guidance
  Sensors; HSTNIC --- {\it Hubble Space Telescope's} Near Infrared
  Camera and Multi-Object Spectrometer; HSTWPC --- {\it Hubble Space
  Telescope's} Wide Field Planetary Camera 2; lkydet --- detection via
  lucky imaging; lkyorb --- orbit from lucky imaging measurements;
  radorb --- orbit from radial velocity measurements; radvel ---
  detection via radial velocity, but no SB type indicated; SB (1, 2,
  3) --- spectroscopic multiple, either single-lined, double-lined, or
  triple-lined; spkdet --- detection via speckle interferometry;
  spkorb --- orbit from speckle interferometry measurements; visdet
  --- detection via visual astrometry; visorb --- orbit from visual
  astrometry measurements}

\tablerefs{
(1) this work; 
(2) \citet{Allen(2008)}; 
(3) \citet{AlShukri(1996)};
(4) \citet{Balega(2007)}; 
(5) \citet{Balega(2013)}; 
(6) \citet{Bartlett(2017)}; 
(7) \citet{Benedict(2000b)}; 
(8) \citet{Benedict(2001)}; 
(9) \citet{Benedict(2016)};
(10) \citet{Bergfors(2010)};
(11) \citet{Bessel(1990)};
(12) \citet{Bessell(1991)};
(13) \citet{Beuzit(2004)}; 
(14) \citet{Biller(2006)}; 
(15) \citet{Blake(2008)};
(16) \citet{Bonfils(2013)}; 
(17) \citet{Bonnefoy(2009)}; 
(18) \citet{Bowler(2015)}; 
(19) \citet{Burningham(2009)}; 
(20) \citet{Chaname(2004)}; 		    
(21) \citet{Cortes-Contreras(2014)}; 	    
(22) \citet{Cvetkovic(2015)}; 		    
(23) \citet{Daemgen(2007)}; 		    
(24) \citet{Dahn(1988)}; 		    
(25) \citet{Davison(2014)}; 		    
(26) \citet{Dawson(2005)}; 		    
(27) \citet{Delfosse(1999c)}; 		    
(28) \citet{Delfosse(1999d)}; 		    
(29) \citet{Diaz(2007)}; 		    
(30) \citet{Dieterich(2012)}; 		    
(31) \citet{Docobo(2006a)}; 		    
(32) \citet{Doyle(1990)};		    
(33) \citet{Duquennoy(1988b)}; 		    
(34) \citet{Femenia(2011)}; 		    
(35) \citet{Forveille(2005)}; 		    
(36) \citet{Freed(2003)}; 		    
(37) \citet{Fu(1997)}; 			    
(38) \citet{Gizis(1998b)}; 		    
(39) \citet{Gizis(2002)}; 		    
(40) \citet{Golimowski(2004)}; 		    
(41) \citet{Harlow(1996)}; 		    
(42) \citet{Harrington(1985)}; 		    
(43) \citet{Hartkopf(2012)}; 		    
(44) \citet{Heintz(1985)};		    
(45) \citet{Heintz(1987)};		    
(46) \citet{Heintz(1990)}; 		    
(47) \citet{Heintz(1991)};		    
(48) \citet{Heintz(1992a)}; 		    
(49) \citet{Heintz(1993)}; 		    
(50) \citet{Heintz(1994)}; 		    
(51) \citet{Henry(1999)};		    
(52) \citet{Henry(2006)}; 		    
(53) \citet{Henry(2018)};                   
(54) \citet{Herbig(1965)};		     
(55) \citet{Horch(2010)}; 		     
(56) \citet{Horch(2011a)}; 		     
(57) \citet{Horch(2012a)};		     
(58) \citet{Horch(2015)};		     
(59) \citet{Ireland(2008)}; 		     
(60) \citet{Janson(2012)};		     
(61) \citet{Janson(2014a)}; 		     
(62) \citet{Janson(2014b)}; 		     
(63) \citet{Jao(2003)}; 		     
(64) \citet{Jao(2009)}; 		     
(65) \citet{Jao(2011)}; 		     
(66) \citet{Jenkins(2009)}; 		     
(67) \citet{Jodar(2013)}; 		     
(68) \citet{Kohler(2012)}; 		     
(69) \citet{Kurster(2009)}; 		     
(70) \citet{Lampens(2007)}; 		     
(71) \citet{Law(2006b)}; 		     
(72) \citet{Law(2008)};			     
(73) \citet{Leinert(1994)};		     
(74) \citet{Lepine(2009)}; 		     
(75) \citet{Lindegren(1997)}; 		     
(76) \citet{Luyten(1979a)};                  
(77) \citet{Malo(2014)}; 		  
(78) \citet{Martin(2000a)}; 		  
(79) \citet{Martinache(2007)}; 		  
(80) \citet{Martinache(2009)};		  
(81) \citet{Mason(2009a)}; 		  
(82) \citet{Mason(2018)};                 
(83) \citet{McAlister(1987c)}; 
(84) \citet{Montagnier(2006)}; 	       
(85) \citet{Nidever(2002)}; 	       
(86) \citet{Pravdo(2004)}; 	       
(87) \citet{Pravdo(2006)}; 	       
(88) \citet{Reid(2001a)}; 	       
(89) \citet{Reid(2002)};	       
(90) \citet{Reiners(2010)}; 	       
(91) \citet{Reiners(2012)}; 	       
(92) \citet{Riddle(1971)}; 	       
(93) \citet{Riedel(2010)}; 	       
(94) \citet{Riedel(2014)}; 	       
(95) \citet{Riedel(2018)};             
(96) \citet{Salim(2003)}; 	        
(97) \citet{Schneider(2011)}; 	       
(98) \citet{Scholz(2010b)}; 	       
(99) \citet{Segransan(2000)}; 	       
(100) \citet{Shkolnik(2010)}; 	        
(101) \citet{Shkolnik(2012)}; 	        
(102) \citet{Siegler(2005)}; 	        
(103) \citet{Skrutskie(2006)};          
(104) \citet{Tokovinin(2012c)};      
(105) \citet{vanBiesbroeck(1974)};   
(106) \citet{vanDessel(1993)}; 	     
(107) \citet{Wahhaj(2011)}; 	     
(108) \citet{Ward-Duong(2015)};
(109) \citet{Weis(1991a)};
(110) \citet{Weis(1993)};
(111) \citet{Weis(1996)};      
(112) \citet{Winters(2011)}; 	   
(113) \citet{Winters(2017)}; 	   
(114) \citet{Winters(2018a)}; 	   
(115) \citet{Woitas(2003)}; 	   
(116) \citet{Worley(1998)};
(117) \citet{Zechmeister(2009b)}.  
}
\end{deluxetable*}

\subsection{Searches at Separations $\leq$ 2\arcsec}
\label{subsec:sub_arc}

In addition to the blinking and CCD imaging searches, investigations
for companions at separations smaller than 2\arcsec~were possible
using a variety of techniques, as detailed below in Sections \S
\ref{subsubsec:elevation}--\ref{subsubsec:hip}.  The availability of
accurate parallaxes for all stars and of $VRIJHK$ photometry for most
stars made possible the identification of overluminous red dwarfs that
could be harboring unresolved stellar companions.  Various subsets of
the sample were also probed using long-term astrometric data for stars
observed during RECONS' astrometry program, as well as via data
reduction flags indicating astrometric signatures of unseen companions
for stars observed by {\it Hipparcos}.  



\subsubsection{Overluminosity via Photometry: Elevation Above the Main Sequence}
\label{subsubsec:elevation}


 
 
  \begin{figure}
  \hspace{-1.0cm}
  \includegraphics[scale=0.40,angle=90]{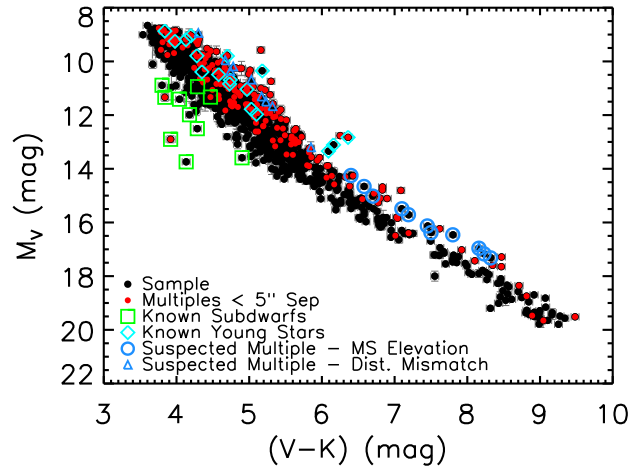}
  \caption{Observational HR Diagram for 1120 M dwarf primaries, with
    $M_V$ plotted versus $(V-K)$ color.  All primaries are plotted as
    black points.  Overplotted are known close multiples with
    separations less than 5\arcsec~having blended photometry (red
    points), known subdwarfs (open green squares), and known young
    objects (open cyan diamonds). Error bars are shown in gray, and
    are smaller than the points, in most cases. The large $K$
    magnitude errors for four objects (GJ0408, GJ0508.2, LHS3472, and
    LP876-026AB) have been omitted for clarity. As expected, known
    multiples with merged photometry are often elevated above the
    middle of the distribution.  The 11 stars suspected to be new
    unresolved multiples due to their elevated positions relative to
    the main sequence are indicated with open blue circles.  The ten
    stars suspected to be new unresolved multiples due to their
    distance mismatches from Figure \ref{fig:ccd_trig} are indicated
    with open blue triangles. Note that the candidate multiples
    detected by main sequence elevation are mostly mid-to-late type M
    dwarfs, while the suspected multiples identified by the distance
    mismatch technique are primarily early-type M
    dwarfs. \label{fig:abs_color}}
  \end{figure}

Accurate parallaxes and $V$ and $K$ magnitudes for stars in the sample
allow the plotting of the observational HR Diagram shown in Figure
\ref{fig:abs_color}, where $M_V$ and the $(V-K)$ color are used as
proxies for luminosity and temperature, respectively.  Unresolved
companions that contribute significant flux to the photometry cause
targets to be overluminous, placing them above the main sequence.
Known multiples with separations $<$ 5\arcsec~\footnote{This
  5\arcsec~separation appears to be the boundary where photometry for
  multiple systems from the literature --- specifically from Bessell
  and Weis --- becomes blended.  For photometry available from the
  SAAO group (e.g., Kilkenny, Koen), the separation is
  $\sim$10\arcsec~because they use large apertures when calculating
  photometric values.} are evident as points clearly elevated above
the presumed single stars on the main sequence, and merge with a few
young objects. Subdwarfs are located below and to the left of the
singles, as they are old, metal-poor, and underluminous at a given
color. Eleven candidate multiples lying among the sequence of known
multiples have been identified by eye via this HR Diagram.  These
candidates are listed in Table \ref{tab:suspects} and are marked in
Figures \ref{fig:abs_color} and \ref{fig:ccd_trig}.  Note that these
candidates are primarily mid-to-late M dwarfs. Known young stars and
subdwarfs were identified during the literature search and are listed
in Tables \ref{tab:young} and \ref{tab:subdwarfs}, along with their
identifying characteristics. More details on these young and old
systems are given in Sections \ref{subsubsec:young} and
\ref{subsubsec:old}.


\subsubsection{Overluminosity via Photometry: Trigonometric \& CCD Distance Mismatches}
\label{subsubsec:mismatches}

 
  
  
  \begin{figure}
  \includegraphics[scale=0.42,angle=90]{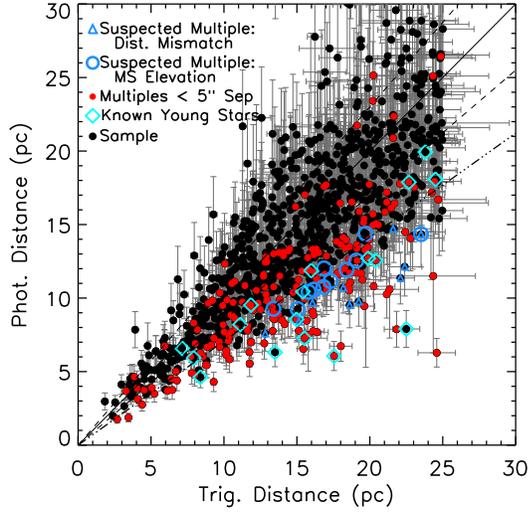}
  \caption{A comparison of distance estimates from $VRIJHK$ photometry
    vs.~distances using $\pi_{trig}$ for 1091 of the M dwarf primaries
    in the sample. The 29 stars with photometric distances $>$ 30 pc
    are not included in this plot. Errors on the distances are noted
    in gray. The diagonal solid line represents 1:1 agreement in
    distances, while the dashed lines indicate the 15\% uncertainties
    associated with the CCD distance estimates from
    \citet{Henry(2004)}.  The dash-dot line traces the location where
    the trigonometric distance exceeds the photometric estimate by a
    factor of $\sqrt{2}$, corresponding to an equal-luminosity/mass
    pair of stars. Known unresolved multiples with blended photometry
    are indicated with red points.  The 11 candidate unresolved
    multiples from the HR Diagram in Figure \ref{fig:abs_color} are
    enclosed with open blue circles.  The ten new candidates that may
    be unresolved multiples from this plot are enclosed with open blue
    triangles. \label{fig:ccd_trig}}
  \end{figure}

Because both $VRI$ and 2MASS $JHK$ photometry are now available for
nearly the entire sample, photometric distances based on CCD
photometry ({\it ccddist}) were estimated and compared to the accurate
trigonometric distances ({\it trigdist}) available from the
parallaxes.  Although similar in spirit to the HR Diagram test
discussed above that uses $V$ and $K$ photometry, all of the $VRIJHK$
photometry is used for each star to estimate the {\it ccddist} via the
technique described in \citet{Henry(2004)}, thereby leveraging
additional information.  As shown in Figure \ref{fig:ccd_trig},
suspected multiples that would otherwise have been missed due to the
inner separation limit (2\arcsec) of our main imaging survey can be
identified due to mismatches in the two distances.  For example, an
unresolved equal magnitude binary would have an estimated {\it
  ccddist} closer by a factor of $\sqrt{2}$ compared to its measured
{\it trigdist}. Unresolved multiples with more than two
  components, e.g., a triple system, could be even more overluminous,
  as could young, multiple systems. By contrast, cool subdwarfs are
  underluminous and therefore, their photometric distances are
  overestimated.

With this method, 50 candidate multiples were revealed with {\it
  ccddists} that were $\sqrt{2}$ or more times closer than their {\it
  trigdists}.  Of these, 40 were already known to have at least one
close companion (36 stars), or to be young (four stars), verifying the
technique.  The remaining ten are new candidates, and are listed in
Table \ref{tab:suspects}. 



\begin{deluxetable}{lccccc}
\centering
\setlength{\tabcolsep}{0.03in}
\tablewidth{0pt}
\tabletypesize{\tiny}
\tablecaption{Suspected Multiple Systems}
\tablehead{\colhead{Name}                &
	   \colhead{\# Stars}            &
 	   \colhead{RA}                  &
	   \colhead{DEC}                 &
           \colhead{Flag}                &
	   \colhead{Reference}           \\
	   \colhead{   }                 &
	   \colhead{   }                 &
	   \colhead{(hh:mm:ss)}          &
           \colhead{(dd:mm:ss)}          &
           \colhead{     }               &
	   \colhead{     }               }

\startdata                
GJ1006A         & 3?  & 00 16 14.62 &$+$19 51 37.6  &  dist     & 1     \\  
HIP006365       & 2?  & 01 21 45.39 &$-$46 42 51.8  &  X        & 3     \\  
LHS1288         & 2?  & 01 42 55.78 &$-$42 12 12.5  &  X        & 3     \\  
GJ0091          & 2?  & 02 13 53.62 &$-$32 02 28.5  &  X        & 3     \\  
GJ0143.3        & 2?  & 03 31 47.14 &$+$14 19 17.9  &  X        & 3     \\  
BD-21-01074A    & 4?  & 05 06 49.47 &$-$21 35 03.8  &  dist     & 1     \\
GJ0192          & 2?  & 05 12 42.22 &$+$19 39 56.5  &  X        & 3     \\  
GJ0207.1        & 2?  & 05 33 44.81 &$+$01 56 43.4  &  possSB   & 4     \\  
SCR0631-8811    & 2?  & 06 31 31.04 &$-$88 11 36.6  &  elev     & 1     \\    
LP381-004       & 2?  & 06 36 18.25 &$-$40 00 23.8  &  G        & 3     \\  
SCR0702-6102    & 2?  & 07 02 50.36 &$-$61 02 47.7  &  elev,pb? & 1,1   \\ 
LP423-031       & 2?  & 07 52 23.93 &$+$16 12 15.0  &  elev     & 1     \\    
SCR0757-7114    & 2?  & 07 57 32.55 &$-$71 14 53.8  &  dist     & 1     \\
GJ1105          & 2?  & 07 58 12.70 &$+$41 18 13.4  &  X        & 3     \\   
LHS2029         & 2?  & 08 37 07.97 &$+$15 07 45.6  &  X        & 3     \\  
LHS0259         & 2?  & 09 00 52.08 &$+$48 25 24.7  &  elev     & 1     \\    
GJ0341          & 2?  & 09 21 37.61 &$-$60 16 55.1  &  possSB   & 4     \\  
GJ0367          & 2?  & 09 44 29.83 &$-$45 46 35.6  &  X        & 3     \\  
GJ0369          & 2?  & 09 51 09.63 &$-$12 19 47.6  &  X        & 3     \\  
GJ0373          & 2?  & 09 56 08.68 &$+$62 47 18.5  &  possSB   & 4     \\  
GJ0377          & 2?  & 10 01 10.74 &$-$30 23 24.5  &  dist     & 1     \\
GJ1136A         & 3?  & 10 41 51.83 &$-$36 38 00.1  &  X,possSB & 3,4   \\
GJ0402          & 2?  & 10 50 52.02 &$+$06 48 29.4  &  X        & 3     \\   
LHS2520         & 2?  & 12 10 05.59 &$-$15 04 16.9  &  dist     & 1     \\ 
GJ0465          & 2?  & 12 24 52.49 &$-$18 14 32.3  &  pb?      & 2     \\   
DEN1250-2121    & 2?  & 12 50 52.65 &$-$21 21 13.6  &  elev     & 1     \\  
GJ0507.1        & 2?  & 13 19 40.13 &$+$33 20 47.7  &  X        & 3     \\  
GJ0540          & 2?  & 14 08 12.97 &$+$80 35 50.1  &  X        & 3     \\  
2MA1507-2000    & 2?  & 15 07 27.81 &$-$20 00 43.3  &  dist,elev& 1     \\ 
G202-016        & 2?  & 15 49 36.28 &$+$51 02 57.3  &  G        & 3     \\  
LHS3129A        & 3?  & 15 53 06.35 &$+$34 45 13.9  &  dist     & 1     \\ 
GJ0620          & 2?  & 16 23 07.64 &$-$24 42 35.2  &  G        & 3     \\  
GJ1203          & 2?  & 16 32 45.20 &$+$12 36 45.9  &  X        & 3     \\  
LP069-457       & 2?  & 16 40 20.65 &$+$67 36 04.9  &  elev     & 1     \\  
LTT14949        & 2?  & 16 40 48.90 &$+$36 18 59.9  &  X        & 3     \\  
HIP083405       & 2?  & 17 02 49.58 &$-$06 04 06.5  &  X        & 3     \\  
LP044-162       & 2?  & 17 57 15.40 &$+$70 42 01.4  &  elev     & 1     \\    
LP334-011       & 2?  & 18 09 40.72 &$+$31 52 12.8  &  X        & 3     \\  
SCR1826-6542    & 2?  & 18 26 46.83 &$-$65 42 39.9  &  elev     & 1     \\ 
LP044-334       & 2?  & 18 40 02.40 &$+$72 40 54.1  &  elev     & 1     \\  
GJ0723          & 2?  & 18 40 17.83 &$-$10 27 55.3  &  X        & 3     \\  
HIP092451       & 2?  & 18 50 26.67 &$-$62 03 03.8  &  possSB   & 4     \\
LHS3445A        & 3?  & 19 14 39.15 &$+$19 19 03.7  &  dist     & 1     \\ 
GJ0756          & 2?  & 19 21 51.42 &$+$28 39 58.2  &  X        & 3     \\  
LP870-065       & 2?  & 20 04 30.79 &$-$23 42 02.4  &  dist     & 1     \\
GJ1250          & 2?  & 20 08 17.90 &$+$33 18 12.9  &  dist     & 1     \\ 
LEHPM2-0783     & 2?  & 20 19 49.82 &$-$58 16 43.0  &  elev     & 1     \\
GJ0791          & 2?  & 20 27 41.65 &$-$27 44 51.9  &  X        & 3     \\  
LHS3564         & 2?  & 20 34 43.03 &$+$03 20 51.1  &  X        & 3     \\  
GJ0811.1        & 2?  & 20 56 46.59 &$-$10 26 54.8  &  X        & 3     \\  
L117-123        & 2?  & 21 20 09.80 &$-$67 39 05.6  &  X        & 3     \\  
HIP106803       & 2?  & 21 37 55.69 &$-$63 42 43.0  &  X        & 3     \\  
LHS3748         & 2?  & 22 03 27.13 &$-$50 38 38.4  &  X        & 3     \\  
G214-014        & 2?  & 22 11 16.96 &$+$41 00 54.9  &  X        & 3     \\  
GJ0899          & 2?  & 23 34 03.33 &$+$00 10 45.9  &  X        & 3     \\  
GJ0912          & 2?  & 23 55 39.77 &$-$06 08 33.2  &  X        & 3     \\ 
\enddata


\tablerefs{(1) this work; (2) \citet{Heintz(1986)}; (3)
  \citet{Lindegren(1997)}; (4) \citet{Reiners(2012)}.} 

\tablecomments{Flag Description: {\bf dist} means that the {\it
    ccddist} is at least $\sqrt{2}$ times closer than the {\it
    trigdist} due to the object's overluminousity; {\bf elev} means
  that the object is elevated above the main sequence in the HR
  Diagram in Figure \ref{fig:abs_color} due to overluminosity; {\bf
    possSB} means that the object has been noted as a possible
  spectroscopic binary by \citet{Reiners(2012)}; {\bf pb?} indicates
  that a possible perturbation was noted. The the single letters are
  {\it Hipparcos} reduction flags as follows: $G$ is an acceleration
  solution where a component might be causing a variation in the
  proper motion; $V$ is for Variability-Induced Movers, where one
  component in an unresolved binary could be causing the photocenter
  of the system to be perturbed; $X$ is for a stochastic solution,
  where no reliable astrometric parameters could be determined, and
  which may indicate an astrometric binary.\label{tab:suspects}}

\end{deluxetable}

\subsubsection{RECONS Perturbations}
\label{subsubsec:pbs} 

\begin{figure}
  \begin{center}
  \includegraphics[scale=0.33,angle=90]{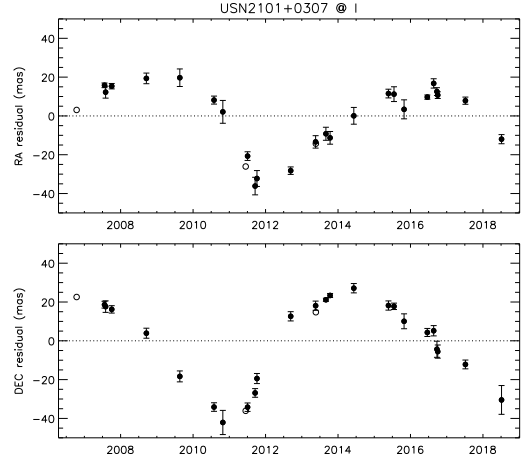}
 
\hspace{0.5cm}
  \caption{Nightly mean astrometric residual plots in RA and DEC for
    USN2101$+$0307. The astrometric signature of the system's proper
    motion and parallax has been removed, leaving clear indication
    that an unseen companion is perturbing the position of the primary
    star.}{\label{fig:usn2101_pb}}
 
  \end{center}
  \end{figure}

A total of 324 red dwarfs in the sample have parallax measurements by
RECONS, with the astrometric coverage spanning 2--16 years.  This
number is slightly higher than the 308 parallaxes listed in Table
\ref{tab:pisource}, due to updated and more accurate RECONS parallax
measurements that improved upon YPC parallaxes with high errors.  The
presence of a companion with a mass and/or luminosity different from
the primary causes a perturbation in the photocenter of the system
that is evident in the astrometric residuals after solving for the
proper motion and parallax.  This is the case for 39 of the observed
systems, which, although sometimes still unseen, are listed as
confirmed companions in Table \ref{tab:multinfo}, where references are
given. Because 13 of these 39 stars with perturbations were detected
during the course of this project, we note them as new discoveries,
although they were first reported in other papers \citep[e.g.,
][]{Bartlett(2017),Jao(2017),Winters(2017),Henry(2018),Riedel(2018)}. A
new companion to USN2101$+$0307 was reported in \citet{Jao(2017)}. We
present here the nightly mean astrometric residual plots in RA and DEC
for this star (shown in in Figure \ref{fig:usn2101_pb}), which
exhibits a perturbation due to its unseen companion. This system is
discussed in more detail in \S \ref{subsec:newcomp}.

This is the only technique used in this companion search that may have
revealed brown dwarf companions. None of the companions have been
  resolved, so it remains uncertain whether the companion is a red or
  brown dwarf. As we noted in \citet{Winters(2017)}, the magnitude of
  the perturbation in the photocenter of the system, $\alpha$, follows
  the relation $\alpha$ = (B $-$ $\beta$)$a$, where B is the
  fractional mass M$_B$/(M$_A$ $+$ M$_B$), $\beta$ is the relative
  flux expressed as $(1 + 10^{0.4\Delta m})^{-1}$, and $a$ is the
  semi-major axis of the relative orbit of the two components
  \citep{vandeKamp(1975)}. The degeneracy between the mass ratio/flux
  difference and the scaling of the photocentric and relative orbits
  results in an uncertainty in the nature of the companion. We are
  able to assume that the companion is a red dwarf if the system is
  overluminous, which is the case for eight of these
  systems. Therefore, we conservatively assume that all the companions
  are red dwarfs. These particular systems are high priority targets
  for high-resolution, speckle observations through our large program
  on the Gemini telescopes that is currently in-progress, with a goal
  of resolving and characterizing the companions.

\subsubsection{Hipparcos Reduction Flags}
\label{subsubsec:hip}

All 460 stars in the sample with {\it Hipparcos} parallaxes were
searched for entries in the Double and Multiple Systems Annex (DMSA)
\citep{Lindegren(1997)} of the original {\it Hipparcos} Catalog to
reveal any evidence of a companion.  Of these 460 stars, 229 have a
parallax measured only by {\it Hipparcos}, while 231 also have a
parallax measurement from another source. Various flags exist in the
DMSA that confirm or infer the presence of a companion: C ---
component solutions where both components are resolved and have
individual parallaxes; G --- acceleration solutions, i.e., due to
variable proper motions, which could be caused by an unseen companion;
O --- orbits from astrometric binaries; V --- Variability-Induced
Movers (VIMs), where the variability of an unresolved companion causes
the photocenter of the system to move or be perturbed; X ---
stochastic solutions, for which no astrometric parameters could be
determined and which may indicate that the star is actually a
short-period astrometric binary.


Most of the AFGK systems observed by {\it Hipparcos} that have flags
in the DMSA have been further examined or re-analyzed by
\citet{Pourbaix(2003), Pourbaix(2004)}, \citet{Platais(2003)},
\citet{Jancart(2005)}, \citet{Frankowski(2007)},
\citet{Horch(2002),Horch(2011a),Horch(2011b),Horch(2015b),Horch(2017)};
however, few of the M dwarf systems have been investigated to date.
Stars with C and O flags were often previously known to be binary, are
considered to be confirmed multiples, and are included in Table
\ref{tab:multinfo}.  We found that G, V, or X flags existed for 31
systems in the survey --- these suspected multiples are listed in
Table \ref{tab:suspects}.

\subsection{Literature Search}
\label{subsec:lit}

Finally, a literature search was carried out by reviewing all 1120
primaries in {\it SIMBAD} and using the available bibliography tool to
search for papers reporting additional companions.  While {\it SIMBAD}
is sometimes incomplete, most publications reporting companions are
included.  Papers that were scrutinized in detail include those
reporting high resolution spectroscopic studies (typically radial
velocity planet searches or rotational velocity results that might
report spectroscopic binaries), parallax papers that might report
perturbations, high resolution imaging results, speckle interferometry
papers, and other companion search papers.  A long list of references
for multiple systems found via the literature search is included in
Table \ref{tab:multinfo}.

In addition, the Washington Double Star Catalogue (WDS), maintained by
Brian Mason\footnote{The primary copy is kept at USNO and a back-up
  copy is housed at Georgia State University.}, was used extensively
to find publications with information on multiple systems. Regarding
the WDS, we note that: (1) not all reported companions are true
physical members of the system with which they are listed, and (2)
only resolved companions (i.e., no spectroscopic or astrometric
binaries) are included in the catalog.  Thus, care had to be taken
when determining the true number of companions in a
system. Information pertaining to a star in the WDS was usually only
accessed after it was already known that the system was multiple and
how many components were present in the system, so this was not really
troublesome.  However, the WDS sometimes had references to
multiplicity publications that {\it SIMBAD} had not listed. Thus, the
WDS proved valuable in identifying references and the separations,
magnitude differences, and other information included in Table
\ref{tab:multinfo}.

Finally, all of the multiple systems were cross-checked against the
{\it Gaia} DR2 through the $Aladin$ interface. Thirty-two known
multiple systems had positional data, but no parallax, while fourteen
known systems were not found. Of the 575 stellar components presented
in this sample, 133 were not found to have separate data points. The
majority of these companions are located at sub-arcsecond angular
separations from their primaries, with a rare few having separations 1
-- 3\arcsec. An additional 15 companions had unique coordinates, but
no individual parallax or proper motion. We anticipate that future
{\it Gaia} data releases will provide some of the currently missing
information for these low-mass multiple systems.

Information for all multiple systems (including brown dwarf
components) is presented in Table \ref{tab:multinfo}, with {\it n-1}
lines for each system, where {\it n} is the total number of components
in the system.  For example, a quadruple system will have three lines
of information that describe the system.  The name is followed by the
number of components in the system and the configuration map of the
components detailed in that line of the Table.  If the line of data
pertains to higher order systems containing component configurations
for sub-systems (e.g., `BC' of a triple system), the number of
components noted will be `0', as the full membership of the system
will already have been noted in the line of data containing the `A'
component.  These data are followed by epoch J2000.0 coordinates, the
angular separation ($\rho$) in arcseconds, the position angle
($\theta$) in degrees measured East of North, the year of the
measurement, the code for the technique used to identify the
component, and the reference.  We assign a separation of
$<$1\arcsec~for all astrometric and spectroscopic binaries (unless
more information is available) and/or to indicate that a companion has
been detected, but not yet resolved.  We note that where orbit
determinations from the literature are reported, the semimajor axis,
$a$, is listed instead of $\rho$.  If $a$ was not reported in the
reference given, it was calculated from the period and the estimated
masses of the components in question via Kepler's Third Law.


The final three columns give a magnitude difference ($\Delta$mag)
between the components indicated by the configuration map, the filter
used to measure this $\Delta$mag, and the reference for this
measurement. Photometry from photographic plates is denoted by
`V*'. In many cases, there are multiple separation and $\Delta$mag
measurements available in the literature from different groups using
different techniques.  An exhaustive list of these results is beyond
the scope of this work; instead, a single recent result for each
system is listed.  In a few cases, the position angles and/or
$\Delta$mag measurements are not available. We discuss how these
systems are treated in \S \ref{subsubsec:deblend}.

\subsubsection{Suspected Companions}
\label{subsubsec:suspects}

Forty-nine singles suspected to be doubles were revealed during this
survey, three of which (GJ~912, GJ~1250, and SCR~1826-6542) have so
far been confirmed with continuing follow-up observations. An
additional five doubles are suspected to be triples, yielding a
current total of 54 suspected additional companions (only one
companion per system in all cases) listed in Table \ref{tab:suspects},
but {\it not} included in Table \ref{tab:multinfo}.\footnote{For
  consistency, companions to GJ~912, GJ~1250, and SCR~1826-6542 are
  included in Table \ref{tab:suspects} and in the 'Suspects' portion
  of the histogram in Figure \ref{fig:seps_hist}.}  Systems in Table
\ref{tab:suspects} are listed with the suspected number of components
followed by a question mark to indicate the system's suspect status,
followed by J2000.0 coordinates, a flag code for the reason a system
is included as having a candidate companion, and the reference.  Notes
to the table give detailed descriptions of the flags.  Among the 56
suspected companions, 31 are from the {\it Hipparcos} DMSA, in which
they are assigned G, V, or X flags.  A number of primaries that were
suspected to be multiple due to either an underestimated {\it ccddist}
or an elevated position on the HR diagram were found through the
literature search to have already been resolved by others and have
been incorporated into Table \ref{tab:multinfo} and included in the
analysis as confirmed companions.  There remain 21 systems in Table
\ref{tab:suspects} with {\it ccddist} values that do not match their
trigonometric parallax distances and/or that are noticeably elevated
above the main sequence that have not yet been confirmed.  A few more
systems had other combinations of indicators that they were multiple,
e.g., an object with a perturbation might also have a distance
mismatch. Six stars were reported as suspected binaries in the
literature. GJ 207.1, GJ 341, GJ 373, GJ 1136A and HIP 92451 were
noted by \citet{Reiners(2012)} as possible spectroscopic binaries, and
GJ 465 was identified by \citet{Heintz(1986)} as a possible
astrometric binary. These are listed in Table \ref{tab:suspects}. We
reiterate that none of these {\it suspected} companions have been
included in any of the analyses of the previous section or that
follow; only {\it confirmed} companions have been used.


\subsubsection{Young Stellar Objects}
\label{subsubsec:young}

Within the solar neighborhood are young moving groups that have
contributed members to the multiplicity sample. Within the studied
collection are 16 confirmed young M dwarfs, nine of which are known to
be multiple, yielding a multiplicity rate of 56$\pm$19\%. We note
that this result is not statistically robust due to the small number
of objects with which it was calculated. Presented in Table
\ref{tab:young} are these known nearby young red dwarfs, with their
astrometric data duplicated from Table \ref{tab:astrdata}. In
addition, the tangential velocity $v_{tan}$ is listed, along with the
youth indicators, the moving group with which they are associated, and
the reference. The youth indicators are as follows: BF --- a {\it bona
  fide} and well-known member of a moving group; Li --- the presence
of lithium; ol --- over-luminous. Any member of a multiple system
where one component exhibits any of these indicators is also assumed
to be young, as in the case of GJ 871.1AB. We note that we do not
  exclude any of the identified young M dwarf members from our sample.
  This would arbitrarily bias our sample, as a comprehensive search
  for these objects in the current sample has not been conducted.



\begin{deluxetable*}{lcccccccclc}
\centering
\setlength{\tabcolsep}{0.03in}
\tablewidth{0pt}
\tabletypesize{\small}
\tablecaption{Young Members \label{tab:young}}
\tablehead{\colhead{Name}                &
	   \colhead{\# objects}          &
	   \colhead{RA}                  &
 	   \colhead{DEC}                 &
           \colhead{$\mu$}               &
	   \colhead{P.A.}                &
	   \colhead{Ref}                 &
           \colhead{$v_{tan}$}            &
	   \colhead{Youth}               &
           \colhead{Moving}              &
	   \colhead{Ref}                 \\
	   \colhead{   }                 &
	   \colhead{   }                 &
	   \colhead{(hh:mm:ss)}          &
           \colhead{(dd:mm:ss)}          &
           \colhead{(\arcsec ~yr$^{-1}$)}&
	   \colhead{(deg)}               &
           \colhead{     }               &
           \colhead{(km s$^{-1}$)}        &
           \colhead{Indicator}           &
           \colhead{Group}               &
	   \colhead{   }                 }

\startdata
LTT~10301AB    &   2      & 00 50 33.23 &$+$24 49 00.9&  0.203     & 101.9    &   5   &  11   &  ol         & Argus     & 3  \\ 
G~80-21        &   1      & 03 47 23.35 &$-$01 58 19.8&  0.323     & 147.8    &   5   &  25   &  Li         & AB Dor    & 3  \\ 
2MA~0414-0906  &   1      & 04 14 17.29 &$-$09 06 54.3&  0.168     & 325.2    &   4   &  19   &  Li         & none      & 3   \\ 
LP~776-25      &   1      & 04 52 24.42 &$-$16 49 22.2&  0.243     & 150.7    &   2   &  18   &  ol         & AB Dor    & 3   \\ 
GJ~2036AB      &   2      & 04 53 31.20 &$-$55 51 37.3&  0.149     & 060.6    &   5   &   8   &  BF         & AB Dor    & 3   \\ 
LP~717-36AB    &   2      & 05 25 41.67 &$-$09 09 12.6&  0.197     & 164.7    &   6   &  19   &  ol         & AB Dor    & 3   \\ 
AP~COL         &   1      & 06 04 52.16 &$-$34 33 36.0&  0.330     & 003.6    &   6   &  13   &  Li         & Argus     & 3   \\ 
CD-35 2722AB   &   2      & 06 09 19.22 &$-$35 49 31.1&  0.057     & 186.4    &   4   &   6   &  BF         & AB Dor    & 3   \\ 
GJ~2060ABC     &   3      & 07 28 51.37 &$-$30 14 49.2&  0.212     & 207.9    &   5   &  15   &  BF         & AB Dor    & 3   \\ 
G~161-71       &   1      & 09 44 54.19 &$-$12 20 54.4&  0.321     & 277.1    &   1   &  21   &  ol         & Argus     & 1   \\ 
GJ~382         &   1      & 10 12 17.67 &$-$03 44 44.4&  0.314     & 219.0    &   5   &  12   &  ol         & AB Dor    & 3   \\ 
TWA~22AB       &   2      & 10 17 26.91 &$-$53 54 26.5&  0.149     & 264.4    &   6   &  12   &  Li         & Beta Pic  & 3   \\ 
GJ~393         &   1      & 10 28 55.56 &$+$00 50 27.6&  0.950     & 219.3    &   5   &  32   &  ol         & AB Dor    & 3   \\ 
GJ~490ABCD     &   4      & 12 57 40.26 &$+$35 13 30.3&  0.307     & 240.9    &   5   &  29   &  BF         & Tuc-Hor   & 3   \\ 
GJ~856AB       &   2      & 22 23 29.08 &$+$32 27 33.1&  0.329     & 129.1    &   5   &  24   &  Li         & AB Dor    & 3   \\ 
GJ~871.1AB     &   2      & 22 44 57.96 &$-$33 15 01.7&  0.230     & 123.1    &   5   &  25   &  (Li)       & Beta Pic  & 3   \\ 
HIP~114066     &   1      & 23 06 04.83 &$+$63 55 33.9&  0.185     & 108.5    &   5   &  21   &  Li         & AB Dor    & 3   \\
\enddata
\tablecomments{The youth indicators are as follows: BF
  --- a bona fide and well-known member of a moving group; Li --- the
  presence of lithium; ol --- over-luminous.}

\tablerefs{(1) \citet{Bartlett(2017)}; (2) \citet{Hog(2000)}; (3)
  \citet{Riedel(2017)}; (4) \citet{Shkolnik(2012)}; (5)
  \citet{vanLeeuwen(2007)}; (6) \citet{Winters(2015)}.}

\end{deluxetable*}


\subsubsection{Old, Cool Subdwarfs}
\label{subsubsec:old}

There are a small number of old halo members, also known as cool
subdwarfs, that happen to be passing through the solar
neighborhood. The objects have been identified either
spectroscopically or stand out on a Reduced Proper Motion Diagram. Out
of the eleven confirmed subdwarf members with trigonometric parallaxes
found within 25 pc, only three are multiple systems, resulting in a
multiplicity rate of 27$\pm$16\%. This is very similar to that of the
M dwarf population as a whole (26.8$\pm$1.4\%; see \S
\ref{subsec:corrections}) and in agreement with a larger sample of 62
K and M subdwarfs has been studied by \citet{Jao(2009)}, who found a
rate of of 26$\pm$6\%.  As with the young stars, with such a small
number of objects, this result is not statistically robust.  The known
subdwarfs in our sample are identified in Table \ref{tab:subdwarfs}.
In addition to the astrometric data for each system that have been
duplicated from Table \ref{tab:astrdata}, the calculated tangential
velocities $v_{tan}$ and spectral types from the literature are
listed. We note that we do not exclude any of the identified M-type
subdwarfs from our study. This would arbitrarily bias our sample, as a
comprehensive search for these objects in the current sample has not
been conducted.


\begin{deluxetable*}{lccccccccc}
\centering
\setlength{\tabcolsep}{0.03in}
\tablewidth{0pt}
\tabletypesize{\small}
\tablecaption{Subdwarf Members \label{tab:subdwarfs}}
\tablehead{\colhead{Name}                &
	   \colhead{\# objects}          &
	   \colhead{RA}                  &
 	   \colhead{DEC}                 &
           \colhead{$\mu$}               &
	   \colhead{P.A.}                &
	   \colhead{Ref}                 &
           \colhead{$v_{tan}$}            &
	   \colhead{Spectral}             &
           \colhead{Ref}                 \\
	   \colhead{   }                 &
	   \colhead{   }                 &
	   \colhead{(hh:mm:ss)}          &
           \colhead{(dd:mm:ss)}          &
           \colhead{(\arcsec ~yr$^{-1}$)}&
	   \colhead{(deg)}               &
           \colhead{     }               &
           \colhead{(km s$^{-1}$)}        &
           \colhead{Type}                &
           \colhead{}                    }

\startdata
LHS~1490          &    1     & 03 02 06.36 &$-$39 50 51.9  & 0.859 & 221.3 &   8   &   58 & M5.0 VI   &   4    \\
GJ~1062           &    1     & 03 38 15.70 &$-$11 29 13.5  & 3.033 & 152.0 &   7   &  230 & M2.5 VI   &   3    \\
LHS~189AB         &    2     & 04 25 38.35 &$-$06 52 37.0  & 1.204 & 145.7 &   2   &  105 & M3.0 VIJ  &   4    \\
LHS~272           &    1     & 09 43 46.16 &$-$17 47 06.2  & 1.439 & 279.2 &   5   &   92 & M3.0 VI   &   3    \\
GJ~455AB          &    2     & 12 02 18.08 &$+$28 35 14.2  & 0.791 & 268.0 &   7   &   74 & M3.5 VIJ  &   3    \\
LHS~2852          &    1     & 14 02 46.67 &$-$24 31 49.7  & 0.506 & 315.6 &   8   &   41 & M2.0 VI   &   3    \\
LHS~375           &    1     & 14 31 38.25 &$-$25 25 32.9  & 1.386 & 269.0 &   7   &  158 & M4.0 VI   &   3    \\
SSSPM~J1444-2019  &    1     & 14 44 20.33 &$-$20 19 25.5  & 3.507 & 236.0 &   6   &  270 & M9.0 VI:  &   6    \\
GJ~2116           &    1     & 15 43 18.33 &$-$20 15 32.9  & 1.173 & 195.3 &   8   &  119 & M2.0 VI   &   1    \\
LHS~3409          &    1     & 18 45 52.24 &$+$52 27 40.7  & 0.843 & 298.0 &   7   &   81 & M4.5 VI   &   3    \\
LHS~64AB          &    2     & 21 07 55.39 &$+$59 43 19.4  & 2.098 & 209.0 &   7   &  238 & M1.5 VIJ  &   3    \\
\enddata

\tablerefs{(1) \citet{Bidelman(1985)}; (2) \citet{Costa(2006)}; (3)
  \citet{Gizis(1997a)}; (4) \citet{Jao(2008)}; (5) \citet{Jao(2011)};
  (6) \citet{Schilbach(2009)}; (7) \citet{vanAltena(1995)}; (8)
  \citet{Winters(2015)}.}

\end{deluxetable*}


\subsubsection{Substellar Companions}
\label{subsubsec:bds}

Because this study focuses on the {\it stellar} companions of M
dwarfs, it was important to determine which companions were stellar
and which were {\it substellar}.  \citet{Dieterich(2014)} found that
the boundary between stars and brown dwarfs is near the L2.0V spectral
type; efforts are underway to determine to what mass this spectral
type corresponds.  As mentioned in \S \ref{sec:sampledef}, $M_V =$
20.0 and $(V-K) =$ 9.5 were used as cutoffs for the faintest and
reddest (and correspondingly least massive) M dwarfs.  Analysis of the
main sequence in the HR diagram created from the RECONS list of stars
and brown dwarfs within 25 pc indicates that $M_V =$ 21.5 and $(V-K)
=$ 10.3 correspond to spectral type L2.0V, and therefore the end of
the stellar main sequence.  Thus, for this large statistical study, we
consider objects fainter or redder than these limits to be substellar
brown dwarfs.

Via the literature search, 18 brown dwarf companions to 15 M dwarf
primaries were identified.  These are noted in Table
\ref{tab:photdata} with a `BD' for the component in the object column
(column 2).  Although no comprehensive searches have been done for
brown dwarfs as companions to the 1120 M dwarfs targeted here
(including ours), we note that as presently known, the rate of M dwarf
primaries with known brown dwarf companions is 1.3$\pm$0.3\%.  This is
a factor of twenty-one less than the stellar multiplicity rate,
considering the stellar and brown dwarf companions detected to date.
While more brown dwarf companions will undoubtedly be found in the
future, it seems that they are genuinely much rarer than stellar
companions.  We note that astrometric detection via a perturbation is
the only technique used in this survey that was sensitive to brown
dwarf companions, and only a few stars (GJ 1215, SCR 1845-6357) have
so far been found to be orbited by brown dwarfs via perturbations in
data from our astrometric survey at the CTIO/SMARTS 0.9m.

We do not include any brown dwarf companions in the analysis that
follows, nor are planetary companions addressed in this work.


\section{Notes on Individual Systems}
\label{sec:individual_systems}

\subsection{New Companions}
\label{subsec:newcomp}

Several new companions were discovered during the surveys, or
confirmed after being noticed during the long-term astrometry program
at the CTIO/SMARTS 0.9m. They are listed in order of RA. In each case,
archival SuperCOSMOS plates were blinked to eliminate the
possibility that new companions were background objects.


$\bullet$ {\bf GJ~84.1C (0205$-$2804)} was found during the imaging
survey separated by 1\farcs5 from GJ 84.1B at a position angle of
299\arcdeg ~and with a $\Delta$$I$ = 0.23 mag, making the system a
triple. The SuperCOSMOS $B$ and $R$ plates, taken September 1977 and
November 1994, respectively, provide a $\Delta$t of 17.2
years. Blinking these two plates using the $CDS$ Aladin interface
showed that the B component, with a proper motion of 0\farcs549
yr$^{-1}$, moved 9\farcs4. Projecting the star's position forward 17
years to the date of the observation (October 2011) indicates that
there was no background star at the position of the B component at
that time. Astrometry from the {\it Gaia} DR2 confirms our discovery,
with reported parallaxes of 38.29$\pm$0.03, 38.51$\pm$0.08, and
38.15$\pm$0.10 mas for the now three components. We therefore consider
this star a new member of the system.

$\bullet$ {\bf 2MA~0936-2610C (0936$-$2610)} During the blinking
  survey, we detected a new companion to this previously known double
star at $\rho$ $=$ 41\farcs8, $\theta$ $=$ 315\arcdeg, making this a
triple system. The $\Delta$$V$ between the AB pair and the C component
is 6.9 mag. There is no {\it Gaia} DR2 parallax available for the AB
pair, yet, but the trigonometric distance of 18.50$\pm$0.05 pc for the
C component \citep[54.04$\pm$0.12 mas;][]{GaiaDR2(2018)} is in
agreement with the ground-based parallax of 53.75$\pm$1.42 mas
\citep[18.6 pc;][]{Riedel(2018)} of the AB pair. We therefore consider
this star a new member of the system.

$\bullet$ {\bf UPM~1710-5300B (1710$-$5300)} was found during the
imaging survey as a companion separated by 1\farcs2 from its primary
at a position angle of 339\arcdeg, with $\Delta$$I$ = 1.2 mag. The
archival $B$ and $R$ plates, taken July 1977 and April 1993,
respectively, provide a $\Delta$t of 15.8 years. Blinking these two
plates showed that the primary component, with a proper motion of
0\farcs207 yr$^{-1}$, moved 4\farcs1. Projecting the star's position
forward 20 years to the date of the observation (April 2013) indicates
that there was no background star of similar brightness at the
position of the primary component at that time. Astrometry from the
{\it Gaia} DR2 confirms our discovery, with reported parallaxes of
45.54$\pm$0.12 and 45.05$\pm$0.15 mas for the two components.  We
consider this star a new binary system.

$\bullet$ {\bf LHS~5348 (1927$-$2811)} This binary system discovery
was previously reported in \citet{Winters(2017)} along with its
parallax measurement; however, we include it here with updated
multiplicity information, as it was discovered during the multiplicity
survey. It is separated by 0\farcs89 from its primary at a position
angle of 283\arcdeg, with $\Delta$$I$ = 2.3 mag. The archival $B$
  and $I$ plates, taken September 1976 and July 1996, respectively,
  provide a $\Delta$t of 19.8 years. Blinking these two plates showed
  that the primary component, with a proper motion of 0\farcs509
  yr$^{-1}$, moved 10\arcsec. Projecting the star's position forward
  17 years to the date of the observation (May 2013) indicates that
  there was no background star of similar brightness at the position
  of the primary component at that time. While there are two data
  points in the {\it Gaia} DR2 at the positions of the two components,
  only one reports a parallax in agreement with the distance of the
  system. The other has no associated parallax measurement. We
  nevertheless consider this a new binary system and anticipate that
  future {\it Gaia} data releases will confirm our discovery.

$\bullet$ {\bf 2MA~1951-3510B (1951$-$3510)} was found during the
  imaging survey as a companion separated by 1\farcs9 from its primary
  at a position angle of 131\arcdeg, with $\Delta$$I$ = 0.3 mag. The
  $B$ and $R$ plates, taken July 1976 and September 1990,
  respectively, provide a $\Delta$t of 14.1 years. Blinking these two
  plates showed that the primary component, with a proper motion of
  0\farcs373 yr$^{-1}$, moved 5\farcs3. Projecting the star's position
  forward 23 years to the date of the observation (May 2013) indicates
  that there was no background star of similar brightness at the
  position of the primary component at that time. Astrometry from the
  {\it Gaia} DR2 confirms our discovery, with reported parallaxes of
  88.20$\pm$0.08 and 88.27$\pm$0.09 mas for the two components
  \citep{GaiaDR2(2018)}. We consider this a new binary system.

$\bullet$ {\bf USN~2101$+$0307 (2101$+$0307)} This object shows a
  clear astrometric perturbation, which was not reported in
  \citet{Jao(2017)}, although it was noted as a binary. As shown in
  Figure \ref{fig:usn2101_pb}, the orbit has not yet wrapped, with
  more than ten years of data available for this system. The
  photometric distance estimate of 13.9$\pm$2.2 pc disagrees
  marginally with the trigonometric distance of 17.7$\pm$0.6 pc. {\bf
    This indicates that the system is slightly overluminous, which
    indicates that the companion contributes some light to the system
    and is therefore likely a star and not a brown dwarf.} The
  companion is not resolved in the {\it Gaia} DR2.


%



\subsection{Interesting Systems}
\label{subsec:interest}

There are some systems that require more detail than that given in
Table \ref{tab:multinfo}, typically those that constitute more than
two components or that are particularly interesting.  These are listed
here with the first four digits of RA and Decl. in sexagesimal hours
and degrees, respectively.  Note that below, we adopt the moniker GJ
to identify all Gl, GL, and GJ stars, also known as ``Gliese'' stars.




{\bf GJ~2005ABC (0024-2708)} This system is a triple, not a quadruple.
Upon further analysis of the $HST$-$FGS$ data, the D component reported in
\citet{Henry(1999)} is a false detection and is not real.


{\bf GJ~1046AB (0219-3646)} The SB companion is a probable brown dwarf
with 168.848 day period \citep{Kurster(2008)}; also described by
\citet{Zechmeister(2009b)}. \citet{Bonfils(2013)} also note that the
companion is a brown dwarf or sub-stellar in nature.

{\bf GJ~109 (0244+2531)} This object is tagged as a VIM
(Variability-Induced Mover) in the {\it Hipparcos} catalog, which
could imply duplicity. However, \citet{Pourbaix(2003)} have shown this
to be an incorrect tag in their re-analysis and re-calculation of the
$(V-I)$ colors. This object is not included as a candidate multiple.



{\bf GJ~165AB (0413+5031)} \citet{Allen(2007)} refer to Gl~165B (what
we here call GJ 165B) as an L4 dwarf; however, this is likely a typo
for GD~165B, which is reported as a bona fide brown dwarf companion to
a white dwarf \citet{Kirkpatrick(1994),McLean(2003)}. The coordinates
of GD~165B are $\alpha$ $=$ 14:24:39.09, $\delta$ $=$ $+$09:17:10.4,
so it is not the same object as GJ~165B. \citet{Kirkpatrick(1994)}
discuss both Gl~65B and GD~165B. GJ~165AB seems to be a possible equal
magnitude binary \citep{Heintz(1992a)}, while \citet{McAlister(1987c)}
provide separation information from speckle observations. These data
are noted in Table \ref{tab:multinfo}.




{\bf LTT~11399 (0419+3629)} Both \citet{Worley(1961)} and
\citet{Worley(1962)} report that this is a binary with a separation of
6\farcs4 at 226$^{\circ}$. \citet{Balega(2007)} observed this star
using speckle interferometry but were unable to resolve it. It is also
marked as having a stochastic solution in
\citet{Lindegren(1997)}. Closer inspection and backtracking of its
proper motion indicate that the alleged component is a background
star. We consider this to be a single star and do not list it in Table
\ref{tab:suspects}.







{\bf GJ~273 (Luyten's Star) (0727$+$0513)} This star was reported in
\citet{Ward-Duong(2015)} as having a close companion; however, based
on previous observations of this object with myriad methods (IR
speckle \citep{Leinert(1997)}, long-term astrometric monitoring
\citep[15.2 yr][]{Gatewood(2008)}, high resolution spectroscopy
\citep{Nidever(2002),Bonfils(2013)}, etc.), we conclude that the
reported companion is likely an unassociated background object. We
treat this object as single.



{\bf GJ~289 (0748+2022)} \citet{Marcy(1989)} note that Gl~289 (GJ~289)
is an SB2, but this is likely a typo for GJ~829, which is noted in
\citet{Marcy(1987)} as being a probable SB2. We did not see reports of
GJ~289 being an SB2 noted anywhere else in the literature. We treat
GJ~289 as a single object.


{\bf GJ~1103 (0751-0000)} \citet{Reiners(2012)} cite
\citet{Delfosse(1998)} for GJ~1103(A) being an SB. However,
\citet{Delfosse(1998)} note only that they exclude GJ~1103 from their
sample due to it being a binary. They do not claim that it is an SB.
In the LHS Catalog \citep{Luyten(1979a)}, where it is listed as
LHS~1951, it is advertised to have a companion LHS~1952 with a
separation of 3\arcsec ~at $\theta$ $=$ 78$^{\circ}$, with component
magnitudes of m$_R$ $=$ 13.0 and 15.5, and m$_{pg}$ $=$ 15.0 and
17.5. We searched for the companion by examining both RECONS
astrometry frames, in which the seeing was sometimes 1\farcs2 or
better; and via the blinking campaign.  No companion was found. We
conclude that GJ~1103 is single.

{\bf GJ~450 (1151$+$3516)} This object was reported as a low
probability binary candidate with a low velocity amplitude by
\citet{Young(1987)} who measured five epochs of precise radial
velocities. However, more recent high resolution observations with
ELODIE and SOPHIE over a range of resolutions (R $=$ 40,000 - 75,000)
make no mention of this object being multiple \citep{Houdebine(2010)},
nor was a companion detected with lucky imaging \citep{Jodar(2013)} or
with infrared adaptive optics observations
\citep{Ward-Duong(2015)}. We therefore consider this object to be
single.

{\bf GJ~452 (1153$-$0722)} \citet{Gould(2004)} report a likely white
dwarf companion at $\rho$ $=$ 9\arcsec~and $\theta$ $=$ 110\arcdeg
~that was detected in 1960 by \citet{Luyten(1980a)}. However, blinking
SuperCOSMOS plate images (epochs 1984 - 1997) with the 300 second
integration taken at the CTIO/SMARTS 0.9m reveals no co-moving
companion. Backtracking the proper motion of the star to epoch 1960
places a background star at the appropriate separation and position
angle of the reported companion. We thus consider this object to be
single.


{\bf GJ~1155AB (1216+0258)} This object was previously thought to have
a white dwarf companion, but \citet{Gianninas(2011)} report that the
white dwarf is actually a misclassified M dwarf. A survey of SDSS
objects by \citet{Kleinman(2004)} did not spectroscopically confirm
the companion as a white dwarf. We therefore consider the companion to
be an M dwarf.

{\bf GJ~465 (1224-1814)} \citet{Heintz(1986)} notes this object may
yet be a long-term binary. We note this system as a candidate multiple
in Table \ref{tab:suspects}.

{\bf GJ~471 (1231+0848)} \citet{Poveda(1994)} report that this object
is a common proper motion companion to the binary GJ~469AB with a
separation of 2490\arcsec. While the weighted mean trigonometric
distances from \citet{vanAltena(1995)} and \citet{vanLeeuwen(2007)}
agree within the error bars (73.13$\pm$1.28 mas for GJ~471 versus
74.77$\pm$3.39 mas for GJ~469AB), their proper motions are
significantly different: 822 mas yr$^{-1}$ at $\theta$ $=$ 231\arcdeg
~for GJ~471 versus 685 mas yr$^{-1}$ at $\theta$ $=$ 248\arcdeg ~for
GJ~469AB. Thus, we consider GJ~471 a single star.

{\bf GJ~477AB (1235-4556)} This object is flagged in the Hipparcos
DMSA \citep{Lindegren(1997)} as having a stochastic solution,
indicating that it is a probable short-period astrometric
binary. \citet{Zechmeister(2009b)} note it as an SB1 using VLT$+$UVES,
with the companion being low-mass or a brown dwarf. We treat the
companion as a stellar component.


{\bf GJ~1167 (1309+2859)} \citet{Jahreiss(2008)} note a B component
with $\mu$ $=$ 0\farcs292 yr$^{-1}$, $\theta$ $=$ 234.9$^{\circ}$, but
then note that the two stars are not physically associated, as the
photometric distance for B is 190 pc while the trigonometric distance
for A is 12 pc. We calculate the {\it ccddist} for A to be 14.2
pc. \citet{Janson(2012)} note that A is single in their survey. We
conclude that GJ~1167`B' is not a CPM companion to GJ~1167A and that
GJ~1167 is a single system.



%

{\bf GJ~541.2AB (1417$+$4526)} This system had too small a proper
motion \citep[47 mas yr$^{-1}$ at a position angle of
  109.8\arcdeg;][]{vanLeeuwen(2007)} to confirm its secondary
component with a separation of 55.2\arcsec ~at a position angle of
204\arcdeg. Our image, taken in 2013, provided 18 years of coverage
when blinked with the 1995 SuperCOSMOS plate, but resulted in the
system moving less than an arcsecond during our blinking
survey. However, the {\it Gaia} DR2 catalog confirms this system as a
binary, reporting parallaxes and proper motions for both components
that agree both with each other and the HIP parallax of 52.60$\pm$1.25
mas: $\pi$$_A =$ 51.43$\pm$0.03 mas and $\mu$$_{A} =$ 47.3 mas
yr$^{-1}$, $\theta$$_{A} =$ 113\arcdeg; $\pi$$_B =$ 51.74$\pm$0.16 mas
and $\mu$$_{B} =$ 47.8 mas yr$^{-1}$, $\theta$$_{B} =$ 113\arcdeg. We
therefore treat this system as binary.




{\bf GJ~680 (1735$-$4840)} A companion at $\rho =$ 3\farcs94,
$\theta =$ 323.7\arcdeg ~with $\Delta$$H$ $=$ 1.93 mag was reported in
\citet{Ward-Duong(2015)}. However, whilst performing PSF photometry in
the crowded field in order to deblend the magnitudes of the primary
and secondary, we noted that the alleged secondary had not moved with
the primary between epochs of photometry taken eighteen months
apart. With a proper motion of 462 mas yr$^{-1}$, the two stars would
have moved a pixel and a half. This was the case with the primary, but
not the alleged secondary. Therefore, we deem the companion a
background star and not physically associated.

{\bf GJ~687 (1736+6820)} \citet{Montet(2014)} cite
\citet{Jenkins(2009)} for the M3.5 companion that they note in their
Table 2. But the spectral type of the {\it primary} is
M3.5. \citet{Jenkins(2009)} do not note any additional information
about a companion.

We find that the WDS lists GJ~687 as the B component to the F5 star
HD~160861. The parallax for the F5 star is 11.20 mas \citep[89
  pc;][]{vanLeeuwen(2007)} which places this object at a much larger
distance than the M dwarf, which has $\pi$ $=$ 220.47 mas \citep[4.5
  pc;][]{vanAltena(1995),vanLeeuwen(2007)}. It appears that the `A'
component, the F5 star (at $\alpha$ $=$ 17:36:42, $\delta$ $=$
$+$68:22:59, compared to $\alpha$ $=$ 17:36:25.9, $\delta$ $=$
$+$68:20:20 for GJ~687) is an SB, for which the measurement by
\citet{McAlister(1987c)} pertains. \citet{Tokovinin(1992b)} notes
GJ~687 as both an astrometric and speckle binary in the table in that
paper named `Long-period spectroscopic binaries'. But this, too, is
likely for the F dwarf. The companion is deemed to be optical, and we
consider GJ~687 to be single.





{\bf LTT~15769 (1945+3223)} This system is listed as double in the
{\it Hipparcos} DMSA \citep{Lindegren(1997)} with $\rho$ $=$
12\farcs76, $\theta$ $=$ 339$^{\circ}$ with a quality code of `D',
indicating an uncertain solution. Blinking the SuperCOSMOS $R$ plate
image with an $I$-band image taken with the Lowell 42in image results
in an epoch difference of nineteen years and reveals the motion of the
primary, but not that of the bright `secondary' with a $\Delta$$H_p$
of 2.3 mag. We thus refute this low probability companion and deem the
system `single.'


{\bf GJ~793 (2030$+$6527)} \citet{Weis(1991a)} lists this object in
the `Rejected Pairs' table (Table 5) and note that the alleged
companion is not listed in either the Luyten or Giclas catalogs. The
SuperCOSMOS photographic plates had an epoch spread of only one year,
so we used the 300-second image taken during our imaging campaign at
the Lowell 42in to blink with the SuperCOSMOS $I-$band archival
image. This resulted in a $\delta$t $=$ 20 yr. No common proper motion
companion was detected. We confirm that this object is single.



{\bf GJ~873 (2246$+$4420)} This object was reported by
\citet{vandeKamp(1972)} to be an astrometric binary, but this
detection was later found to be due to systemmatic errors in the
micrometric separation measurements \citep{Heintz(1976)}.

\citet{Young(1987)} initially report this object as a high probability
SB1, but then note in the appendix to that paper that the detection is
tentative due to the low velocity amplitude of the signal.

\citet{Helminiak(2009)} also investigate this system, citing a `B'
component that they infer is a real binary with spectral type G, but
not associated with the `A' component, our M dwarf. We confirm by
blinking that the two are not physically bound, as the G dwarf binary
has a very different proper motion ($\alpha_{\mu}$ $=$ 8.6 mas,
$\delta_{\mu}$ $=$ -2.0 mas) from that of the M dwarf component
($\alpha_{\mu}$ $=$ -705 mas, $\delta_{\mu}$ $=$ -461 mas), and
thus the two systems do not move together. Also, the $V$ magnitudes
for the two `components' are not physically possible if they are
located at similar distances: the M dwarf has $V =$ 10.22, while the G
dwarf has $V =$ 10.66.

\citet{Tanner(2010)} report two unconfirmed companions found via AO,
but the two candidates are too faint to have 2MASS magnitudes
available. 

Finally, \citet{Docobo(2010)} observed this object using speckle
interferometry on a 6m telescope and did not resolve a companion. They
were able to resolve companions down to angular separations of 22 mas,
corresponding to 0.11 AU at the object's distance of ~5 pc.

We consider this object to be a single star.


\section{Overall Results}
\label{sec:results_all}

We first report the multiplicity and companion rates for the
1120 M dwarfs surveyed, before any corrections are applied.  The {\it
  Multiplicity Rate} (MR) is the percentage of all systems that are
multiple, regardless of whether the system is double, triple or higher
order.  For example, discovering that a member of a binary system has
additional close companion makes the system a triple, but would not
affect the multiplicity rate.  The {\it Companion Rate} (CR) is the
average number of companions per primary in the sample, so higher
order multiples {\it do} affect this rate, as they add a companion to
the statistics.  The equations describing these percentages are given
below, where $N_S$ is the number of Singles, $N_D$ is the number of
Doubles, $N_T$ is the number of Triples, $N_{Qd}$ is the number of
Quadruples, and $N_{Qn}$ is the number of Quintuples, the highest
order multiples so far known among M dwarf primaries within 25 pc.
The denominator in both cases is 1120.

\begin{equation}
MR = 100 \ast \frac{N_D + N_T + N_{Qd} + N_{Qn}}{N_S + N_D + N_T + N_{Qd} + N_{Qn}}
\end{equation}

\begin{equation}
CR = 100 \ast \frac{N_D + 2N_T + 3N_{Qd} + 4N_{Qn}}{N_S + N_D + N_T + N_{Qd} + N_{Qn}}
\end{equation}

\noindent We analyze all companions in relation to the primary of the
system, even if they are members of sub-binaries or -triples.

\subsection{Uncorrected Multiplicity and Companion Rates for Confirmed Companions}
\label{subsec:confirmed}

Among the 1120 M dwarfs searched, there are 265 multiple systems with
310 new and confirmed {\it stellar} companions to their primaries,
resulting in an initial uncorrected multiplicity rate of MR =
23.7$\pm$1.3\% and an uncorrected stellar companion rate of CR =
27.7$\pm$1.3\%.  The ratios of singles:doubles:triples:higher-order
systems is 856:223:37:3, corresponding to 76:20:3:0.3\%.  For
comparison, \citet{Raghavan(2010)} found 56:33:8:3\% for
singles:doubles:triples:higher-order systems for companions (including
brown dwarfs) in a sample of 454 solar-type stars.  If we include the
known brown dwarf companions to M dwarfs, the ratios change only
slightly, to 844:230:41:2:2, corresponding to 75:21:4:0.3\%.


\subsection{Adjustment to the Multiplicity Rate at Small Separations}
\label{subsec:corrections}

Because this survey was not uniformly sensitive to systems with
companions at $\rho$ $<$ 2\arcsec, a correction should be made in
order to determine a final multiplicity rate.  The sample of M dwarfs
within 10 pc appears to be at least 90\% complete based on decades of
RECONS work.  Thus, this volume-limited 10 pc sample provides a
reasonably complete set of stars that can provide insight to the
stellar companion rate at small separations. If effectively all of the
primaries have been targeted by some kind of high resolution
technique, an adjustment based on the stellar multiplicity rate of
those objects for the closest companions can be determined and applied
to the sample of 1120 M dwarfs in the full survey outlined here. We
note that 2\arcsec ~separations correspond to 20 AU at 10 pc, but to
50 AU at 25 pc. Thus, we perform the correction based on a 50 AU
separation.


A literature search for spectroscopic and high resolution imaging
studies targeting M dwarfs within 10 pc was performed to determine the
companion population at small separations.  These two techniques cover
most of the separation phase space for stellar companions at
separations $\rho$ $<$ 50 AU.  It was found that all but two systems
either already had a close companion at $\rho$ $<$ 50 AU, or had been
observed with high resolution techniques, e.g., spectroscopy, {\it
  HST} imaging, speckle interferometry, lucky imaging, or long-term
astrometry.  Because 186 of the 188 M dwarfs within 10 pc have been
searched, we infer that we can use the 10 pc sample to correct for
unresolved companions with $\rho$ $<$ 50 AU at distances 10--25 pc.

 
  \begin{figure}
  \begin{center}
  \includegraphics[scale=0.35,angle=90]{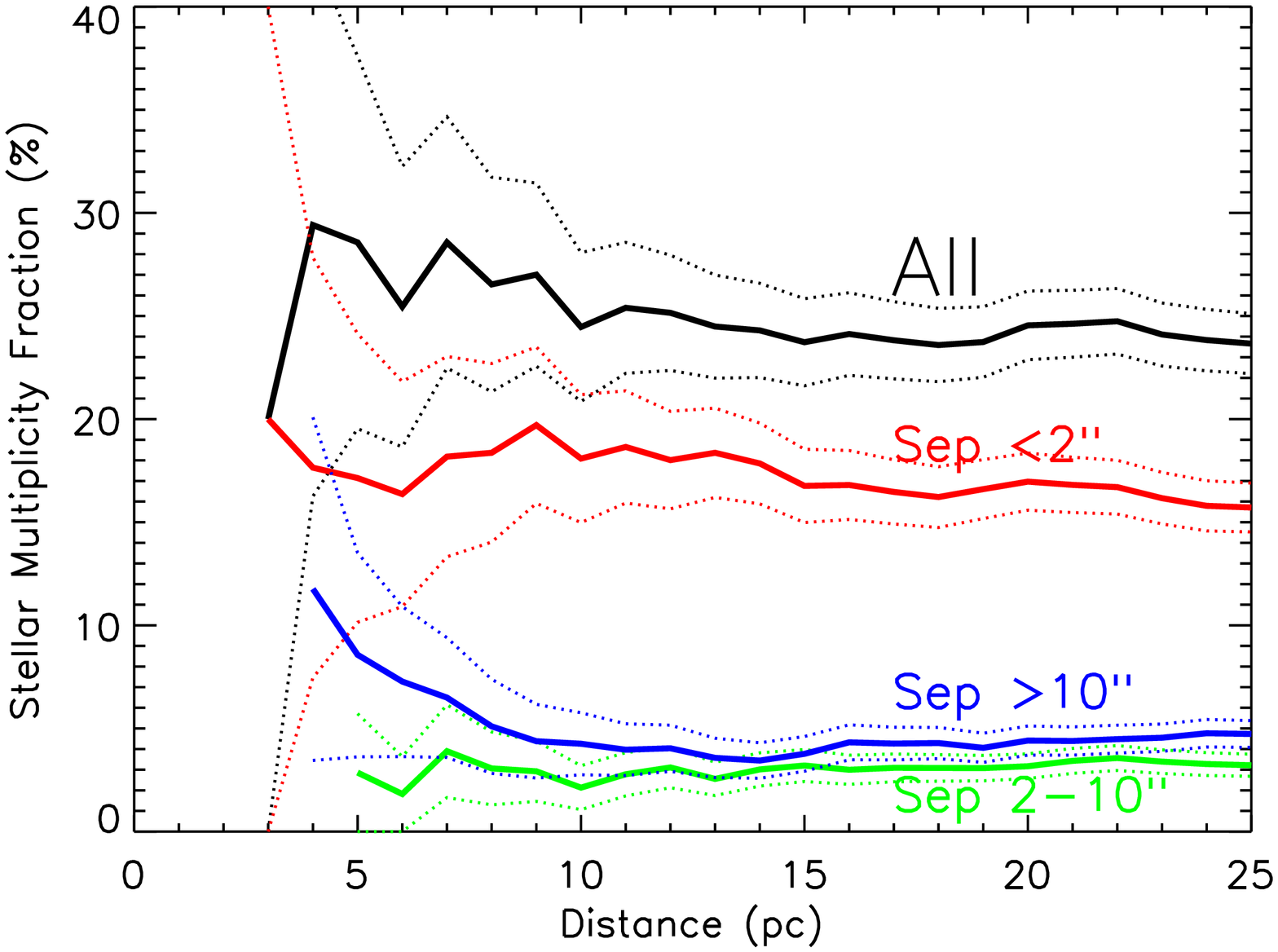}
  \includegraphics[scale=0.35,angle=90]{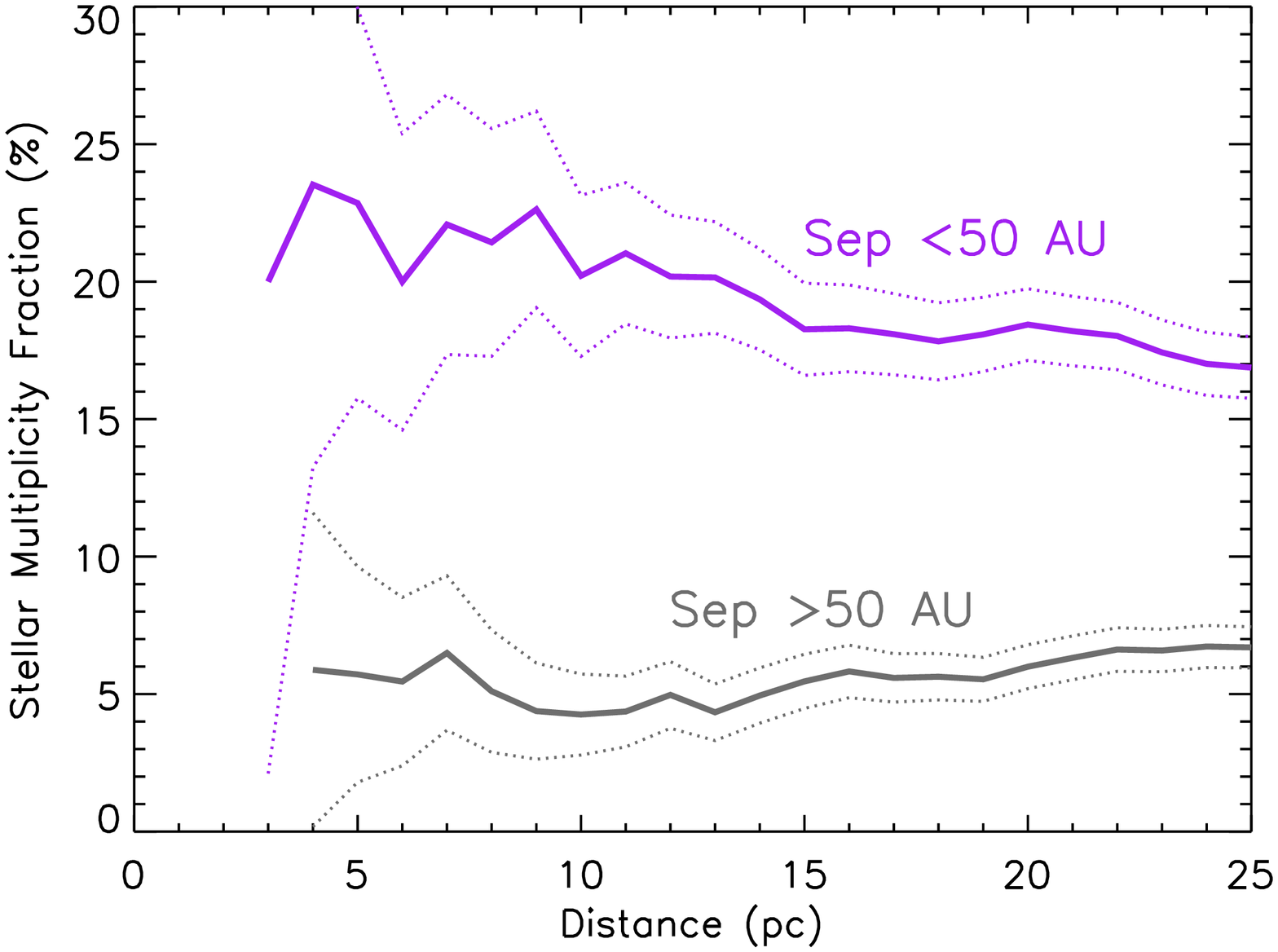}
 
  \caption{The cumulative multiplicity rate at different angular (top)
    and projected linear (bottom) separations for the 265 multiples
    among 1120 M dwarf systems, binned by one parsec and subdivided
    into separations of $\rho$ $<$ 2\arcsec, 2\arcsec $<$ $\rho$ $<$
    10\arcsec, and $\rho$ $>$ 10\arcsec ~(top) and separations of
    $\rho$ $<$ 50 AU and $\rho$ $>$ 50 AU (bottom).  The two rates for
    companions with separations greater than 2\arcsec~remain fairly
    constant from 10 to 25 pc, while the rate for separations smaller
    than 2\arcsec~decreases slightly from $\sim$10 to 25 pc, and
    particularly from 13--25 pc, indicating that a correction is
    warranted. The bottom panel shows the translation of angular to
    projected linear separations, which emphasizes the decreasing
    trend at $\rho$ $<$ 50 AU.  Neither suspected nor substellar
    companions are included in these graphs.  The dotted lines
    indicate the binomial errors on each line. The large scatter and
    errors on the curves at distances less than 10 pc are due to small
    samples of stars.}{\label{fig:run_mult}}
 
  \end{center}
  \end{figure}

Figure \ref{fig:run_mult} presents a graph of the running stellar
multiplicity at different angular and projected linear separations as
a function of distance for the sample of M dwarfs found within 25 pc.
For the 13 M dwarfs with only substellar companions, the system was
considered single for the purpose of the stellar multiplicity
calculation, while the two systems that had both an M dwarf and
a substellar companion were considered multiple.  For higher order
multiple systems with more than one companion, the smallest separation
between the primary and a companion was chosen to mitigate the likely
incompleteness at small $\rho$ in the top panel. All stellar
companions at all separations are included in the bottom panel.

The striking feature of Figure \ref{fig:run_mult} (top) is that most
stellar companions to red dwarfs are found at angular separations less
than 2\arcsec ~(50 AU).  Thus, it is not surprising that the two main
campaigns for detecting companions undertaken here yielded few new
objects, yet those searches needed to be done systematically.  It is
evident that the two sets of companions at separations greater than
2\arcsec~are effectively constant from 10 to 25 pc, indicating that
there are not significant numbers of overlooked companions at large
separations from their primaries.  The multiplicity rates for
companions with angular separations 2--10\arcsec~and $>$ 10\arcsec~are
3.2\% and 4.7\%, respectively, indicating that only 7.9$\pm$0.8\% of
the nearest M dwarfs have companions beyond 2\arcsec.  In contrast,
the curve for companions with separations smaller than
2\arcsec~decreases from 10 pc to 25 pc, implying that more close
companions remain to be found from 10--25 pc.  Many of these are
presumably the candidates discussed in \S \ref {subsubsec:suspects}
and listed in Table \ref{tab:suspects}.

The bottom panel illustrates the cumulative multiplicity rate
subdivided into projected linear separations of less than and greater
than 50 AU. Neither curve is flat. A decreasing trend is evident in
the curve showing separations $<$ 50 AU, illustrating that multiple
systems are missing at those separations. An increasing trend in the
$>$ 50 AU curve hints that multiple systems are missing at large
separations, as well; however, this corresponds to one missing
companion, which is within the errors of our correction (as described
below), so we do not apply a correction for this.

By comparing the multiplicity rate for $\rho$ $<$ 50 AU at 10 pc
(38/188 $=$ 20.2$\pm$2.9\%) to the rate at 10--25 pc (153/932 $=$
16.4$\pm$1.2\%), we find a correction of an additional 3.8$\pm$0.6\%
multiple systems that can be appropriately applied to stars from 10--25 pc.
This corresponds to 35 multiple systems among the 932 stars in this shell,
bringing the total number of M dwarf multiples within 25 pc to 265
$+$ 35 = 300 for 1120 primaries. Thus, the corrected multiplicity rate
for the entire sample is 26.8$\pm$1.4\%, indicating that roughly
three-quarters of M dwarfs are single, compared to only half of
solar-type stars.

We use the same method to calculate the correction to the companion
rate and find a companion rate of 23.9$\pm$3.1\% at 10 pc and a
correction of 5.7$\pm$0.8\%. The correction corresponds to 53 missing
companions to the 932 primaries at 10--25 pc. This brings the total
number of companions to 310 $+$ 53 = 363 for 1120 primaries and
results in a corrected companion rate of 32.4$\pm$1.4\%.

Thus, M dwarfs have a multiplicity rate of 20.2$\pm$1.2\% and a
companion rate of 23.9$\pm$1.4\% at separations less than 50 AU.

We note that we have identified 56 candidate multiple systems, 51 of
which are currently believed to be single; the other five are already
known to have a companion at a large angular separation. An additional
four of these 51 candidates are located within 10 pc.  All four have
been observed with both high resolution spectroscopy and imaging, with
no companions detected.  If any are found to have a companion, the
correction will change, but that appears unlikely to happen.

\subsection{Masses for Red Dwarfs in the Sample}
\label{subsec:mass_est}

In order to perform any quantitative analysis of the multiplicity
results related to stellar mass, a conversion from $M_V$ to mass using
a mass-luminosity relation (MLR) was necessary for each primary and
stellar companion in the sample.  For single M dwarfs or wide binaries
with separate $V$ photometry, this is straightforward.  For multiples
with separations less than roughly 4\arcsec, system photometry was
deblended at $V$ using Point Spread Function (PSF) fits.  For
multiples too close for PSF photometry, estimates of the $\Delta$mags
were made based on the information available in the literature.

\subsubsection{Deblending Photometry}
\label{subsubsec:deblend}

For systems with companions at separations too small (typically
1--4\arcsec) to perform effective aperture correction photometry, PSF
photometry was performed on frames acquired in Arizona and Chile
during the imaging program in order to measure $\Delta$$V$ for each
system.  First, the contribution from the sky background was
calculated from a `blank' part of the image.  The region around the
close pair being analyzed was then cropped to contain only the
relevant pair, and the background subtracted.  A Moffat curve was fit
to the PSF of the primary, the flux determined from the fit, and then
the primary was subtracted from the image, with care taken to minimize
the residual counts from the primary.  Gaussian and Lorentzian curves
were also tested, but it was found that Moffat curves provided the
best fits to PSFs from all of the 1m-class telescopes used in this
program.  A Moffat curve was then fit to the secondary component's PSF
and the flux calculated from the fit. The ratio of the fluxes
  (fr) of the primary and secondary yielded the $\Delta$$V$
  ($\Delta$$V$ $= -2.5log(fr)$) which, when combined with the
  composite $V$ photometry, provided the individual $V$ magnitudes
  ($V_{B} = V_{AB} + 2.5log(1 + 10^{0.4\Delta V}$); $V_{A} = V_{B} -
  \Delta V $) needed to estimate masses for each component in a
  multiple system.

For triples where all three components were closer than 4\arcsec, the
pair with the widest separation was deblended first using the
appropriate $\Delta$$V$ to calculate the deblended $V$ magnitude for
the single and the resulting pair.  Then the $\Delta$$V$ relevant to
the remaining pair was used to calculate individual $V$ magnitudes for
the components of the closest pair.

\begin{figure}
  \begin{center}
  \includegraphics[scale=0.35,angle=90]{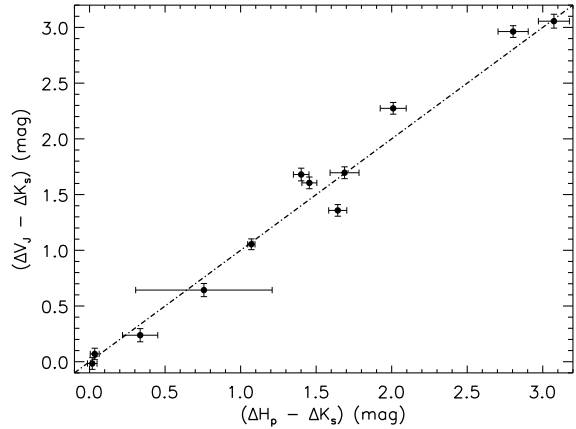}
 
  \caption{Magnitude differences as a function of color for twelve
    common proper motion pairs with $HIP$ parallaxes and individual
    $V_J$, $H_p$, and $K_s$ for each component. The dot-dash line
    indicates a one-to-one relation. Illustrated is the good agreement
    between the $\Delta$$H_p$ and $\Delta$$V_J$ for each
    pair.}{\label{fig:delhpk_delvjk}}
 
  \end{center}
  \end{figure}

For the 96 close multiples with $\Delta$mags reported in the
literature that were not in the $V$-band, the relations reported in
\citet{Riedel(2014)} were used to convert $\Delta$$R_{KC}$,
$\Delta$$I_{KC}$, $\Delta$$i'$, or 2MASS $\Delta$$J$, $\Delta$$H$,
$\Delta$$K$ filters to $\Delta$$V_J$.  Magnitude differences in the
      {\it Hipparcos} $H_p$ filter were considered to be equivalent to
      magnitude differences in $V_J$, as were any visual $\Delta$mags
      reported in the literature, e.g., those from any binary papers
      before $\sim$1995 that used photographic plates. For results
      using the Differential Speckle Survey Instrument \citep[DSSI;
      ][]{Horch(2009)} reported in \citet{Horch(2011a), Horch(2012a),
        Horch(2015)}, $\Delta$562 was adopted to be $\Delta$$V$ and
      $\Delta$692 was adopted to be $\Delta$$R$.  \citet{Horch(2009)}
      state that the 562 and 692 nm DSSI filters' central wavelengths
      are close to those of the $V$ and $R$ of the Johnson $UBVRI$
      system.  For observations from the RIT-Yale Tip-tilt Speckle
      Imager (RYTSI) \citep{Meyer(2006)} reported in
      \citet{Horch(2010), Horch(2012a)}, $\Delta$550 measurements were
      adopted as $\Delta$$V$, $\Delta$698 were adopted as $\Delta$$R$,
      and $\Delta$754 nm measurements were adopted as $\Delta$$I$.
      The $\Delta$814 measurements reported in \citet{Reid(2001a)}
      were assumed to be $\Delta$$I$.  For {\it HST} NICMOS data, the
      mean of the $\Delta$mags in the $F207M$ and $F222M$ filters were
      considered equivalent to $\Delta$$K_s$. We note that we consider
      the $\Delta$mags in these specific filters to be equivalent, not
      the individual component magnitudes themselves. As an example,
      we show in Figure \ref{fig:delhpk_delvjk} the good agreement
      between the $\Delta$$H_p$ and $\Delta$$V_J$ magnitudes.

 
We note that for pairs with $\Delta$$V$ values larger than 3.0, the
mass of the primary was calculated using the observed $V$ magnitude as
if it were single, as a companion with that $\Delta$$V$ contributes
negligible flux ($\sim$6\% or less) to the system.  Although the
$\Delta$$V$s are often not well-defined at differences this large, the
$\Delta$$V$ was simply added to the $V$ magnitude of the primary and
the mass estimated for the companion using that $V$.

The 43 pairs with unknown magnitude differences between the components
(i.e., typically those with separations $<$1\arcsec) required
estimates of the $\Delta$mag. For double-lined spectroscopic binaries
(SB2s), the secondary component would need to contribute enough light
to be able to see its spectral lines, making the system overluminous
and underestimating the photometric distance. Therefore, if the
{\it trigdist} was more than 1.4 times more distant than
the {\it ccddist}, we adopted a $\Delta$mag $=$ 0.5; if the
trigonometric distance {\it trigdist} was 1.3 - 1.4 times more
  distant than the photometric distance {\it ccddist}, we adopted a
$\Delta$mag $=$ 1.0. Single-lined spectroscopic binaries (SB1s) and
unresolved astrometric detections were treated identically. If the
{\it trigdist} was larger than 1.3 times the {\it ccddist}, we
adopted a $\Delta$mag $=$ 2.0, inferring that light from the secondary
component contributed to the photometry of the system; and if the {\it
  trigdist} was $<$ 1.3 times the {\it ccddist}, we adopted a
$\Delta$mag $=$ 3.0, as the companion did not appear to contribute
light to the system. These estimates were all done for the filter in
which the observations were done or reported; for example, an object
being observed in the $I-$band that was noted to have an astrometric
perturbation was assigned a $\Delta$$I$ $=$ 3, which was then
converted to $\Delta$$V$, as long as its two distances agreed.

The $\Delta$$V$ and deblended $V$ magnitudes for the individual
components of multiple systems are given in Table \ref{tab:photdata},
with a note if any of the aforementioned assumptions or conversions
were made.  This is the case for slightly less than half (138) of the
pairs.  We note that while some of the current $\Delta$$V$ estimates
are imperfect, the only way to aquire masses of these systems'
components is to measure their orbits, which was not a goal of this
project.

\subsubsection{Estimating Masses}
\label{subsubsec:mass_cal}

A robust mass-luminosity relation (MLR) for red dwarfs has recently
been provided by \citet{Benedict(2016)}, using extensive sets of
$HST$-$FGS$ and radial velocity data.  We use their results on 47
stars with masses 0.07--0.62 M$_{\odot}$ (average errors of 0.005
M$_{\odot}$) to estimate masses for the red dwarfs in the survey
outlined here. Using their mass-$M_V$ relation, which has a scatter of
0.017 M$_{\odot}$, the massive end of the M dwarf spectral sequence
for which we have adopted $M_V$ $=$ 8.8 corresponds to 0.63
M$_{\odot}$. The least massive M dwarf with $M_V$ $=$ 20.0 results in
a mass of 0.075 M$_{\odot}$, consistent with the lowest mass M dwarf
in \citet{Benedict(2016)}, GJ1245C with mass $=$ 0.076 $\pm$ 0.001
M$_{\odot}$.


%

For the 24 systems with orbits presented in the literature that
reported masses for individual components, these were used `as is', as
long as the masses were true masses and not minimum masses, i.e.,
$M\sin^3i$. In the nine cases where $M\sin^3i$ was reported, we
  used the composite $V-$band photometry for the system, in
  combination with the mass ratio ($M_2\sin^3i$/$M_1\sin^3i$), to
  estimate the $\Delta$$V$ between the components. While the
  inclination of the system is not known, it is not needed, as the
  mass ratio remains the same, regardless of inclination. However, the
  relation between the magnitude difference and mass ratio of the
  components is not linear (e.g., see \citealt{Benedict(2016)}) and
  will change with primary mass. We first assumed an initial
  $\Delta$$V$ to deblend the photometry of the components. We then
  used the parallax to calculate the absolute $V-$band photometry of
  each component. Finally, we used the MLR to estimate individual
  masses from the deblended photometry of the components and
  calculated the mass ratio. We iterated these steps, revising the
  $\Delta$$V$, until the estimated mass ratio matched that reported in
  the literature for the system.  The resulting $\Delta$$V$ estimates
and masses are listed in Table \ref{tab:photdata}.

Twenty-nine of the objects in our survey had masses presented in
\citet{Benedict(2016)} that were used to define the MLR.  This
provides an opportunity to assess the accuracy of the mass estimates
assigned here.  We find a mean absolute deviation between the measured
masses and our estimates of 10.5\%, differences that can be attributed
to cosmic scatter, as discussed in \citet{Benedict(2016)}.  Mass
estimates for all components are listed in Table \ref{tab:photdata}.

\subsubsection{Mass Ratios}
\label{subsec:mass_est}

 
  \begin{figure}
  \begin{center}
  \includegraphics[scale=0.35,angle=90]{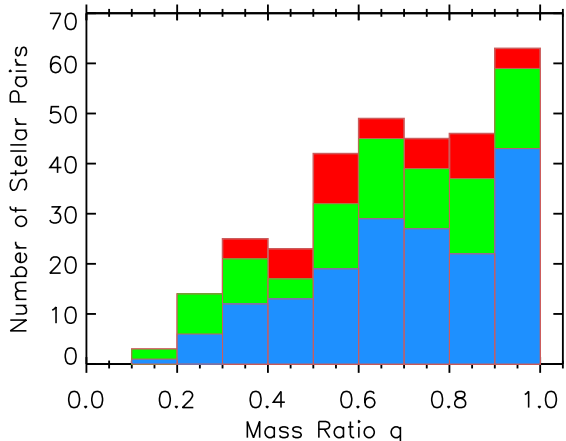}
  \caption{A histogram of the distribution of the mass ratios
    (M$_2$/M$_1$, M$_3$/M$_1$, etc.) of the 310 stellar pairs in the
    sample.  In blue are plotted pairs for which no conversion or
    assumption to $\Delta$$V$ was made, green indicates a conversion
    to $\Delta$$V$ from another filter, and red indicates that an
    assumption was made regarding the $\Delta$mag. The distribution is
    likely incomplete at mass ratios less than 0.5 and uniform from
    0.5 to equal mass ratios of 1.0, as the red shaded portions of the
    histogram will likely shift left-ward in the future once more
      accurate mass measurements are available. We note that no
    corrections due to `missing' multiple systems have been
    incorporated into this distribution, nor are brown dwarf
    companions included. \label{fig:q_star_hist}}
 
  \end{center}
  \end{figure}

With mass estimates for all the stars in multiple systems in hand, we
can evaluate the mass ratios of all companions relative to their
primaries.  Figure \ref{fig:q_star_hist} shows the distribution of the
mass ratios (q $=$ M$_{comp}$/M$_{pri}$) for the 310 pairs in the
sample.  All companions were analyzed in relation to the primary of
the system.  There are 225 binaries, as well as 37 triples that result
in 74 pairs of objects, one quadruple system yielding three pairs, and
the two quintuple systems yielding eight pairs.  In the instances
where $\Delta$mags reported in the literature for hierarchical systems
was other than with respect to the `A' component, these data were
calculated by first deblending the pair in question, calculating
individual magnitudes, estimating individual masses, and then
calculating the mass ratios in relation to the primary.


As shown in Figure \ref{fig:q_star_hist}, most of the pairs in the
sample have mass ratios larger than 0.5 with a distribution that may
very well be flat from q = 0.5--1.0. Once accurate mass
  determinations are available for the pairs represented in red in the
  histogram (primarily spectroscopic and astrometric binaries), the
  distribution may shift left-ward somewhat, as we have assumed a
  conservative $\Delta$ V of 3 mag for some of these targets.  For q
$<$ 0.5, there are a few incompleteness effects.  First, we have
excluded from the analysis the 18 known brown dwarf companions, which
affect mass ratio evaluations more for M dwarfs than for any other
type of star.  Second, we show in \S \ref{subsec:biases} that
primaries with masses 0.075--0.30 M$_{\odot}$ likely have companions
that have eluded detection.  These missing companions might fill in
various parts of the distribution, but only for q $>$ 0.25 (0.075
M$_{\odot}$/0.30 M$_{\odot}$), where 0.25 is the smallest q value
possible in that mass regime when considering only {\it stellar}
companions.

In order to assess whether the trend from q = 0.5--1.0 is real or due
to a bias in the way the $\Delta$$V$ values were calculated (as
described above), the histogram has been color-coded in order to
identify any trends: blue represents pairs with $\Delta$mag in $V$
(172 pairs), green represents pairs for which a conversion from a
$\Delta$mag other than $\Delta$$V$ was made (96 pairs), and red
represents pairs for which an assumption had to be made about the
$\Delta$mag between the two components, e.g., unresolved spectroscopic
and astrometric binaries (43 pairs). We do not see any strong
systematic trend that correlates with the assumptions or conversions
that were used for the mass estimates.

\subsection{The Separation Distribution for the Nearby M Dwarfs}
\label{subsec:sep_distrib}

 
  \begin{figure}
  \includegraphics[scale=0.35,angle=270]{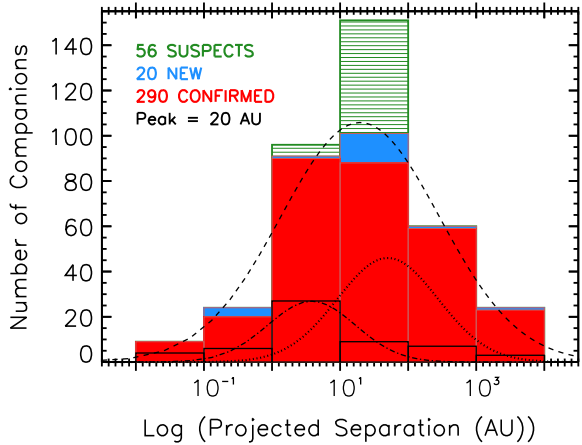}
  \vspace{0.25cm}
  \caption{A histogram of the distribution of the projected
    separations of all stellar companions from their red dwarf
    primaries, in log form. The 290 confirmed (in red), 20 new (in
    blue), and 56 suspected companions (in green) are indicated. The
    dashed curve is a Gaussian that has been fit to the distribution
    of confirmed and new (but not suspected) companions and has a peak
    at 20 AU, with $\sigma$$_{loga}$ = 1.16. The dot-dash line is a
    fit to the 56 M dwarf pairs within 10 pc, denoted by the black
    outline, indicating a peak at 4 AU. For comparison, the dotted
    line indicates the fit for solar-type stars from
    \citet{Raghavan(2010)}, which peaks at 51 AU or log P = 5.03, with
    $\sigma$$_{logP}$ = 2.28 yr. \label{fig:seps_hist}}
  \end{figure}

Figure \ref{fig:seps_hist} illustrates the projected separations of
the 290 confirmed, 20 new, and 56 suspected stellar companions from
their primaries. A Gaussian curve has been fit to the distribution of
confirmed and new (but not suspected) companions as a function of
log-separation, providing a reasonable fit to the distribution as
known in the current dataset. Exoploration of fitting a skewed
Gaussian curve to the data resulted in a skew value very close to
zero, with a peak at 20 AU. Therefore, we justify our use of a normal
Gaussian curve in this analysis. In an even larger sample that has
been completely searched for companions at small separations, a
different distribution may prove to be more appropriate.



The peak in the separation distribution of our sample falls at 20 AU,
with a broad spread.  This distribution peaks at larger projected
separations than the one presented in \citet{Janson(2014a)} for
mid-type M dwarfs (6 AU, their Figure 3). However, their search was
for companions at angular separations 0\farcs08--6\arcsec~from their
primaries at inferred distances within 52 pc (corresponding to 4--312
AU), whereas our study searched for companions at angular separations
out to 300\arcsec, corresponding to separations as large as 7500 AU at
the survey horizon of 25 pc. In addition, their results are based on a
sample containing 91 pairs in 85 multiples (from 79 binaries and six
triples), compared to our 310 pairs in 265 multiples.  Similarly, the
distribution peak that we find is also at much larger separations than
the 5.3 AU peak reported for stars 0.1 $\lesssim$ M/M$_{\odot}$
$\lesssim$ 0.5 by \citet{Duchene(2013)}, i.e.~M dwarfs. However,
  we recognize that our search is not complete at small separations
  and that the upper limits we have assumed for the separations of
  systems without measurements are likely overestimates. Therefore, we
  show the distribution of the 56 M dwarf pairs within 10 pc outlined
  in black. The dot-dash line indicates a peak at 4 AU for the 10 pc
  sub-sample, which is very close to the peaks reported above in
  \citet{Janson(2014a),Duchene(2013)}. Comparing our sample to
solar-type stars, the peak for M dwarfs at 20 AU (dashed line) is
roughly one-half the distance for the peak at 51 AU for the 259
companions found to 454 stars (dotted line) by \citet{Raghavan(2010)}.
In summary, stellar companions to M dwarfs are most often found
  at separations 4--20 AU, i.e., similar to those of the gas giant
  planets in our Solar System, and at least half the distance of
  stellar companions to more massive, solar-type stars.

The distribution shown in Figure \ref{fig:seps_hist} represents the
projected separations observed, but two possible shifts in the
distribution for these confirmed companions should be noted.
Incorporating the correction of a factor of 1.26 from
\citet{Fischer(1992)} for those systems for which no orbital
information is known (all but 33 of the systems presented here), would
shift the distribution to larger separations.  In the opposite sense,
we adopted projected separations of 1\arcsec~for unresolved systems,
which is the case for 33 of the confirmed and new companions, plus all
of the 56 suspected systems; these generally should be considered
upper limits.  Consequently, the distribution shown in Figure
\ref{fig:seps_hist} will shift leftward to smaller separations once
all known close companions have orbital semimajor axes determined.


\section{Discussion}
\label{sec:meaning}

\subsection{Understanding How Primary Mass Determines Companion Types, Separations, and the Multiplicity Rate}
\label{subsec:md_mult}

Exploring how the multiplicity of M dwarf systems depends on various
characteristics provides hints about star formation processes.
Because the target sample includes more than 1000 stars and more than
300 pairs, we can evaluate trends in multiplicity as functions of
mass, and separation.

\subsubsection{Mass Ratio as a Function of Primary Mass}
\label{subsubsec:q_pri_mass}
 
  \begin{figure}

  \hspace{-0.8cm}
  \includegraphics[scale=0.38,angle=90]{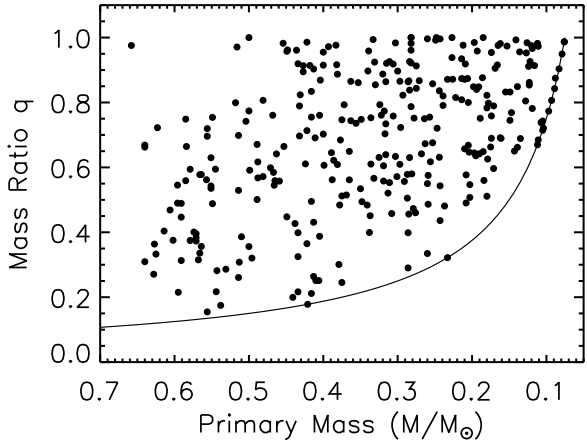}
  \caption{The mass ratios of 310 pairs as a function of primary
    mass. A trend of mass ratios increasing to unity for low mass
    primaries is noted. The solid line indicates the mass ratio
    boundary relative to the lowest mass star considered here for this
    survey, with M $=$ 0.075 M$_{\odot}$.  \label{fig:q_pri_mass}}
  \end{figure}

Figure \ref{fig:q_pri_mass} shows the mass ratios of all 310 pairs in
the sample, as a function of primary mass. We note that the
distribution is fairly uniform, with no preference for equal-mass
systems. There appear to be a dearth of equal mass companions to the
more massive M dwarfs (masses 0.52 -- 0.62 M$_{\odot}$). The apparent
trend of mass ratios converging to unity with the decrease in primary
mass is expected because we have set a hard limit on companion masses
by only including stellar companions, and there is a decreasing amount
of mass phase space available as the primary's mass approaches this
stellar boundary.

As noted in \S \ref{subsubsec:bds}, brown dwarf companions have been
excluded from the analysis. The percentage of M dwarfs with known
brown dwarf companions is 1.3\%, a fraction too low to fill in the
open region on the graph where low mass primaries have no secondaries
at large mass ratios. We note that this rate is consistent with the
number of solar-type stars with brown dwarf companions in the sample
studied by \citet{Raghavan(2010)}: 7/454 $=$ 1.5$\pm$0.6\%.
Furthermore, as shown by \cite{Dieterich(2012)}, there are not many
brown dwarfs found around small stars.

\subsubsection{Separation as a Function of Primary Mass}
\label{subsubsec:sep_pri_mass}

\begin{figure}
  \hspace{-0.8cm}
  \includegraphics[scale=0.38,angle=90]{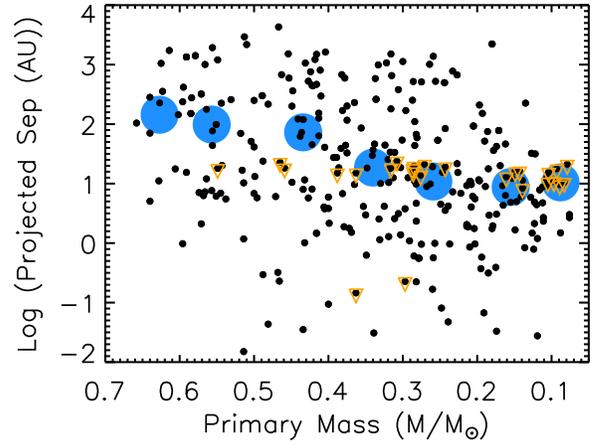}
  \caption{The log of the projected separation in AU as a function of
    primary mass for the 310 stellar pairs. The more massive M dwarf
    primaries seem to lack companions at very close separations. A
    weak trend of decreasing projected separation with primary mass is
    seen, as emphasized by the large blue points indicating the log of
    the median projected linear separation, with the median primary
    mass binned by 0.10 M$_{\odot}$. Pairs with estimated upper limits
    on their separations have been indicated with inverted orange
    triangles. \label{fig:sep_pri_mass}}
  \end{figure}

Next, we examine the explicit dependence of separation on primary
mass, as shown in Figure \ref{fig:sep_pri_mass}.  Of note is the trend
of decreasing projected separation with primary mass. In orange are
highlighted those pairs with upper estimates on their
separations. Because these pairs are generally spectroscopic and
  astrometric binaries, they typically have angular separations very
  much less than the assumed 1\arcsec. Specifically, the astrometric
  pairs in the sample with measured orbits have separations
  0\farcs0184 -- 0\farcs239, while the spectroscopic binaries in the
  sample with measured orbits have separations 0\farcs001 --
  0\farcs346. Thus, they will likely move to smaller separations once
higher resolution observations are obtained and/or their orbits are
measured.

 
The overall trends of Figures \ref{fig:q_pri_mass} and
\ref{fig:sep_pri_mass} illustrate that {\bf lower mass M dwarfs tend
  to have stellar companions at larger mass ratios and smaller separations
  than more massive M dwarfs.}

\subsubsection{The Multiplicity Rate as a Function of Primary Mass}
\label{subsubsec:mr_by_mass}

It is known that the multiplicity rate decreases with the mass of the
primary star for spectral types O through G \citep{Duchene(2013)}, and
one of the primary goals of our survey is to determine if that trend
continues as a function of primary mass through the M dwarfs.


 
  \begin{figure}
  \begin{center}
  \includegraphics[scale=0.35,angle=90]{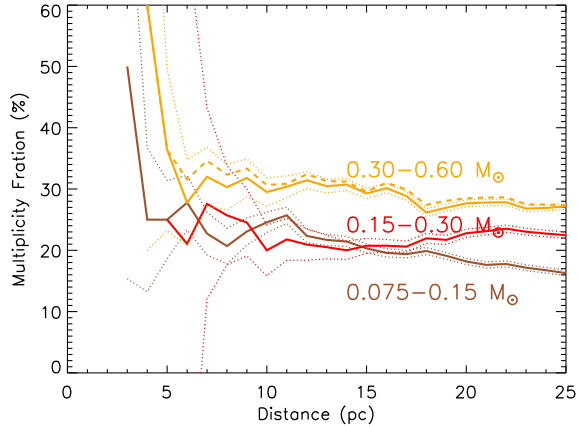}
  \caption{Cumulative stellar multiplicity rate by primary mass. Shown
    are the running multiplicity rates of the three mass subsets
    binned by one parsec, with the masses of the primaries calculated
    from deblended photometry: primary masses 0.30--0.60 M$_{\odot}$
    (orange line), with an uncorrected multiplicity rate at 25 pc of
    28.2$\pm$2.1\%; primary masses 0.15--0.30 M$_{\odot}$ (red line)
    with an uncorrected multiplicity rate at 25 pc of 21.4$\pm$2.0\%;
    and primary masses 0.075--0.15 M$_{\odot}$ (brown line), with an
    uncorrected multiplicity rate at 25 pc of 16.0$\pm$2.5\%. The
    dashed orange line indicates the addition of the 28 systems that
    have a primary mass larger than 0.60 M$_{\odot}$. The dotted lines
    on each curve indicate the Poisson errors. The large scatter and
    errors on the curves at distances less than 10 pc are due to small
    number of statistics. Neither suspected nor substellar companions
    are included. The largest mass bin has a higher multiplicity rate
    than the two smaller mass bins. It appears likely that multiple
    systems are missing from the smallest mass bin at distances 18 to
    25 pc, as the curve decreases at those
    distances.  \label{fig:run_mass}}
  \end{center}
  \end{figure}

Figure \ref{fig:run_mass} illustrates the multiplicity rate dependence
for three mass regimes subdivided into mass bins that span factors of
two --- 0.30--0.60 M$_{\odot}$, 0.15--0.30 M$_{\odot}$, and
0.075--0.15 M$_{\odot}$ --- as a function of distance for the target
stars. No corrections have been applied for undetected companions in
any of the three subsamples. The dashed orange line shows the
contribution of the 28 systems that have a primary mass larger than
0.60 M$_{\odot}$, of which ten are multiple. It is evident that the
highest mass bin of stars has the largest multiplicity rate at 15 pc
(31.1$\pm$3.4\%), roughly 10\% greater than for lower masses
(19.4$\pm$2.8\% and 19.8$\pm$3.6\% for the mid- and low-mass bins,
respectively) at the same distance horizon.  These values are
consistent with Figure 3 in \citet{Janson(2012)} and Figure 7 in
\citet{Janson(2014a)}, although their sample only extends to spectral
type M6, so they do not address our lowest mass bin.  It also appears
that there are slight dropoffs in detected multiples for the highest
and lowest mass bins from 15 to 25 pc, whereas the multiplicity rate
for the medium-mass bin remains fairly flat.  The dropoff for the
highest mass stars is only 2\%, a shift that we deem insignificant.
However, the dropoff for the lowest mass stars is 4\% from 15 to 25 pc
and in fact is 9\% from 11 to 25 pc.  This is likely because the
lowest mass primaries are the most difficult to study, so some
companions have yet to be detected. Nonetheless, the overall situation
is clear: {\bf high-mass M dwarfs have more stellar companions than
  low-mass M dwarfs.}


 
  \begin{figure}
  
  \hspace{-0.5cm}
  {\includegraphics[scale=0.37,angle=90]{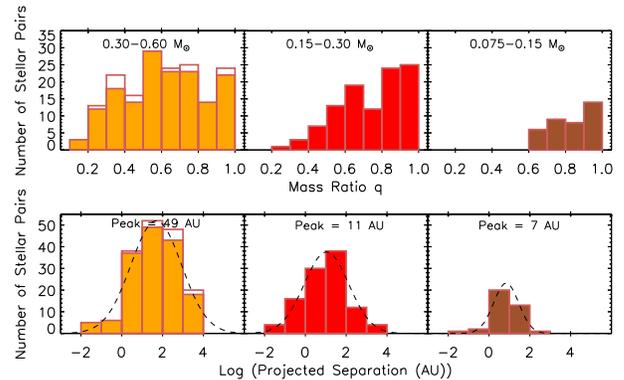}}
 
  \caption{Mass ratios and log-separation distributions for subsamples
    of M dwarfs as a function of primary masses. Mass ratios for
    primaries with masses 0.30--0.60 M$_{\odot}$ (in orange, top left)
    with the ten multiple systems with primaries more massive than
    0.60 M$_{\odot}$ shown unfilled; for primaries with masses
    0.15--0.30 M$_{\odot}$ (in red, top middle); mass ratios for
    primaries with masses 0.075--0.15 M$_{\odot}$ (in brown, top
    right). The mass ratio ranges shrink as a function of decreasing
    primary mass in part due to the imposed lower stellar companion
    mass limit of 0.075 M$_{\odot}$, although some effect due to
    gravitational binding energy is likely. Distribution of the
    projected separations for companions to stars with 0.30--0.60
    M$_{\odot}$ (bottom left), with the ten multiple systems with
    primaries more massive than 0.60 M$_{\odot}$ shown unfilled; for
    companions to stars with 0.15--0.30 M$_{\odot}$ (bottom middle);
    for companions to stars with 0.075 $-$ 0.15 M$_{\odot}$ (bottom
    right). The axis scales are the same for both trio of plots to
    highlight the differences between each mass subsample.  The peaks
    of the projected separation distributions shift to smaller
    separations with decreasing primary mass
    subset. \label{fig:mass_ratio_subsets}}
 
  \end{figure}

With our large sample, it is also possible to evaluate the mass ratio
and separation distributions by mass subset.  Figure
\ref{fig:mass_ratio_subsets} presents the mass ratios (top row of
three plots) and the log of the projected separations in AU (bottom
row of three plots, discussed below) of the 310 stellar components by
mass subset.

It is evident that the number of multiples in each subset decreases
with decreasing primary mass, as shown more explicitly in Figure
\ref{fig:run_mass}. Of note is the wide range in the mass ratios for
the most massive primaries, indicating that such stars tend to form
with companions filling nearly the entire suite of possible
masses. There is also a general shift to larger mass ratios with
decreasing primary masses. This trend is somewhat expected because we
have set a hard limit on companion masses by only including stellar
companions, and there is a decreasing amount of mass phase space
available as the primary's mass approaches this stellar/sub-stellar
boundary; as noted in \S \ref{subsubsec:bds}, brown dwarf companions
have been excluded from the analysis. Thus, it appears that lower mass
red dwarfs may only form and/or gravitationally retain companions when
they are of comparable mass and at small separations.  {\bf We
  conclude that M dwarfs have nearly every type of lower mass
  companion star.}

We next evaluate how the mass of the primary drives the separations at
which companions are found.  The bottom three panels of Figure
\ref{fig:mass_ratio_subsets} show the log-projected separation
distributions for the three mass subsets.  Again, the axis scales are
the same between plots, and no substellar or suspected companions are
included in these histograms.  As in the top row of plots, the number
of multiples decreases as a function of primary mass.  It is evident
that the distribution peaks move to smaller separations as a function
of decreasing primary mass, following the trend from solar-type stars
to M dwarfs as a whole.  Comparison of our distribution peak for the
mid-mass sub-sample, which corresponds most closely to the mid-M
sample surveyed in \citet{Janson(2014a)}, shows a peak within the
range of the projected separations (11 AU) of the peak reported from
their survey (3--10 AU).

\subsection{Understanding How Tangential Velocity Influences the Multiplicity Rate and Companion Separations}
\label{subsec:age_vtan}

 
 
  \begin{figure}
  \hspace{-0.5cm}
  \includegraphics[scale=0.35,angle=90]{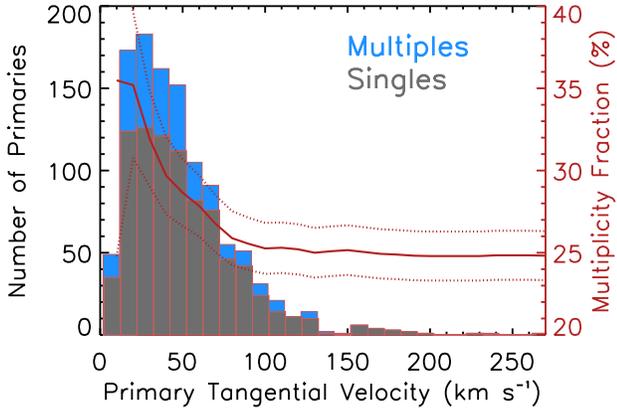}
  \caption{Histogram of primary tangential velocity. The tangential
    velocity of the primary (or single) component in each system is
    plotted in gray, with the primaries of confirmed multiple systems
    indicated in blue. Overplotted in red is the curve of the running
    multiplicity rate as a function of $v_{tan}$, indicating that
    slower moving objects tend to have companions slightly more
    often. Poisson errors on the multiplicity fraction are indicated by
    dotted lines. \label{fig:vtan_hist}}
  \end{figure}

Because it is known that the tangential velocity, $v_{tan}$, of a star
generally increases with age due to gravitational kicks from objects
in the Milky Way (usually from Giant Molecular Clouds), cool subdwarfs
will generally have larger velocities \citep[e.g., $v_{tan}$ $>$ 200
  km s$^{-1}$;][]{Jao(2017)} than young stars (estimated to be
$v_{tan}$ $<$ 35 km s$^{-1}$).\footnote{This is a 1-$\sigma$ deviation
  from the tangential velocity of the oldest nearby young moving group
  AB Doradus. We consider this a reasonable maximum $v_{tan}$ rate for
  young stars after comparing the total space motions of the ten
  nearby young moving groups listed in Table 1 of
  \citet{Mamajek(2016)} with ages $\lesssim$ 150 Myr. The mean for all
  ten (including AB Doradus) is 25 km s$^{-1}$.}  Thus, we investigate
M dwarf multiplicity as a function of $v_{tan}$, which we use as a
proxy for age.  Figure \ref{fig:vtan_hist} shows the $v_{tan}$
distribution of the observed sample, using the $v_{tan}$ of the
primary component, where singles are shown in gray and multiples in
blue.  There are noticeably more multiple systems with $v_{tan}$ $<$
50 km s$^{-1}$ than at larger $v_{tan}$.  The overplotted red curve
gives the running multiplicity rate as a function of $v_{tan}$,
showing a general decrease in multiplicity with increasing
$v_{tan}$. The apparent drop in the MR at $v_{tan}$ $\approx$ 15 km
s$^{-1}$ is due to the incompleteness of the sample at low proper
motions and corresponding $v_{tan}$ values, and is not a real
trend. This implies that as small stars age and experience kicks in
their travels through the Galaxy, they lose companions.  Alternately,
these older stars may have experienced different multiplicity
formation rates at the outset, either because of different (presumably
lower) metallicities, or different star formation environments.


 
  \begin{figure}
  \begin{center}
  \includegraphics[scale=0.35,angle=90]{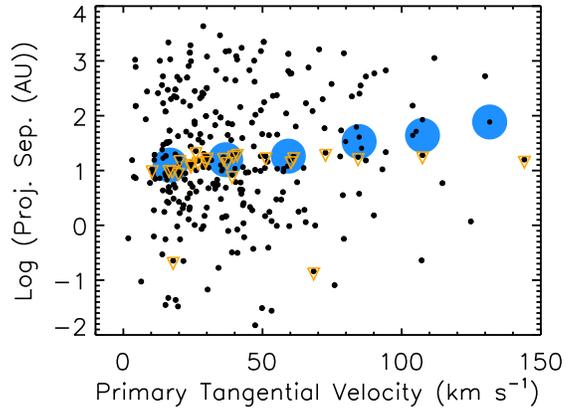}
  \caption{The log of the projected separation in AU as a function of
    tangential velocity for the 310 stellar pairs. Not shown is the
    one subdwarf binary with $v_{tan}$ $>$ 150 km s$^{-1}$. A weak
    trend of increasing projected separation with increasing
    tangential velocity is seen. This is emphasized by the large blue
    points indicating the log of the median projected linear
    separation as a function of the median tangential velocity, in
    bins of 25 km s$^{-1}$. Pairs with estimated upper limits on their
    separations have been indicated with inverted orange
    triangles. \label{fig:vtan_seps}}
  \end{center}
  \end{figure}


Figure \ref{fig:vtan_seps} further illustrates how multiplicity
changes with $v_{tan}$, showing the log of the projected separation as
a function of $v_{tan}$.  The blue open circles represent the log of
the median projected separation in 25 km s$^{-1}$ bins, illustrating a
weak trend of increasing projected separation with increasing
$v_{tan}$.  The pairs for which the separations are upper limits are
more numerous at the slower end of the plot, hinting that the median
$v_{tan}$ may decrease even further there, once true separations are
available for those close pairs.  {\bf Thus, the overall trends are
  that faster moving stars have fewer companions, and that the
  separations of multiples with higher velocities tend to be larger
  than their slower-moving counterparts.}

\subsection{The Luminosity and Mass Distributions}
\label{subsec:lf_mf}

 
  \begin{figure}
  \centering

  \includegraphics[scale=0.35,angle=90]{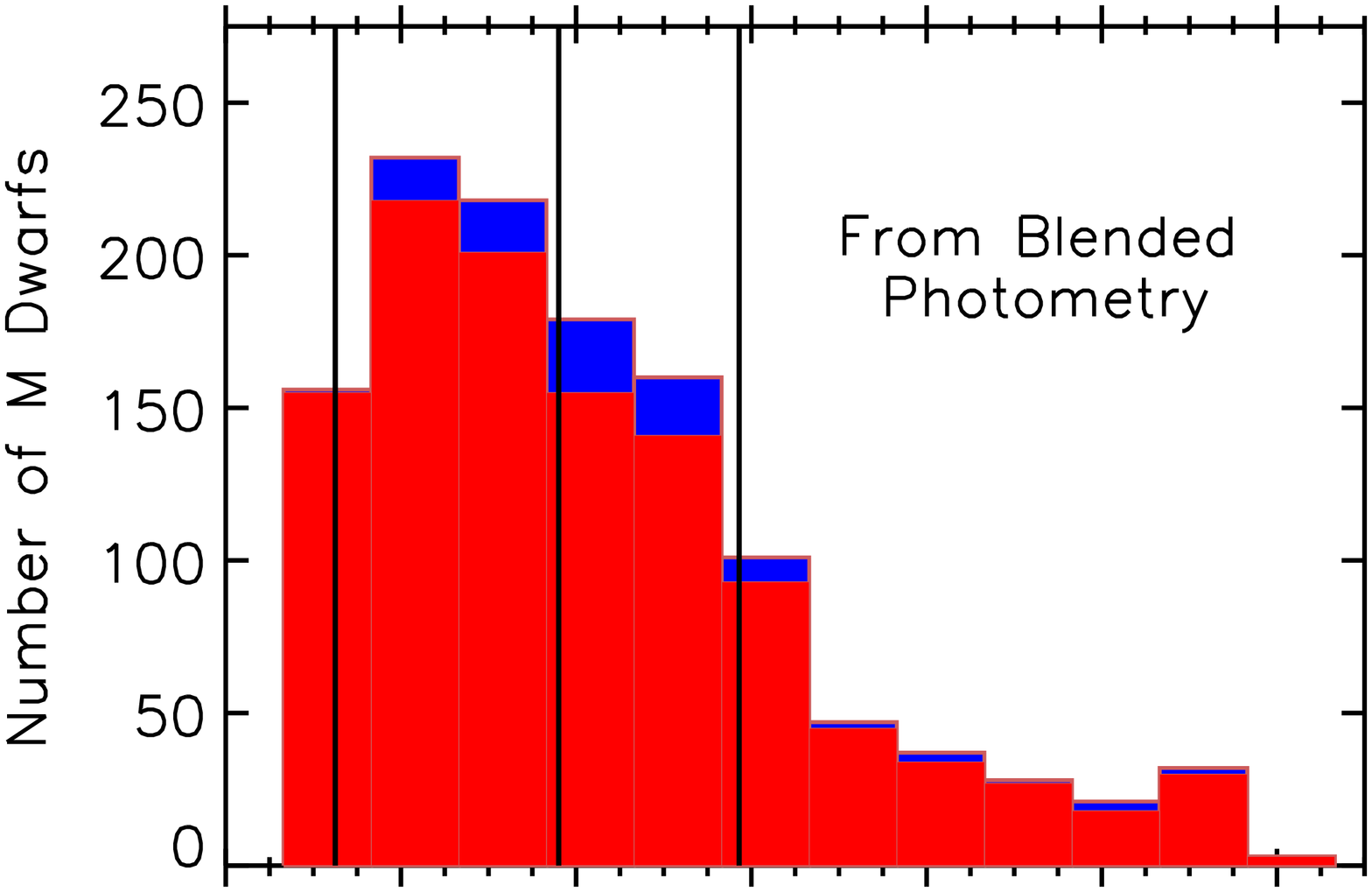}

  \vspace{-1.6cm}
  \includegraphics[scale=0.35,angle=90]{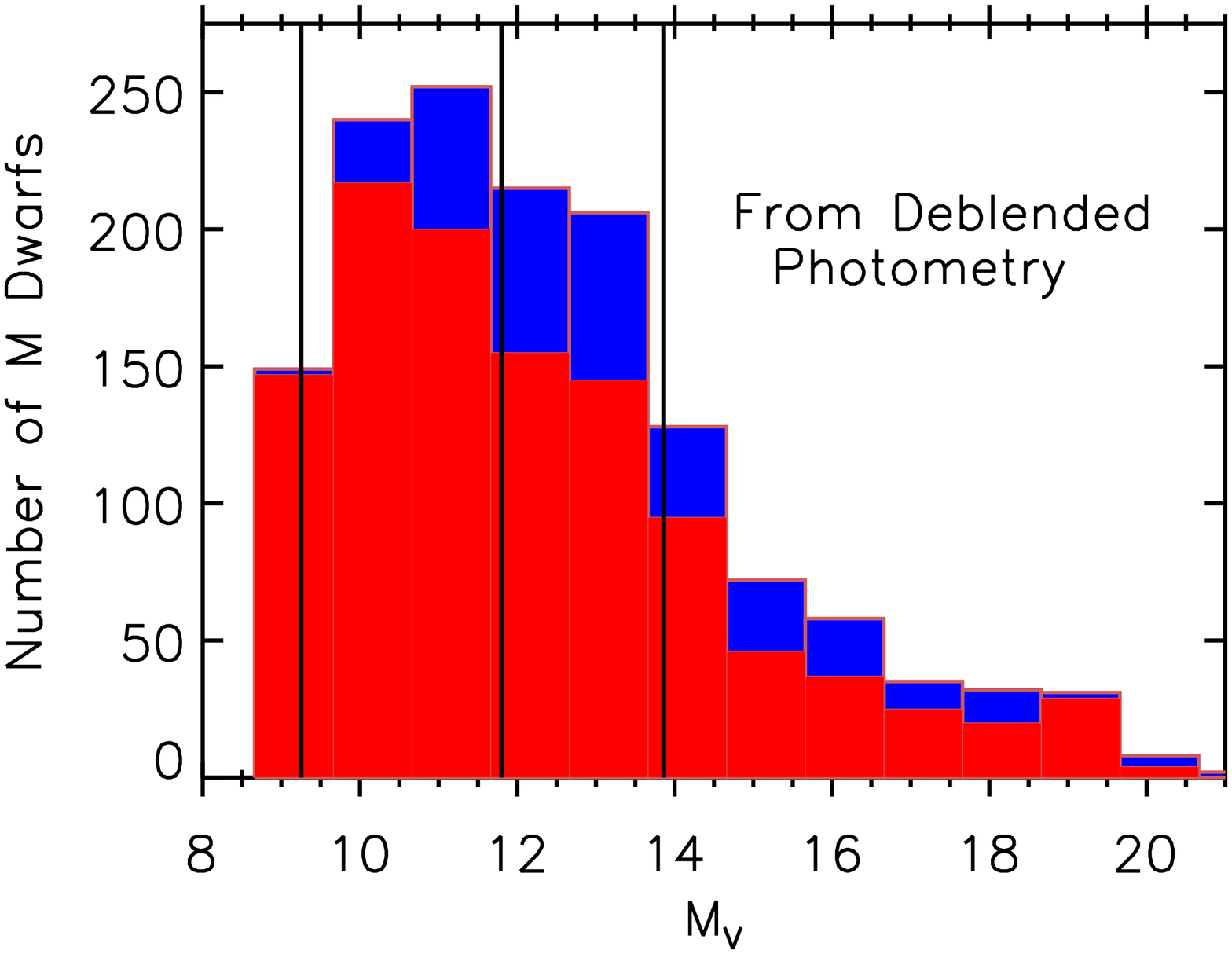}

  \caption{Luminosity Distributions. ({\it top}) The luminosity
    distribution for the 1214 M dwarfs in the sample with individual
    photometry. $M_V$ has been calculated from the blended
    photometry. ({\it bottom}) The luminosity distribution for all
    1432 M dwarf primaries and secondaries in the sample with $M_V$
    calculated from deblended photometry. Primaries are plotted in
    red; companions in blue. The vertical lines denote the
    subdivisions by factors of two in mass explored throughout the
    study. The contributions from the companions in the deblended
    luminosity distribution are greater than in the blended luminosity
    distribution. \label{fig:lum_fcn_db}}
  \end{figure}
 
 
  \begin{figure}
  \centering

  \includegraphics[scale=0.35,angle=90]{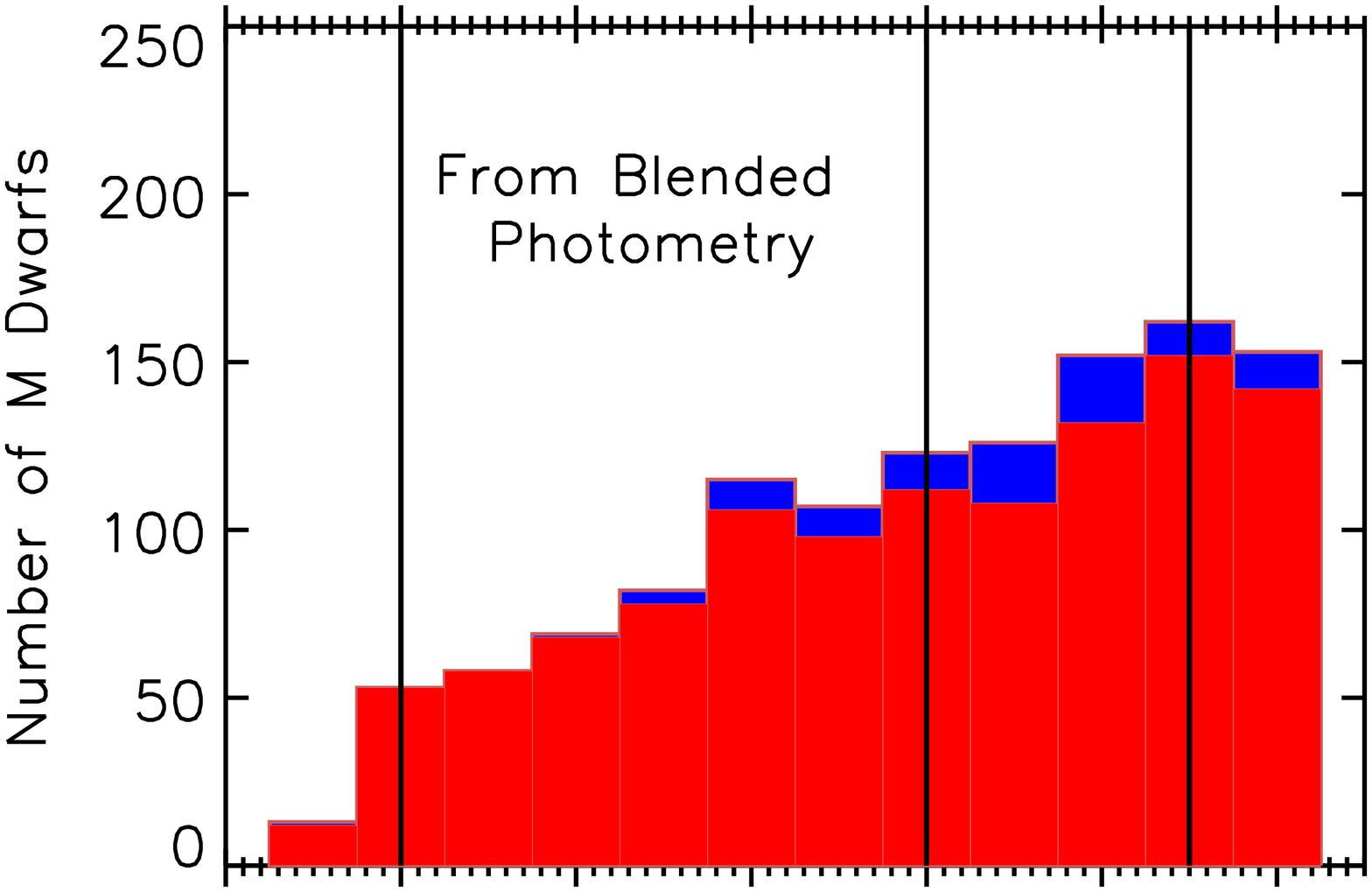}

  \vspace{-1.6cm}
  \includegraphics[scale=0.35,angle=90]{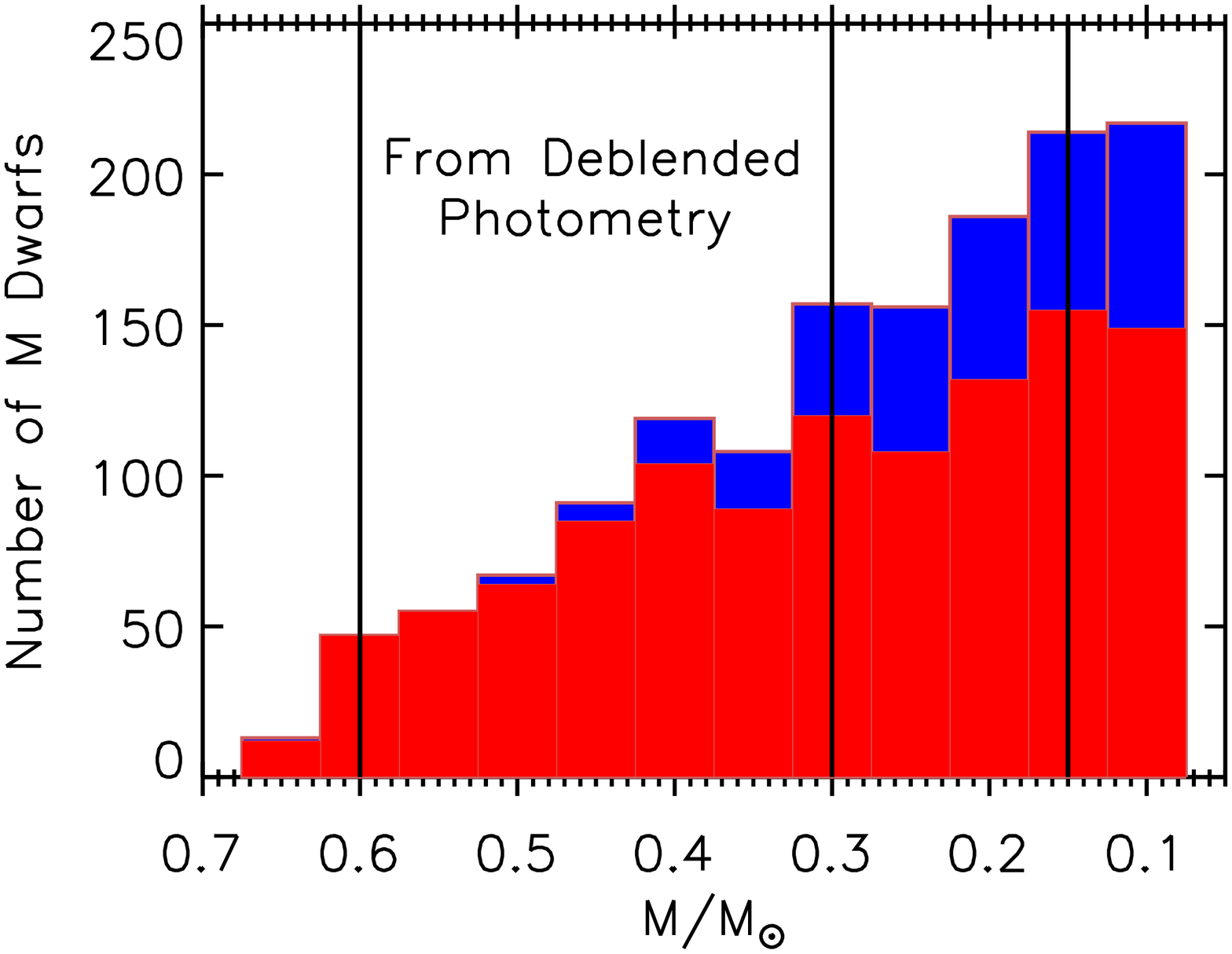}

  \caption{Mass Distributions. ({\it top}) The mass distribution for
    the 1214 M dwarfs in the sample with masses estimated from blended
    photometry. ({\it bottom}) The mass distribution for all 1432 M
    dwarfs in the sample with masses estimated from deblended
    photometry. Primaries are plotted in red; stellar companions in
    blue. The vertical lines denote the subdivisions by factors of two
    in mass explored throughout the study. There is a noticeable
    difference in the shapes of the distributions, as the
    contributions from the low mass companions contribute to the rise
    of the mass distribution to the end of the main
    sequence. \label{fig:mass_fcn_db}}
  \end{figure}

Shown in Figure \ref{fig:lum_fcn_db} are luminosity distributions for
our sample in $M_V$, calculated both before (top) and after (bottom)
deblending `joint' photometry with contributions from close
companions. Primaries are indicated in red, while companions are shown
in blue. The vertical lines at $M_V$ $=$ 9.25, 11.80, and 13.86
correspond to masses of 0.60, 0.30, and 0.15 M$_{\odot}$ and indicate
the factors of two in mass used to further divide the 25 pc sample for
the analysis presented in \ref{subsec:md_mult}. It is evident
that a substantial number of companions were hiding in the blended
photometry of the M dwarf primaries.


Figure \ref{fig:mass_fcn_db} illustrates the mass distribution for the
1214 M dwarfs with individual photometry with masses calculated from
blended photometry (top) and the distribution for all 1432 M dwarf
components (primaries plus all secondaries) found within 25 pc in the
survey after deblending photometry (bottom). Both histograms show a
gentle, but steady rise to the end of the stellar main sequence, what
we have defined to be 0.075 M$_{\odot}$. This trend is emphasized by
the additions of the companions in each histogram, as they are all
lower masses than their primaries, by definition. Given that
  there are missing M dwarf systems within 25 pc that are
  preferentially of low mass, the mass distribution will ultimately be
  even steeper than shown.

\subsubsection{Mass Contributions from Primaries and Hidden Companions}
\label{subsubsec:comps}

We now consider the contributions to the Galactic mass budget by M
dwarfs. Without any prior knowledge of unresolved companions, a naive
estimate of the mass of the 1214 M dwarfs in the sample with
indivdually measured photometry, from which masses are estimated,
yields 381 M$_{\odot}$. This includes the 1120 M dwarf primaries plus
their 94 well-separated secondaries.

However, of the 265 primaries in the sample of multiples, 210 have 257
companions located at angular separations of less than 2\arcsec~from
either their primaries or their widely separated secondaries,
resulting in `joint', or blended, $V$ magnitudes. After deblending, we
find that the 1120 primaries contribute 336 M$_{\odot}$. The deblended
companions have added 45 $M_{\odot}$ to the total M dwarf mass,
indicating that 11\% was hidden as unresolved stars. We find that at
least 17\% (66 M$_{\odot}$/402 M$_{\odot}$) of M dwarf mass is found
in companions, with unresolved companions contributing at least 68\%
(45 M$_{\odot}$/66 M$_{\odot}$) of the companion mass and donating
11\% (45 M$_{\odot}$/402 M$_{\odot}$) to the total mass budget of M
dwarfs. We emphasize that these values are all lower limits, as this
collection of M dwarfs has not been thoroughly canvassed for
companions at separations of less than 2\arcsec, where most M dwarf
companions are found.

\subsection{Comparisons to Previous Work on Other Stellar Populations}

We now put the results from this survey in perspective by making
comparisons to results from other M dwarf multiplicity surveys, as
well as to surveys of more massive stars. We then discuss how
unresolved companions affect the red dwarf luminosity and mass
distributions. Finally, directions for future work will be outlined.

\subsubsection{Comparison to Other M Dwarf Surveys}

As noted in \S \ref{sec:intro}, previous surveys have been done to
determine M dwarf multiplicity, but most have studied samples on the
order of a hundred stars. Some of the surveys \citep{Skrutskie(1989),
  Delfosse(1999c)} did not report a multiplicity rate in their
results, so they will not be addressed. Others explored the regions
around M dwarfs in search of different types of objects (brown dwarfs
in the case of \citet{Dieterich(2012)} and Jovian mass planets in the
case of \citet{Endl(2006)}) or at different separation regimes
(\citet{Dhital(2010)} and \citet{Law(2010)} explored only the wide
binary rate) and are thus not relevant to the present comparison.  For
example, searches for substellar objects can provide lower limits for
the types of companions found around M dwarfs, but stellar companions
are not always reported. \citet{Law(2006b), Law(2008)} probed
different sample sizes of late-type M dwarfs using lucky imaging and
report multiplicity rates that are different from each other by a
factor of two, but still within their large errors. We note that the
uncorrected multiplicity rate calculated here for the lowest mass bin
of M dwarfs --- 16.0$\pm$2.5\% --- agrees with that reported in
\citet{Law(2008)}: 13.6 $^{+6.5}_{-4}$\%.


A number of the other samples studied for M dwarf multiplicity
determination were volume-limited.  \citet{Henry(1990)} searched the 5
pc sample of M dwarfs, while \citet{Henry(1991)} and
\citet{Simons(1996)} extended the volume searched to 8
pc. \citet{Fischer(1992)} searched a varied sample of M dwarfs within
20 pc. The samples of \citet{Bergfors(2010)}, \citet{Janson(2012)},
and \citet{Janson(2014a)} were all within 52 pc, but most distances
were photometric parallaxes.

We find that our multiplicity rate result agrees with most of the more
recent surveys. \citet{Bergfors(2010)}, \citet{Janson(2012)}, and
\citet{Janson(2014a)} report MRs of 32\%, 27\%, and 21--27\%,
respectively. Our results also agree with the earlier studies of
\citet{Henry(1990)} and \citet{Henry(1991)} (34\% and 20\%) within the
errors, but are smaller than the studies of \citet{Fischer(1992)} and
\citet{Simons(1996)} (42\% and 40\%). It is likely that some of the
earlier studies simply did not have enough targets from which to
calculate accurate results with low statistical errors.

The only other sizeable survey that was volume-limited and had
trigonometric parallaxes available was that of
\citet{Ward-Duong(2015)}; however, their sample included late K dwarfs
and did not include any late M dwarfs. We find a slightly larger
multiplicity rate than the 23.5$\pm$3.2\% of \citet
{Ward-Duong(2015)}, although results agree within the
errors. Examination of the sample studied here reveals an additional
308 M dwarfs with parallaxes from sources other than
\citet{vanLeeuwen(2007)} that place them within 15 pc, 247 of which
are within the color-limits of their sample (3.65 $<$ $(V-K)$
$\lesssim$ 6.8).\footnote{This red limit has been estimated from the
  color-color diagram in Figure 1 in their paper, as it is not
  specified.}

Due to all of the targets in our multiplicity sample having accurate
trigonometric parallaxes, the study presented here has a number of
advantages over ones conducted by others. All of the targets
considered were reliably known to be within 25 pc. Because we measured
$VRI$ photometry for almost all targets lacking it, we were able to
use a homogeneous set of data on the same photometric system, combined
with the existing parallaxes, to calculate $M_V$ and thus, estimate
masses. Most other surveys were forced to use less accurate types of
distances to draw conclusions from their data. We were also able to
calculate projected separations that were more accurate than those of
others, as our sample has trigonometric distances. Finally, our survey
was comprehensive in two search regimes, while also able to infer the
presence of candidate companions using other methods.

\subsubsection{Comparison to More Massive Stars}
\label{subsubsec:massive_mult}

Listed in Table \ref{tab:other_mult} are the multiplicity statistics
for stars of other main sequence spectral types, along with the
percentages of all stars by that spectral type. While brown dwarfs are
not main sequence objects, they have been included for comparison. The
percentage of stars that they comprise has been purposely left blank,
as they are not stars, and in fact, the size of the brown dwarf
population is not well constrained.

While the multiplicity rate decreases as a function of primary mass,
it is evident that the number of stars increases with decreasing
mass. Massive stars of types OBA are the most rare, accounting for
fewer than 1\% of all stars \citep{Binney(1998)}, while solar-type FGK
stars make up $\sim$21\% of all stars \citep{Binney(1998)}. The M
dwarfs make up 75\% of all stars; thus, their multiplicity statistics
have the largest impact. The K dwarf multiplicity rate is the most
uncertain, with no comprehensive multiplicity search having yet been
done for that spectral type, although efforts to remedy this are
currently underway by members of the RECONS group. The thorough study
presented here provides an anchor for the statistics at the low end of
the stellar main sequence, enabling a complete picture of stellar
multiplicity.


\begin{deluxetable}{crcccc}
\centering
\setlength{\tabcolsep}{0.03in}
\tablewidth{0pt}
\tabletypesize{\small}
\tablecaption{Multiplicity of Main Sequence Stars \label{tab:other_mult}}
\tablehead{\colhead{Spectral}            &
	   \colhead{\% of}               &
 	   \colhead{Ref}                 &
	   \colhead{Mult.}               &
           \colhead{Comp}                &
	   \colhead{Ref}                \\
	   \colhead{Type}                &
	   \colhead{Stars}               &
	   \colhead{   }                 &
           \colhead{Rate}            &
           \colhead{Rate}            &
	   \colhead{   }                 }

\startdata
O             &   $<$0.1     &  2       &   $>$80       & 130           & 5,3     \\
B             &    0.1       &  2       & $>$70         & 100           & 6,3     \\
A             &    0.6       &  2       & $>$70         & 100           & 6,3     \\
F             &    3.3       &  2       &   50$\pm$4    &  75           & 6       \\
G             &    7.8       &  2       &   46$\pm$2    &  75           & 6       \\
K             &   10.2       &  2       &   41$\pm$3    &  56           & 6       \\
M             &   75.0       &  4       &   26.8$\pm$1.4&  32.4$\pm$1.4 & 1       \\
L,T           &   \nodata    &          &   22          &  22           & 3       \\
\enddata

\tablecomments{The columns indicate the spectral type of object, the
  percentage of stars that each spectral type comprises, along with
  the reference. Next, the Multiplicity Rate, the Companion Rate and
  the reference are listed.}

\tablerefs{(1) this work; (2) \citet{Binney(1998)}; (3)
  \citet{Duchene(2013)}; (4) \citet{Henry(2006)}; (5)
  \citet{Mason(2009a)}; (6) \citet{Raghavan(2010)}.}

\end{deluxetable}
 
 
  \begin{figure}
  \centering
  {\includegraphics[scale=0.35,angle=90]{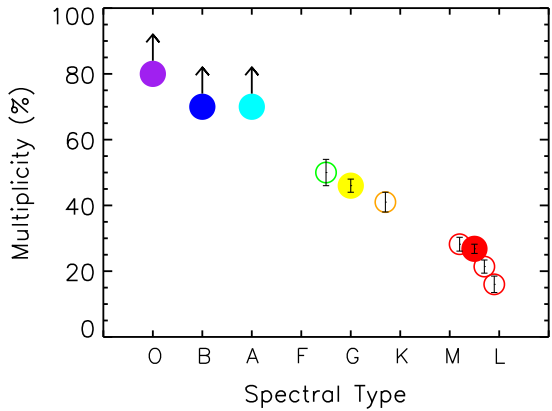}}
  \caption{Multiplicity rate as a function of spectral type. Shown is
    the MR for dwarf stars, with the rates for M dwarfs presented here
    in red. Open red points are the uncorrected MRs for the three mass
    bins explored throughout this paper, while the solid point is the
    total corrected MR for all M dwarfs. Values for stars more massive
    than M dwarfs are taken from the literature, as listed in Table
    \ref{tab:other_mult}. The open green and orange points are the
    blue and red subsamples from \citet{Raghavan(2010)}, while the
    solid yellow point is the average reported in that paper. The
    arrows indicate the MRs that are likely lower limits. We do not
    include the L dwarfs here. Clearly, the MR is a function of
    decreasing mass. \label{fig:all_multi}}
  \end{figure}

Figure \ref{fig:all_multi} indicates the multiplicity rates for dwarf
stars, with values taken from the literature for the more massive main
sequence stars. The clear decrease in multiplicity with decreasing
primary mass is evident.


From this comprehensive picture of stellar multiplicity, we can
determine the multiplicity rate of all star systems. Consider one
million stars. Table \ref{tab:total_mult} duplicates the percentages
of stars for each main sequence spectral type and the multiplicity
rate for each of those spectral types from Table
\ref{tab:other_mult}. In addition, however, is listed the number of
stars per one million that each spectral type would contribute and how
many of those would be multiple. The extra three percent of stars not
shown in the second column are made up of giants, supergiants, and
white dwarfs \citep{Binney(1998)}. Based on the numbers presented, we
can conclude that the multiplicity rate of all main sequence star
systems is 31$\pm$0.05\%, and that therefore, most stellar systems are
single.


\begin{deluxetable}{crrcr}
\centering
\setlength{\tabcolsep}{0.03in}
\tablewidth{0pt}
\tabletypesize{\small}
\tablecaption{Multiplicity Example \label{tab:total_mult}}
\tablehead{\colhead{Spectral}            &
	   \colhead{\% of}               &
 	   \colhead{\# of}               &
	   \colhead{Mult.}               &
           \colhead{\# Mult.}            \\
	   \colhead{Type}                &
	   \colhead{Stars}               &
	   \colhead{Stars}               &
           \colhead{Rate}            &
           \colhead{     }               }

\startdata
O             &  0.003       &      30  &    80      &      24     \\
B             &    0.1       &   1,000  &    70      &     700     \\
A             &    0.6       &   6,000  &    70      &   4,200     \\
F             &    3.3       &  33,000  &    50      &  16,500     \\
G             &    7.8       &  78,000  &    46      &  35,880     \\
K             &   10.2       & 102,000  &    41      &  41,820     \\
M             &   75.0       & 750,000  &    26.8    & 201,000     \\
\hline                                                          
              &   97.0       & 970,030  &    31      & 300,124     \\
\enddata

\end{deluxetable}

\section{Conclusions}
\label{sec:conclusions}

\subsection{Summary of Results}
\label{subsec:reddot_summary}

$\bullet$ We report 20 new and 56 suspected companions to M dwarfs
within 25 pc;

$\bullet$ We find a corrected multiplicity rate of 26.8$\pm$1.4\%
  and a corrected companion rate of 32.4$\pm$1.4\% for M dwarfs;

$\bullet$ We find that M dwarfs have multiplicity and companion
  rates of 20.2$\pm$1.2\% and 23.9$\pm$1.4\% at projected linear
  separations $<$ 50 AU;

$\bullet$ We find the uncorrected multiplicity rate of the three mass
subsets (0.30 --- 0.60 M/M$_{\odot}$, 0.15 --- 0.30 M/M$_{\odot}$, and
0.075 --- 0.15 M/M$_{\odot}$) to be 28.2$\pm$2.1\%, 21.4$\pm$2.0\%,
and 16.0$\pm$2.5\%;

$\bullet$ We find a uniformity in the mass ratio distribution, with no
apparent preference in companion mass for the nearby M dwarfs;

$\bullet$ The distribution of the projected separations of the
companions peaks at 4 -- 20 AU for M dwarfs, i.e., the scale of the
gas giants in our Solar System;

$\bullet$ A weak trend of decreasing projected separation with primary
mass was found;

$\bullet$ A possible relation between multiplicity and tangential
velocity was found, indicating that older, faster moving M dwarfs tend
to have fewer companions at smaller separations as a population than
their younger counterparts;

$\bullet$ We find that at least 17\% of M dwarf mass is contained in
companions, with 11\% of the total mass budget made up of `hidden'
stellar companions;

$\bullet$ Finally, we demonstrate that the mass distribution of our
volume-limited sample rises to the end of the stellar main sequence.


\subsection{What Is Yet to Come}
\label{subsec:future_work}

While the multiplicity survey presented here was comprehensive for
stellar companions to M dwarfs with separations 2 -- 300\arcsec, much
work remains to be done. Currently underway are high-resolution
speckle imaging and radial velocity studies to probe within
2\arcsec~of these nearby M dwarfs in order to complete our
understanding of M dwarf multiplicity at all separation regimes. The
results from these ongoing surveys will provide the separation and
delta-magnitude measurements needed for a more thorough understanding
of the characteristics of these multiple systems, e.g., the mass ratio
and separation distributions. In addition, the radial velocities being
measured provide the third velocity component needed to calculate
precise $UVW$ space motions. These space motions will allow further
exploration of the possible trend of stellar multiplicity with
tangential velocity.


{\it Gaia} will have five years or more of exquisite astrometric
measurements that will enable the detection of binaries. Any
astrometric binary orbits should provide inclinations, and thus the
dynamical masses of each component when combined with ground-based
spectroscopic orbits. In addition, {\it Gaia} should reveal many of
the very low-mass stars that have escaped detection to-date, providing
a more complete picture of the nearby M dwarf population that we will
study in the future.


\section{Acknowledgments}

We thank the anonymous referee for a rigorous review that greatly
improved the manuscript. JGW is especially grateful to Jonathan Irwin,
Willie Torres, and Eric Mamajek for illuminating and clarifying
discussions and suggestions, and to Douglas Gies, Harold McAlister,
Russel White, Sebasti{\`e}n L{\`e}pine and David Charbonneau for
constructive comments. We thank Brian Mason for access to the
Washington Double Star Catalog, a copy of which is housed at Georgia
State University. JGW extends a heartfelt thanks to David Fanning for
the availability of his IDL Coyote Graphics System.

We thank the members of the SMARTS Consortium, who have enabled the
operations of the small telescopes at CTIO since 2003, as well as
observers and observer support at CTIO, specifically Arturo Gomez,
Mauricio Rojas, Hernan Tirado, Joselino Vasquez, Alberto Miranda, and
Edgardo Cosgrove. We are indebted to the support staff and astronomers
at Lowell Observatory for their assistance, particularly Len Bright,
Larry Wasserman, Brian Skiff, and Ted Dunham.

Data products from the Two Micron All Sky Survey, which is a joint
project of the University of Massachusetts and the Infrared Processing
and Analysis Center/California Institute of Technology, funded by the
National Aeronautics and Space Administration (NASA) and the NSF have
been used extensively, as have the SIMBAD database and the Aladin and
Vizier interfaces, operated at CDS, Strasbourg, France.  This work has
made ample use of the Smithsonian Astrophysical Observatory/NASA
Astrophysics Data System. This work has made use of data from the
European Space Agency (ESA) mission {\it Gaia}
(\url{https://www.cosmos.esa.int/gaia}), processed by the {\it Gaia}
Data Processing and Analysis Consortium (DPAC,
\url{https://www.cosmos.esa.int/web/gaia/dpac/consortium}). Funding
for the DPAC has been provided by national institutions, in particular
the institutions participating in the {\it Gaia} Multilateral
Agreement.

This work was made possible by National Science Foundation (NSF)
grants 09-08402, 0507711, 1109445, and 141206, Sigma Xi
Grants-In-Aid-of-Research, the generous budget allotment of Georgia
State University that made possible access to the SMARTS telescopes at
CTIO and Lowell, and the Georgia State University Dissertation
grant. JGW was supported by a grant from the John Templeton Foundation
for a portion of the time that it took to finalize these results. The
opinions expressed here are those of the authors and do not
necessarily reflect the views of the John Templeton Foundation.

\software{IRAF \citep{Tody(1986),Tody(1993)}, SExtractor
  \citep{Bertin(1996)}, PSFEx \citep{Bertin(2011)}.}


\bibliographystyle{apj}
\bibliography{/home/jwinters/references/masterref}

\begin{thebibliography}{}
\expandafter\ifx\csname natexlab\endcsname\relax\def\natexlab#1{#1}\fi

\bibitem[{{Al-Shukri} {et~al.}(1996){Al-Shukri}, {McAlister}, {Hartkopf},
  {Hutter}, \& {Franz}}]{AlShukri(1996)}
{Al-Shukri}, A.~M., {McAlister}, H.~A., {Hartkopf}, W.~I., {Hutter}, D.~J., \&
  {Franz}, O.~G. 1996, \aj, 111, 393

\bibitem[{{Allen} {et~al.}(2007){Allen}, {Koerner}, {McElwain}, {Cruz}, \&
  {Reid}}]{Allen(2007)}
{Allen}, P.~R., {Koerner}, D.~W., {McElwain}, M.~W., {Cruz}, K.~L., \& {Reid},
  I.~N. 2007, \aj, 133, 971

\bibitem[{{Allen} \& {Reid}(2008)}]{Allen(2008)}
{Allen}, P.~R., \& {Reid}, I.~N. 2008, \aj, 135, 2024

\bibitem[{{Andrei} {et~al.}(2011){Andrei}, {Smart}, {Penna}, {d'Avila},
  {Bucciarelli}, {Camargo}, {Crosta}, {Dapr{\`a}}, {Goldman}, {Jones},
  {Lattanzi}, {Nicastro}, {Pinfield}, {da Silva Neto}, \&
  {Teixeira}}]{Andrei(2011)}
{Andrei}, A.~H., {Smart}, R.~L., {Penna}, J.~L., {et~al.} 2011, \aj, 141, 54

\bibitem[{{Anglada-Escud{\'e}} {et~al.}(2012){Anglada-Escud{\'e}}, {Boss},
  {Weinberger}, {Thompson}, {Butler}, {Vogt}, \& {Rivera}}]{Anglada(2012)}
{Anglada-Escud{\'e}}, G., {Boss}, A.~P., {Weinberger}, A.~J., {et~al.} 2012,
  \apj, 746, 37

\bibitem[{{Balega} {et~al.}(2013){Balega}, {Balega}, {Gasanova}, {Dyachenko},
  {Maksimov}, {Malogolovets}, {Rastegaev}, \& {Shkhagosheva}}]{Balega(2013)}
{Balega}, I.~I., {Balega}, Y.~Y., {Gasanova}, L.~T., {et~al.} 2013,
  Astrophysical Bulletin, 68, 53

\bibitem[{{Balega} {et~al.}(2007){Balega}, {Balega}, {Maksimov},
  {Malogolovets}, {Rastegaev}, {Shkhagosheva}, \& {Weigelt}}]{Balega(2007)}
{Balega}, I.~I., {Balega}, Y.~Y., {Maksimov}, A.~F., {et~al.} 2007,
  Astrophysical Bulletin, 62, 339

\bibitem[{{Barbieri} {et~al.}(1996){Barbieri}, {De Marchi}, {Nota}, {Corrain},
  {Hack}, {Ragazzoni}, \& {Macchetto}}]{Barbieri(1996)}
{Barbieri}, C., {De Marchi}, G., {Nota}, A., {et~al.} 1996, \aap, 315, 418

\bibitem[{{Bartlett} {et~al.}(2017){Bartlett}, {Lurie}, {Riedel}, {Ianna},
  {Jao}, {Henry}, {Winters}, {Finch}, \& {Subasavage}}]{Bartlett(2017)}
{Bartlett}, J.~L., {Lurie}, J.~C., {Riedel}, A., {et~al.} 2017, \aj, 154, 151

\bibitem[{{Benedict} {et~al.}(2000){Benedict}, {McArthur}, {Franz},
  {Wasserman}, \& {Henry}}]{Benedict(2000b)}
{Benedict}, G.~F., {McArthur}, B.~E., {Franz}, O.~G., {Wasserman}, L.~H., \&
  {Henry}, T.~J. 2000, \aj, 120, 1106

\bibitem[{{Benedict} {et~al.}(1999){Benedict}, {McArthur}, {Chappell}, {Nelan},
  {Jefferys}, {van Altena}, {Lee}, {Cornell}, {Shelus}, {Hemenway}, {Franz},
  {Wasserman}, {Duncombe}, {Story}, {Whipple}, \& {Fredrick}}]{Benedict(1999)}
{Benedict}, G.~F., {McArthur}, B., {Chappell}, D.~W., {et~al.} 1999, \aj, 118,
  1086

\bibitem[{{Benedict} {et~al.}(2001){Benedict}, {McArthur}, {Franz},
  {Wasserman}, {Henry}, {Takato}, {Strateva}, {Crawford}, {Ianna}, {McCarthy},
  {Nelan}, {Jefferys}, {van Altena}, {Shelus}, {Hemenway}, {Duncombe}, {Story},
  {Whipple}, {Bradley}, \& {Fredrick}}]{Benedict(2001)}
{Benedict}, G.~F., {McArthur}, B.~E., {Franz}, O.~G., {et~al.} 2001, \aj, 121,
  1607

\bibitem[{{Benedict} {et~al.}(2002){Benedict}, {McArthur}, {Forveille},
  {Delfosse}, {Nelan}, {Butler}, {Spiesman}, {Marcy}, {Goldman}, {Perrier},
  {Jefferys}, \& {Mayor}}]{Benedict(2002c)}
{Benedict}, G.~F., {McArthur}, B.~E., {Forveille}, T., {et~al.} 2002, \apjl,
  581, L115

\bibitem[{{Benedict} {et~al.}(2016){Benedict}, {Henry}, {Franz}, {McArthur},
  {Wasserman}, {Jao}, {Cargile}, {Dieterich}, {Bradley}, {Nelan}, \&
  {Whipple}}]{Benedict(2016)}
{Benedict}, G.~F., {Henry}, T.~J., {Franz}, O.~G., {et~al.} 2016, \aj, 152, 141

\bibitem[{{Bergfors} {et~al.}(2010){Bergfors}, {Brandner}, {Janson}, {Daemgen},
  {Geissler}, {Henning}, {Hippler}, {Hormuth}, {Joergens}, \&
  {K{\"o}hler}}]{Bergfors(2010)}
{Bergfors}, C., {Brandner}, W., {Janson}, M., {et~al.} 2010, \aap, 520, A54

\bibitem[{{Bertin}(2011)}]{Bertin(2011)}
{Bertin}, E. 2011, in Astronomical Society of the Pacific Conference Series,
  Vol. 442, Astronomical Data Analysis Software and Systems XX, ed. I.~N.
  {Evans}, A.~{Accomazzi}, D.~J. {Mink}, \& A.~H. {Rots}, 435

\bibitem[{{Bertin} \& {Arnouts}(1996)}]{Bertin(1996)}
{Bertin}, E., \& {Arnouts}, S. 1996, \aaps, 117, 393

\bibitem[{{Bessel}(1990)}]{Bessel(1990)}
{Bessel}, M.~S. 1990, \aaps, 83, 357

\bibitem[{{Bessell}(1991)}]{Bessell(1991)}
{Bessell}, M.~S. 1991, \aj, 101, 662

\bibitem[{{Bessell} \& {Weis}(1987)}]{Bessell(1987)}
{Bessell}, M.~S., \& {Weis}, E.~W. 1987, \pasp, 99, 642

\bibitem[{{Beuzit} {et~al.}(2004){Beuzit}, {S{\'e}gransan}, {Forveille},
  {Udry}, {Delfosse}, {Mayor}, {Perrier}, {Hainaut}, {Roddier}, {Roddier}, \&
  {Mart{\'{\i}}n}}]{Beuzit(2004)}
{Beuzit}, J.-L., {S{\'e}gransan}, D., {Forveille}, T., {et~al.} 2004, \aap,
  425, 997

\bibitem[{{Bidelman}(1985)}]{Bidelman(1985)}
{Bidelman}, W.~P. 1985, \apjs, 59, 197

\bibitem[{{Biller} \& {Close}(2007)}]{Biller(2007)}
{Biller}, B.~A., \& {Close}, L.~M. 2007, \apjl, 669, L41

\bibitem[{{Biller} {et~al.}(2006){Biller}, {Kasper}, {Close}, {Brandner}, \&
  {Kellner}}]{Biller(2006)}
{Biller}, B.~A., {Kasper}, M., {Close}, L.~M., {Brandner}, W., \& {Kellner}, S.
  2006, \apjl, 641, L141

\bibitem[{{Binney} \& {Merrifield}(1998)}]{Binney(1998)}
{Binney}, J., \& {Merrifield}, M. 1998, {Galactic Astronomy} (Princeton
  University Press)

\bibitem[{{Blake} {et~al.}(2008){Blake}, {Charbonneau}, {White}, {Torres},
  {Marley}, \& {Saumon}}]{Blake(2008)}
{Blake}, C.~H., {Charbonneau}, D., {White}, R.~J., {et~al.} 2008, \apjl, 678,
  L125

\bibitem[{{Bonfils} {et~al.}(2013){Bonfils}, {Delfosse}, {Udry}, {Forveille},
  {Mayor}, {Perrier}, {Bouchy}, {Gillon}, {Lovis}, {Pepe}, {Queloz}, {Santos},
  {S{\'e}gransan}, \& {Bertaux}}]{Bonfils(2013)}
{Bonfils}, X., {Delfosse}, X., {Udry}, S., {et~al.} 2013, \aap, 549, A109

\bibitem[{{Bonnefoy} {et~al.}(2009){Bonnefoy}, {Chauvin}, {Dumas}, {Lagrange},
  {Beust}, {Desort}, {Teixeira}, {Ducourant}, {Beuzit}, \&
  {Song}}]{Bonnefoy(2009)}
{Bonnefoy}, M., {Chauvin}, G., {Dumas}, C., {et~al.} 2009, \aap, 506, 799

\bibitem[{{Bowler} {et~al.}(2015){Bowler}, {Liu}, {Shkolnik}, \&
  {Tamura}}]{Bowler(2015)}
{Bowler}, B.~P., {Liu}, M.~C., {Shkolnik}, E.~L., \& {Tamura}, M. 2015, \apjs,
  216, 7

\bibitem[{{Burningham} {et~al.}(2009){Burningham}, {Pinfield}, {Leggett},
  {Tinney}, {Liu}, {Homeier}, {West}, {Day-Jones}, {Huelamo}, {Dupuy}, {Zhang},
  {Murray}, {Lodieu}, {Barrado Y Navascu{\'e}s}, {Folkes}, {Galvez-Ortiz},
  {Jones}, {Lucas}, {Calderon}, \& {Tamura}}]{Burningham(2009)}
{Burningham}, B., {Pinfield}, D.~J., {Leggett}, S.~K., {et~al.} 2009, \mnras,
  395, 1237

\bibitem[{{Chanam{\'e}} \& {Gould}(2004)}]{Chaname(2004)}
{Chanam{\'e}}, J., \& {Gould}, A. 2004, \apj, 601, 289

\bibitem[{{Cortes-Contreras} {et~al.}(2014){Cortes-Contreras}, {Caballero}, \&
  {Montes}}]{Cortes-Contreras(2014)}
{Cortes-Contreras}, M., {Caballero}, J.~A., \& {Montes}, D. 2014, The
  Observatory, 134, 348

\bibitem[{{Costa} {et~al.}(2005){Costa}, {M{\'e}ndez}, {Jao}, {Henry},
  {Subasavage}, {Brown}, {Ianna}, \& {Bartlett}}]{Costa(2005)}
{Costa}, E., {M{\'e}ndez}, R.~A., {Jao}, W.-C., {et~al.} 2005, \aj, 130, 337

\bibitem[{{Costa} {et~al.}(2006){Costa}, {M{\'e}ndez}, {Jao}, {Henry},
  {Subasavage}, \& {Ianna}}]{Costa(2006)}
---. 2006, \aj, 132, 1234

\bibitem[{{Cvetkovi{\'c}} {et~al.}(2015){Cvetkovi{\'c}}, {Pavlovi{\'c}}, \&
  {Boeva}}]{Cvetkovic(2015)}
{Cvetkovi{\'c}}, Z., {Pavlovi{\'c}}, R., \& {Boeva}, S. 2015, \aj, 149, 150

\bibitem[{{Daemgen} {et~al.}(2007){Daemgen}, {Siegler}, {Reid}, \&
  {Close}}]{Daemgen(2007)}
{Daemgen}, S., {Siegler}, N., {Reid}, I.~N., \& {Close}, L.~M. 2007, \apj, 654,
  558

\bibitem[{{Dahn} {et~al.}(1988){Dahn}, {Harrington}, {Kallarakal}, {Guetter},
  {Luginbuhl}, {Riepe}, {Walker}, {Pier}, {Vrba}, {Monet}, \&
  {Ables}}]{Dahn(1988)}
{Dahn}, C.~C., {Harrington}, R.~S., {Kallarakal}, V.~V., {et~al.} 1988, \aj,
  95, 237

\bibitem[{{Dahn} {et~al.}(2002){Dahn}, {Harris}, {Vrba}, {Guetter}, {Canzian},
  {Henden}, {Levine}, {Luginbuhl}, {Monet}, {Monet}, {Pier}, {Stone}, {Walker},
  {Burgasser}, {Gizis}, {Kirkpatrick}, {Liebert}, \& {Reid}}]{Dahn(2002)}
{Dahn}, C.~C., {Harris}, H.~C., {Vrba}, F.~J., {et~al.} 2002, \aj, 124, 1170

\bibitem[{{Davison} {et~al.}(2014){Davison}, {White}, {Jao}, {Henry}, {Bailey},
  {Quinn}, {Cantrell}, {Riedel}, {Subasavage}, {Winters}, \&
  {Crockett}}]{Davison(2014)}
{Davison}, C.~L., {White}, R.~J., {Jao}, W.-C., {et~al.} 2014, \aj, 147, 26

\bibitem[{{Davison} {et~al.}(2015){Davison}, {White}, {Henry}, {Riedel}, {Jao},
  {Bailey}, {Quinn}, {Cantrell}, {Subasavage}, \& {Winters}}]{Davison(2015)}
{Davison}, C.~L., {White}, R.~J., {Henry}, T.~J., {et~al.} 2015, \aj, 149, 106

\bibitem[{{Dawson} \& {De Robertis}(2005)}]{Dawson(2005)}
{Dawson}, P.~C., \& {De Robertis}, M.~M. 2005, \pasp, 117, 1

\bibitem[{{Deacon} \& {Hambly}(2001)}]{Deacon(2001)}
{Deacon}, N.~R., \& {Hambly}, N.~C. 2001, \aap, 380, 148

\bibitem[{{Deacon} {et~al.}(2005{\natexlab{a}}){Deacon}, {Hambly}, \&
  {Cooke}}]{Deacon(2005b)}
{Deacon}, N.~R., {Hambly}, N.~C., \& {Cooke}, J.~A. 2005{\natexlab{a}}, \aap,
  435, 363

\bibitem[{{Deacon} {et~al.}(2005{\natexlab{b}}){Deacon}, {Hambly}, {Henry},
  {Subasavage}, {Brown}, \& {Jao}}]{Deacon(2005a)}
{Deacon}, N.~R., {Hambly}, N.~C., {Henry}, T.~J., {et~al.} 2005{\natexlab{b}},
  \aj, 129, 409

\bibitem[{{Delfosse} {et~al.}(1999{\natexlab{a}}){Delfosse}, {Forveille},
  {Beuzit}, {Udry}, {Mayor}, \& {Perrier}}]{Delfosse(1999c)}
{Delfosse}, X., {Forveille}, T., {Beuzit}, J.-L., {et~al.} 1999{\natexlab{a}},
  \aap, 344, 897

\bibitem[{{Delfosse} {et~al.}(1998){Delfosse}, {Forveille}, {Perrier}, \&
  {Mayor}}]{Delfosse(1998)}
{Delfosse}, X., {Forveille}, T., {Perrier}, C., \& {Mayor}, M. 1998, \aap, 331,
  581

\bibitem[{{Delfosse} {et~al.}(1999{\natexlab{b}}){Delfosse}, {Forveille},
  {Udry}, {Beuzit}, {Mayor}, \& {Perrier}}]{Delfosse(1999d)}
{Delfosse}, X., {Forveille}, T., {Udry}, S., {et~al.} 1999{\natexlab{b}}, \aap,
  350, L39

\bibitem[{{Dhital} {et~al.}(2010){Dhital}, {West}, {Stassun}, \&
  {Bochanski}}]{Dhital(2010)}
{Dhital}, S., {West}, A.~A., {Stassun}, K.~G., \& {Bochanski}, J.~J. 2010, \aj,
  139, 2566

\bibitem[{{D{\'{\i}}az} {et~al.}(2007){D{\'{\i}}az}, {Gonz{\'a}lez},
  {Cincunegui}, \& {Mauas}}]{Diaz(2007)}
{D{\'{\i}}az}, R.~F., {Gonz{\'a}lez}, J.~F., {Cincunegui}, C., \& {Mauas},
  P.~J.~D. 2007, \aap, 474, 345

\bibitem[{{Dieterich} {et~al.}(2012){Dieterich}, {Henry}, {Golimowski},
  {Krist}, \& {Tanner}}]{Dieterich(2012)}
{Dieterich}, S.~B., {Henry}, T.~J., {Golimowski}, D.~A., {Krist}, J.~E., \&
  {Tanner}, A.~M. 2012, \aj, 144, 64

\bibitem[{{Dieterich} {et~al.}(2014){Dieterich}, {Henry}, {Jao}, {Winters},
  {Hosey}, {Riedel}, \& {Subasavage}}]{Dieterich(2014)}
{Dieterich}, S.~B., {Henry}, T.~J., {Jao}, W.-C., {et~al.} 2014, \aj, 147, 94

\bibitem[{{Docobo} {et~al.}(2006){Docobo}, {Tamazian}, {Balega}, \&
  {Melikian}}]{Docobo(2006a)}
{Docobo}, J.~A., {Tamazian}, V.~S., {Balega}, Y.~Y., \& {Melikian}, N.~D. 2006,
  \aj, 132, 994

\bibitem[{{Docobo} {et~al.}(2010){Docobo}, {Tamazian}, {Balega}, \&
  {Melikian}}]{Docobo(2010)}
---. 2010, \aj, 140, 1078

\bibitem[{{Doyle} \& {Butler}(1990)}]{Doyle(1990)}
{Doyle}, J.~G., \& {Butler}, C.~J. 1990, \aap, 235, 335

\bibitem[{{Duch{\^e}ne} \& {Kraus}(2013)}]{Duchene(2013)}
{Duch{\^e}ne}, G., \& {Kraus}, A. 2013, \araa, 51, 269

\bibitem[{{Dupuy} \& {Liu}(2012)}]{Dupuy(2012)}
{Dupuy}, T.~J., \& {Liu}, M.~C. 2012, \apjs, 201, 19

\bibitem[{{Duquennoy} \& {Mayor}(1988)}]{Duquennoy(1988b)}
{Duquennoy}, A., \& {Mayor}, M. 1988, \aap, 200, 135

\bibitem[{{Endl} {et~al.}(2006){Endl}, {Cochran}, {K{\"u}rster}, {Paulson},
  {Wittenmyer}, {MacQueen}, \& {Tull}}]{Endl(2006)}
{Endl}, M., {Cochran}, W.~D., {K{\"u}rster}, M., {et~al.} 2006, \apj, 649, 436

\bibitem[{{Fabricius} \& {Makarov}(2000)}]{Fabricius(2000)}
{Fabricius}, C., \& {Makarov}, V.~V. 2000, \aaps, 144, 45

\bibitem[{{Faherty} {et~al.}(2012){Faherty}, {Burgasser}, {Walter}, {Van der
  Bliek}, {Shara}, {Cruz}, {West}, {Vrba}, \&
  {Anglada-Escud{\'e}}}]{Faherty(2012)}
{Faherty}, J.~K., {Burgasser}, A.~J., {Walter}, F.~M., {et~al.} 2012, \apj,
  752, 56

\bibitem[{{Falin} \& {Mignard}(1999)}]{Falin(1999)}
{Falin}, J.~L., \& {Mignard}, F. 1999, \aaps, 135, 231

\bibitem[{{Femen{\'{\i}}a} {et~al.}(2011){Femen{\'{\i}}a}, {Rebolo},
  {P{\'e}rez-Prieto}, {Hildebrandt}, {Labadie}, {P{\'e}rez-Garrido},
  {B{\'e}jar}, {D{\'{\i}}az-S{\'a}nchez}, {Vill{\'o}}, {Oscoz}, {L{\'o}pez},
  {Rodr{\'{\i}}guez}, \& {Piqueras}}]{Femenia(2011)}
{Femen{\'{\i}}a}, B., {Rebolo}, R., {P{\'e}rez-Prieto}, J.~A., {et~al.} 2011,
  \mnras, 413, 1524

\bibitem[{{Fischer} \& {Marcy}(1992)}]{Fischer(1992)}
{Fischer}, D.~A., \& {Marcy}, G.~W. 1992, \apj, 396, 178

\bibitem[{{Forveille} {et~al.}(2005){Forveille}, {Beuzit}, {Delorme},
  {S{\'e}gransan}, {Delfosse}, {Chauvin}, {Fusco}, {Lagrange}, {Mayor},
  {Montagnier}, {Mouillet}, {Perrier}, {Udry}, {Charton}, {Gigan}, {Conan},
  {Kern}, \& {Michet}}]{Forveille(2005)}
{Forveille}, T., {Beuzit}, J.-L., {Delorme}, P., {et~al.} 2005, \aap, 435, L5

\bibitem[{{Frankowski} {et~al.}(2007){Frankowski}, {Jancart}, \&
  {Jorissen}}]{Frankowski(2007)}
{Frankowski}, A., {Jancart}, S., \& {Jorissen}, A. 2007, \aap, 464, 377

\bibitem[{{Freed} {et~al.}(2003){Freed}, {Close}, \& {Siegler}}]{Freed(2003)}
{Freed}, M., {Close}, L.~M., \& {Siegler}, N. 2003, \apj, 584, 453

\bibitem[{{Fu} {et~al.}(1997){Fu}, {Hartkopf}, {Mason}, {McAlister},
  {Dombrowski}, {Westin}, \& {Franz}}]{Fu(1997)}
{Fu}, H.-H., {Hartkopf}, W.~I., {Mason}, B.~D., {et~al.} 1997, \aj, 114, 1623

\bibitem[{{Gaia Collaboration} {et~al.}(2018){Gaia Collaboration}, {Brown},
  {Vallenari}, {Prusti}, {de Bruijne}, {Babusiaux}, \&
  {Bailer-Jones}}]{GaiaDR2(2018)}
{Gaia Collaboration}, {Brown}, A.~G.~A., {Vallenari}, A., {et~al.} 2018, ArXiv
  e-prints, arXiv:1804.09365

\bibitem[{{Gaia Collaboration} {et~al.}(2016){Gaia Collaboration}, {Prusti},
  {de Bruijne}, {Brown}, {Vallenari}, {Babusiaux}, {Bailer-Jones}, {Bastian},
  {Biermann}, {Evans}, \& et~al.}]{Gaia(2016a)}
{Gaia Collaboration}, {Prusti}, T., {de Bruijne}, J.~H.~J., {et~al.} 2016,
  \aap, 595, A1

\bibitem[{{Gatewood}(2008)}]{Gatewood(2008)}
{Gatewood}, G. 2008, \aj, 136, 452

\bibitem[{{Gatewood} \& {Coban}(2009)}]{Gatewood(2009)}
{Gatewood}, G., \& {Coban}, L. 2009, \aj, 137, 402

\bibitem[{{Gatewood} {et~al.}(2003){Gatewood}, {Coban}, \&
  {Han}}]{Gatewood(2003)}
{Gatewood}, G., {Coban}, L., \& {Han}, I. 2003, \aj, 125, 1530

\bibitem[{{Gatewood} {et~al.}(1993){Gatewood}, {de Jonge}, \&
  {Stephenson}}]{Gatewood(1993)}
{Gatewood}, G., {de Jonge}, K.~J., \& {Stephenson}, B. 1993, \pasp, 105, 1101

\bibitem[{{Gershberg} {et~al.}(1999){Gershberg}, {Katsova}, {Lovkaya},
  {Terebizh}, \& {Shakhovskaya}}]{Gershberg(1999A)}
{Gershberg}, R.~E., {Katsova}, M.~M., {Lovkaya}, M.~N., {Terebizh}, A.~V., \&
  {Shakhovskaya}, N.~I. 1999, \aaps, 139, 555

\bibitem[{{Gianninas} {et~al.}(2011){Gianninas}, {Bergeron}, \&
  {Ruiz}}]{Gianninas(2011)}
{Gianninas}, A., {Bergeron}, P., \& {Ruiz}, M.~T. 2011, \apj, 743, 138

\bibitem[{{Gizis}(1997)}]{Gizis(1997a)}
{Gizis}, J.~E. 1997, \aj, 113, 806

\bibitem[{{Gizis}(1998)}]{Gizis(1998b)}
---. 1998, \aj, 115, 2053

\bibitem[{{Gizis} {et~al.}(2002){Gizis}, {Reid}, \& {Hawley}}]{Gizis(2002)}
{Gizis}, J.~E., {Reid}, I.~N., \& {Hawley}, S.~L. 2002, \aj, 123, 3356

\bibitem[{{Golimowski} {et~al.}(2004){Golimowski}, {Leggett}, {Marley}, {Fan},
  {Geballe}, {Knapp}, {Vrba}, {Henden}, {Luginbuhl}, {Guetter}, {Munn},
  {Canzian}, {Zheng}, {Tsvetanov}, {Chiu}, {Glazebrook}, {Hoversten},
  {Schneider}, \& {Brinkmann}}]{Golimowski(2004)}
{Golimowski}, D.~A., {Leggett}, S.~K., {Marley}, M.~S., {et~al.} 2004, \aj,
  127, 3516

\bibitem[{{Gould} \& {Chanam{\'e}}(2004)}]{Gould(2004)}
{Gould}, A., \& {Chanam{\'e}}, J. 2004, \apjs, 150, 455

\bibitem[{{Graham}(1982)}]{Graham(1982)}
{Graham}, J.~A. 1982, \pasp, 94, 244

\bibitem[{{Gray} {et~al.}(2003){Gray}, {Corbally}, {Garrison}, {McFadden}, \&
  {Robinson}}]{Gray(2003)}
{Gray}, R.~O., {Corbally}, C.~J., {Garrison}, R.~F., {McFadden}, M.~T., \&
  {Robinson}, P.~E. 2003, \aj, 126, 2048

\bibitem[{{Harlow}(1996)}]{Harlow(1996)}
{Harlow}, J.~J.~B. 1996, \aj, 112, 2222

\bibitem[{{Harrington} \& {Dahn}(1980)}]{Harrington(1980)}
{Harrington}, R.~S., \& {Dahn}, C.~C. 1980, \aj, 85, 454

\bibitem[{{Harrington} {et~al.}(1985){Harrington}, {Kallarakal}, {Christy},
  {Dahn}, {Riepe}, {Guetter}, {Ables}, {Hewitt}, {Vrba}, \&
  {Walker}}]{Harrington(1985)}
{Harrington}, R.~S., {Kallarakal}, V.~V., {Christy}, J.~W., {et~al.} 1985, \aj,
  90, 123

\bibitem[{{Harrington} {et~al.}(1993){Harrington}, {Dahn}, {Kallarakal},
  {Guetter}, {Riepe}, {Walker}, {Pier}, {Vrba}, {Luginbuhl}, {Harris}, \&
  {Ables}}]{Harrington(1993)}
{Harrington}, R.~S., {Dahn}, C.~C., {Kallarakal}, V.~V., {et~al.} 1993, \aj,
  105, 1571

\bibitem[{{Hartkopf} {et~al.}(2012){Hartkopf}, {Tokovinin}, \&
  {Mason}}]{Hartkopf(2012)}
{Hartkopf}, W.~I., {Tokovinin}, A., \& {Mason}, B.~D. 2012, \aj, 143, 42

\bibitem[{{Hawley} {et~al.}(1996){Hawley}, {Gizis}, \& {Reid}}]{Hawley(1996)}
{Hawley}, S.~L., {Gizis}, J.~E., \& {Reid}, I.~N. 1996, \aj, 112, 2799

\bibitem[{{Heintz}(1976)}]{Heintz(1976)}
{Heintz}, W.~D. 1976, \mnras, 175, 533

\bibitem[{{Heintz}(1985)}]{Heintz(1985)}
---. 1985, \apjs, 58, 439

\bibitem[{{Heintz}(1986)}]{Heintz(1986)}
---. 1986, \aj, 92, 446

\bibitem[{{Heintz}(1987)}]{Heintz(1987)}
---. 1987, \apjs, 65, 161

\bibitem[{{Heintz}(1990)}]{Heintz(1990)}
---. 1990, \apjs, 74, 275

\bibitem[{{Heintz}(1991)}]{Heintz(1991)}
---. 1991, \aj, 101, 1071

\bibitem[{{Heintz}(1992)}]{Heintz(1992a)}
---. 1992, \apjs, 83, 351

\bibitem[{{Heintz}(1993)}]{Heintz(1993)}
---. 1993, \aj, 105, 1188

\bibitem[{{Heintz}(1994)}]{Heintz(1994)}
---. 1994, \aj, 108, 2338

\bibitem[{{He{\l}miniak} {et~al.}(2009){He{\l}miniak}, {Konacki}, {Kulkarni},
  \& {Eisner}}]{Helminiak(2009)}
{He{\l}miniak}, K.~G., {Konacki}, M., {Kulkarni}, S.~R., \& {Eisner}, J. 2009,
  \mnras, 400, 406

\bibitem[{{Henry}(1991)}]{Henry(1991)}
{Henry}, T.~J. 1991, PhD thesis, Arizona Univ., Tucson.

\bibitem[{{Henry} {et~al.}(1999){Henry}, {Franz}, {Wasserman}, {Benedict},
  {Shelus}, {Ianna}, {Kirkpatrick}, \& {McCarthy}}]{Henry(1999)}
{Henry}, T.~J., {Franz}, O.~G., {Wasserman}, L.~H., {et~al.} 1999, \apj, 512,
  864

\bibitem[{{Henry} {et~al.}(1997){Henry}, {Ianna}, {Kirkpatrick}, \&
  {Jahreiss}}]{Henry(1997)}
{Henry}, T.~J., {Ianna}, P.~A., {Kirkpatrick}, J.~D., \& {Jahreiss}, H. 1997,
  \aj, 114, 388

\bibitem[{{Henry} {et~al.}(2006){Henry}, {Jao}, {Subasavage}, {Beaulieu},
  {Ianna}, {Costa}, \& {M{\'e}ndez}}]{Henry(2006)}
{Henry}, T.~J., {Jao}, W.-C., {Subasavage}, J.~P., {et~al.} 2006, \aj, 132,
  2360

\bibitem[{{Henry} \& {McCarthy}(1990)}]{Henry(1990)}
{Henry}, T.~J., \& {McCarthy}, Jr., D.~W. 1990, \apj, 350, 334

\bibitem[{{Henry} {et~al.}(2004){Henry}, {Subasavage}, {Brown}, {Beaulieu},
  {Jao}, \& {Hambly}}]{Henry(2004)}
{Henry}, T.~J., {Subasavage}, J.~P., {Brown}, M.~A., {et~al.} 2004, \aj, 128,
  2460

\bibitem[{{Henry} {et~al.}(2018){Henry}, {Jao}, {Winters}, {Dieterich},
  {Finch}, {Ianna}, {Riedel}, {Silverstein}, {Subasavage}, \&
  {Vrijmoet}}]{Henry(2018)}
{Henry}, T.~J., {Jao}, W.-C., {Winters}, J.~G., {et~al.} 2018, \aj, 155, 265

\bibitem[{{Herbig} \& {Moorhead}(1965)}]{Herbig(1965)}
{Herbig}, G.~H., \& {Moorhead}, J.~M. 1965, \apj, 141, 649

\bibitem[{{Hershey} \& {Taff}(1998)}]{Hershey(1998)}
{Hershey}, J.~L., \& {Taff}, L.~G. 1998, \aj, 116, 1440

\bibitem[{{H{\o}g} {et~al.}(2000){H{\o}g}, {Fabricius}, {Makarov}, {Urban},
  {Corbin}, {Wycoff}, {Bastian}, {Schwekendiek}, \& {Wicenec}}]{Hog(2000)}
{H{\o}g}, E., {Fabricius}, C., {Makarov}, V.~V., {et~al.} 2000, \aap, 355, L27

\bibitem[{{Holman} \& {Wiegert}(1999)}]{Holman(1999)}
{Holman}, M.~J., \& {Wiegert}, P.~A. 1999, \aj, 117, 621

\bibitem[{{Horch} {et~al.}(2012){Horch}, {Bahi}, {Gaulin}, {Howell}, {Sherry},
  {Baena Gall{\'e}}, \& {van Altena}}]{Horch(2012a)}
{Horch}, E.~P., {Bahi}, L.~A.~P., {Gaulin}, J.~R., {et~al.} 2012, \aj, 143, 10

\bibitem[{{Horch} {et~al.}(2010){Horch}, {Falta}, {Anderson}, {DeSousa},
  {Miniter}, {Ahmed}, \& {van Altena}}]{Horch(2010)}
{Horch}, E.~P., {Falta}, D., {Anderson}, L.~M., {et~al.} 2010, \aj, 139, 205

\bibitem[{{Horch} {et~al.}(2011{\natexlab{a}}){Horch}, {Gomez}, {Sherry},
  {Howell}, {Ciardi}, {Anderson}, \& {van Altena}}]{Horch(2011a)}
{Horch}, E.~P., {Gomez}, S.~C., {Sherry}, W.~H., {et~al.} 2011{\natexlab{a}},
  \aj, 141, 45

\bibitem[{{Horch} {et~al.}(2002){Horch}, {Robinson}, {Meyer}, {van Altena},
  {Ninkov}, \& {Piterman}}]{Horch(2002)}
{Horch}, E.~P., {Robinson}, S.~E., {Meyer}, R.~D., {et~al.} 2002, \aj, 123,
  3442

\bibitem[{{Horch} {et~al.}(2011{\natexlab{b}}){Horch}, {van Altena}, {Howell},
  {Sherry}, \& {Ciardi}}]{Horch(2011b)}
{Horch}, E.~P., {van Altena}, W.~F., {Howell}, S.~B., {Sherry}, W.~H., \&
  {Ciardi}, D.~R. 2011{\natexlab{b}}, The Astronomical Journal, 141, 180

\bibitem[{{Horch} {et~al.}(2015{\natexlab{a}}){Horch}, {van Belle}, {Davidson},
  {Ciastko}, {Everett}, \& {Bjorkman}}]{Horch(2015b)}
{Horch}, E.~P., {van Belle}, G.~T., {Davidson}, Jr., J.~W., {et~al.}
  2015{\natexlab{a}}, \aj, 150, 151

\bibitem[{{Horch} {et~al.}(2009){Horch}, {Veillette}, {Baena Gall{\'e}},
  {Shah}, {O'Rielly}, \& {van Altena}}]{Horch(2009)}
{Horch}, E.~P., {Veillette}, D.~R., {Baena Gall{\'e}}, R., {et~al.} 2009, \aj,
  137, 5057

\bibitem[{{Horch} {et~al.}(2015{\natexlab{b}}){Horch}, {van Altena},
  {Demarque}, {Howell}, {Everett}, {Ciardi}, {Teske}, {Henry}, \&
  {Winters}}]{Horch(2015)}
{Horch}, E.~P., {van Altena}, W.~F., {Demarque}, P., {et~al.}
  2015{\natexlab{b}}, \aj, 149, 151

\bibitem[{{Horch} {et~al.}(2017){Horch}, {Casetti-Dinescu}, {Camarata},
  {Bidarian}, {van Altena}, {Sherry}, {Everett}, {Howell}, {Ciardi}, {Henry},
  {Nusdeo}, \& {Winters}}]{Horch(2017)}
{Horch}, E.~P., {Casetti-Dinescu}, D.~I., {Camarata}, M.~A., {et~al.} 2017,
  \aj, 153, 212

\bibitem[{{Hosey} {et~al.}(2015){Hosey}, {Henry}, {Jao}, {Dieterich},
  {Winters}, {Lurie}, {Riedel}, \& {Subasavage}}]{Hosey(2015)}
{Hosey}, A.~D., {Henry}, T.~J., {Jao}, W.-C., {et~al.} 2015, \aj, 150, 6

\bibitem[{{Houdebine}(2010)}]{Houdebine(2010)}
{Houdebine}, E.~R. 2010, \mnras, 407, 1657

\bibitem[{{Howell}(2000)}]{Howell(2000)}
{Howell}, S.~B. 2000, {Handbook of CCD Astronomy} (Cambridge University Press)

\bibitem[{{Howell}(2012)}]{Howell(2012)}
---. 2012, \pasp, 124, 263

\bibitem[{{Ianna} {et~al.}(1996){Ianna}, {Patterson}, \& {Swain}}]{Ianna(1996)}
{Ianna}, P.~A., {Patterson}, R.~J., \& {Swain}, M.~A. 1996, \aj, 111, 492

\bibitem[{{Ireland} {et~al.}(2008){Ireland}, {Kraus}, {Martinache}, {Lloyd}, \&
  {Tuthill}}]{Ireland(2008)}
{Ireland}, M.~J., {Kraus}, A., {Martinache}, F., {Lloyd}, J.~P., \& {Tuthill},
  P.~G. 2008, \apj, 678, 463

\bibitem[{{Jahrei{\ss}} {et~al.}(2008){Jahrei{\ss}}, {Meusinger}, {Scholz}, \&
  {Stecklum}}]{Jahreiss(2008)}
{Jahrei{\ss}}, H., {Meusinger}, H., {Scholz}, R.-D., \& {Stecklum}, B. 2008,
  \aap, 484, 575

\bibitem[{{Jancart} {et~al.}(2005){Jancart}, {Jorissen}, {Babusiaux}, \&
  {Pourbaix}}]{Jancart(2005)}
{Jancart}, S., {Jorissen}, A., {Babusiaux}, C., \& {Pourbaix}, D. 2005, \aap,
  442, 365

\bibitem[{{Janson} {et~al.}(2014{\natexlab{a}}){Janson}, {Bergfors},
  {Brandner}, {Kudryavtseva}, {Hormuth}, {Hippler}, \&
  {Henning}}]{Janson(2014a)}
{Janson}, M., {Bergfors}, C., {Brandner}, W., {et~al.} 2014{\natexlab{a}},
  \apj, 789, 102

\bibitem[{{Janson} {et~al.}(2012){Janson}, {Hormuth}, {Bergfors}, {Brandner},
  {Hippler}, {Daemgen}, {Kudryavtseva}, {Schmalzl}, {Schnupp}, \&
  {Henning}}]{Janson(2012)}
{Janson}, M., {Hormuth}, F., {Bergfors}, C., {et~al.} 2012, \apj, 754, 44

\bibitem[{{Janson} {et~al.}(2014{\natexlab{b}}){Janson}, {Bergfors},
  {Brandner}, {Bonnefoy}, {Schlieder}, {K{\"o}hler}, {Hormuth}, {Henning}, \&
  {Hippler}}]{Janson(2014b)}
{Janson}, M., {Bergfors}, C., {Brandner}, W., {et~al.} 2014{\natexlab{b}},
  \apjs, 214, 17

\bibitem[{{Jao} {et~al.}(2008){Jao}, {Henry}, {Beaulieu}, \&
  {Subasavage}}]{Jao(2008)}
{Jao}, W.-C., {Henry}, T.~J., {Beaulieu}, T.~D., \& {Subasavage}, J.~P. 2008,
  \aj, 136, 840

\bibitem[{{Jao} {et~al.}(2003){Jao}, {Henry}, {Subasavage}, {Bean}, {Costa},
  {Ianna}, \& {M{\'e}ndez}}]{Jao(2003)}
{Jao}, W.-C., {Henry}, T.~J., {Subasavage}, J.~P., {et~al.} 2003, \aj, 125, 332

\bibitem[{{Jao} {et~al.}(2005){Jao}, {Henry}, {Subasavage}, {Brown}, {Ianna},
  {Bartlett}, {Costa}, \& {M{\'e}ndez}}]{Jao(2005)}
---. 2005, \aj, 129, 1954

\bibitem[{{Jao} {et~al.}(2014){Jao}, {Henry}, {Subasavage}, {Winters}, {Gies},
  {Riedel}, \& {Ianna}}]{Jao(2014)}
---. 2014, \aj, 147, 21

\bibitem[{{Jao} {et~al.}(2011){Jao}, {Henry}, {Subasavage}, {Winters},
  {Riedel}, \& {Ianna}}]{Jao(2011)}
---. 2011, \aj, 141, 117

\bibitem[{{Jao} {et~al.}(2017){Jao}, {Henry}, {Winters}, {Subasavage},
  {Riedel}, {Silverstein}, \& {Ianna}}]{Jao(2017)}
{Jao}, W.-C., {Henry}, T.~J., {Winters}, J.~G., {et~al.} 2017, \aj, 154, 191

\bibitem[{{Jao} {et~al.}(2009){Jao}, {Mason}, {Hartkopf}, {Henry}, \&
  {Ramos}}]{Jao(2009)}
{Jao}, W.-C., {Mason}, B.~D., {Hartkopf}, W.~I., {Henry}, T.~J., \& {Ramos},
  S.~N. 2009, \aj, 137, 3800

\bibitem[{{Jenkins} {et~al.}(2009){Jenkins}, {Ramsey}, {Jones}, {Pavlenko},
  {Gallardo}, {Barnes}, \& {Pinfield}}]{Jenkins(2009)}
{Jenkins}, J.~S., {Ramsey}, L.~W., {Jones}, H.~R.~A., {et~al.} 2009, \apj, 704,
  975

\bibitem[{{J{\'o}dar} {et~al.}(2013){J{\'o}dar}, {P{\'e}rez-Garrido},
  {D{\'{\i}}az-S{\'a}nchez}, {Vill{\'o}}, {Rebolo}, \&
  {P{\'e}rez-Prieto}}]{Jodar(2013)}
{J{\'o}dar}, E., {P{\'e}rez-Garrido}, A., {D{\'{\i}}az-S{\'a}nchez}, A.,
  {et~al.} 2013, \mnras, 429, 859

\bibitem[{{Khovritchev} {et~al.}(2013){Khovritchev}, {Izmailov}, \&
  {Khrutskaya}}]{Khovritchev(2013)}
{Khovritchev}, M.~Y., {Izmailov}, I.~S., \& {Khrutskaya}, E.~V. 2013, \mnras,
  435, 1083

\bibitem[{{Kirkpatrick} \& {McCarthy}(1994)}]{Kirkpatrick(1994)}
{Kirkpatrick}, J.~D., \& {McCarthy}, Jr., D.~W. 1994, \aj, 107, 333

\bibitem[{{Kleinman} {et~al.}(2004){Kleinman}, {Harris}, {Eisenstein},
  {Liebert}, {Nitta}, {Krzesi{\'n}ski}, {Munn}, {Dahn}, {Hawley}, {Pier},
  {Schmidt}, {Silvestri}, {Smith}, {Szkody}, {Strauss}, {Knapp}, {Collinge},
  {Mukadam}, {Koester}, {Uomoto}, {Schlegel}, {Anderson}, {Brinkmann}, {Lamb},
  {Schneider}, \& {York}}]{Kleinman(2004)}
{Kleinman}, S.~J., {Harris}, H.~C., {Eisenstein}, D.~J., {et~al.} 2004, \apj,
  607, 426

\bibitem[{{Koen} {et~al.}(2002){Koen}, {Kilkenny}, {van Wyk}, {Cooper}, \&
  {Marang}}]{Koen(2002)}
{Koen}, C., {Kilkenny}, D., {van Wyk}, F., {Cooper}, D., \& {Marang}, F. 2002,
  \mnras, 334, 20

\bibitem[{{Koen} {et~al.}(2010){Koen}, {Kilkenny}, {van Wyk}, \&
  {Marang}}]{Koen(2010)}
{Koen}, C., {Kilkenny}, D., {van Wyk}, F., \& {Marang}, F. 2010, \mnras, 403,
  1949

\bibitem[{{K{\"o}hler} {et~al.}(2012){K{\"o}hler}, {Ratzka}, \&
  {Leinert}}]{Kohler(2012)}
{K{\"o}hler}, R., {Ratzka}, T., \& {Leinert}, C. 2012, \aap, 541, A29

\bibitem[{{Kraus} {et~al.}(2016){Kraus}, {Ireland}, {Huber}, {Mann}, \&
  {Dupuy}}]{Kraus(2016)}
{Kraus}, A.~L., {Ireland}, M.~J., {Huber}, D., {Mann}, A.~W., \& {Dupuy}, T.~J.
  2016, \aj, 152, 8

\bibitem[{{K{\"u}rster} {et~al.}(2008){K{\"u}rster}, {Endl}, \&
  {Reffert}}]{Kurster(2008)}
{K{\"u}rster}, M., {Endl}, M., \& {Reffert}, S. 2008, \aap, 483, 869

\bibitem[{{K{\"u}rster} {et~al.}(2009){K{\"u}rster}, {Zechmeister}, {Endl}, \&
  {Meyer}}]{Kurster(2009)}
{K{\"u}rster}, M., {Zechmeister}, M., {Endl}, M., \& {Meyer}, E. 2009, The
  Messenger, 136, 39

\bibitem[{{Lampens} {et~al.}(2007){Lampens}, {Strigachev}, \&
  {Duval}}]{Lampens(2007)}
{Lampens}, P., {Strigachev}, A., \& {Duval}, D. 2007, \aap, 464, 641

\bibitem[{{Landolt}(1992)}]{Landolt(1992)}
{Landolt}, A.~U. 1992, \aj, 104, 372

\bibitem[{{Landolt}(2007)}]{Landolt(2007)}
---. 2007, \aj, 133, 2502

\bibitem[{{Landolt}(2009)}]{Landolt(2009)}
---. 2009, \aj, 137, 4186

\bibitem[{{Landolt}(2013)}]{Landolt(2013)}
---. 2013, \aj, 146, 131

\bibitem[{{Law} {et~al.}(2010){Law}, {Dhital}, {Kraus}, {Stassun}, \&
  {West}}]{Law(2010)}
{Law}, N.~M., {Dhital}, S., {Kraus}, A., {Stassun}, K.~G., \& {West}, A.~A.
  2010, \apj, 720, 1727

\bibitem[{{Law} {et~al.}(2006){Law}, {Hodgkin}, \& {Mackay}}]{Law(2006b)}
{Law}, N.~M., {Hodgkin}, S.~T., \& {Mackay}, C.~D. 2006, \mnras, 368, 1917

\bibitem[{{Law} {et~al.}(2008){Law}, {Hodgkin}, \& {Mackay}}]{Law(2008)}
---. 2008, \mnras, 384, 150

\bibitem[{{Leinert} {et~al.}(1997){Leinert}, {Henry}, {Glindemann}, \&
  {McCarthy}}]{Leinert(1997)}
{Leinert}, C., {Henry}, T., {Glindemann}, A., \& {McCarthy}, Jr., D.~W. 1997,
  \aap, 325, 159

\bibitem[{{Leinert} {et~al.}(1994){Leinert}, {Weitzel}, {Richichi}, {Eckart},
  \& {Tacconi-Garman}}]{Leinert(1994)}
{Leinert}, C., {Weitzel}, N., {Richichi}, A., {Eckart}, A., \&
  {Tacconi-Garman}, L.~E. 1994, \aap, 291, L47

\bibitem[{{L{\`e}pine} \& {Shara}(2005)}]{Lepine(2005a)}
{L{\`e}pine}, S., \& {Shara}, M.~M. 2005, \aj, 129, 1483

\bibitem[{{L{\`e}pine} {et~al.}(2009){L{\`e}pine}, {Thorstensen}, {Shara}, \&
  {Rich}}]{Lepine(2009)}
{L{\`e}pine}, S., {Thorstensen}, J.~R., {Shara}, M.~M., \& {Rich}, R.~M. 2009,
  \aj, 137, 4109

\bibitem[{{Lindegren} {et~al.}(1997){Lindegren}, {Mignard}, {S{\"o}derhjelm},
  {Badiali}, {Bernstein}, {Lampens}, {Pannunzio}, {Arenou}, {Bernacca},
  {Falin}, {Froeschl{\'e}}, {Kovalevsky}, {Martin}, {Perryman}, \&
  {Wielen}}]{Lindegren(1997)}
{Lindegren}, L., {Mignard}, F., {S{\"o}derhjelm}, S., {et~al.} 1997, \aap, 323,
  L53

\bibitem[{{Lurie} {et~al.}(2014){Lurie}, {Henry}, {Jao}, {Quinn}, {Winters},
  {Ianna}, {Koerner}, {Riedel}, \& {Subasavage}}]{Lurie(2014)}
{Lurie}, J.~C., {Henry}, T.~J., {Jao}, W.-C., {et~al.} 2014, \aj, 148, 91

\bibitem[{{Luyten}(1979{\natexlab{a}})}]{Luyten(1979a)}
{Luyten}, W.~J. 1979{\natexlab{a}}, {LHS Catalogue. Second edition.}
  (University of Minnesota)

\bibitem[{{Luyten}(1979{\natexlab{b}})}]{Luyten(1979b)}
---. 1979{\natexlab{b}}, {NLTT catalogue. Volume\_I. +90\_\_to\_+30\_.
  Volume.\_II. +30\_\_to\_0\_.} (University of Minnesota)

\bibitem[{{Luyten}(1980{\natexlab{a}})}]{Luyten(1980a)}
---. 1980{\natexlab{a}}, {NLTT Catalogue. Volume\_III. 0\_\_to -30\_.}
  (University of Minnesota)

\bibitem[{{Luyten}(1980{\natexlab{b}})}]{Luyten(1980b)}
---. 1980{\natexlab{b}}, {NLTT Catalogue. Volume\_IV. -30\_\_to\_-90\_.}
  (University of Minnesota)

\bibitem[{{Malo} {et~al.}(2014){Malo}, {Artigau}, {Doyon}, {Lafreni{\`e}re},
  {Albert}, \& {Gagn{\'e}}}]{Malo(2014)}
{Malo}, L., {Artigau}, {\'E}., {Doyon}, R., {et~al.} 2014, \apj, 788, 81

\bibitem[{{Mamajek}(2016)}]{Mamajek(2016)}
{Mamajek}, E.~E. 2016, in IAU Symposium, Vol. 314, Young Stars \& Planets Near
  the Sun, ed. J.~H. {Kastner}, B.~{Stelzer}, \& S.~A. {Metchev}, 21--26

\bibitem[{{Mamajek} {et~al.}(2013){Mamajek}, {Bartlett}, {Seifahrt}, {Henry},
  {Dieterich}, {Lurie}, {Kenworthy}, {Jao}, {Riedel}, {Subasavage}, {Winters},
  {Finch}, {Ianna}, \& {Bean}}]{Mamajek(2013)}
{Mamajek}, E.~E., {Bartlett}, J.~L., {Seifahrt}, A., {et~al.} 2013, \aj, 146,
  154

\bibitem[{{Marcy} \& {Benitz}(1989)}]{Marcy(1989)}
{Marcy}, G.~W., \& {Benitz}, K.~J. 1989, \apj, 344, 441

\bibitem[{{Marcy} {et~al.}(1987){Marcy}, {Lindsay}, \& {Wilson}}]{Marcy(1987)}
{Marcy}, G.~W., {Lindsay}, V., \& {Wilson}, K. 1987, \pasp, 99, 490

\bibitem[{{Martin} \& {Mignard}(1998)}]{Martin(1998a)}
{Martin}, C., \& {Mignard}, F. 1998, \aap, 330, 585

\bibitem[{{Mart{\'{\i}}n} {et~al.}(2000){Mart{\'{\i}}n}, {Koresko}, {Kulkarni},
  {Lane}, \& {Wizinowich}}]{Martin(2000a)}
{Mart{\'{\i}}n}, E.~L., {Koresko}, C.~D., {Kulkarni}, S.~R., {Lane}, B.~F., \&
  {Wizinowich}, P.~L. 2000, \apjl, 529, L37

\bibitem[{{Martinache} {et~al.}(2007){Martinache}, {Lloyd}, {Ireland},
  {Yamada}, \& {Tuthill}}]{Martinache(2007)}
{Martinache}, F., {Lloyd}, J.~P., {Ireland}, M.~J., {Yamada}, R.~S., \&
  {Tuthill}, P.~G. 2007, \apj, 661, 496

\bibitem[{{Martinache} {et~al.}(2009){Martinache}, {Rojas-Ayala}, {Ireland},
  {Lloyd}, \& {Tuthill}}]{Martinache(2009)}
{Martinache}, F., {Rojas-Ayala}, B., {Ireland}, M.~J., {Lloyd}, J.~P., \&
  {Tuthill}, P.~G. 2009, \apj, 695, 1183

\bibitem[{{Mason} {et~al.}(2009){Mason}, {Hartkopf}, {Gies}, {Henry}, \&
  {Helsel}}]{Mason(2009a)}
{Mason}, B.~D., {Hartkopf}, W.~I., {Gies}, D.~R., {Henry}, T.~J., \& {Helsel},
  J.~W. 2009, \aj, 137, 3358

\bibitem[{{Mason} {et~al.}(2018){Mason}, {Hartkopf}, {Miles}, {Subasavage},
  {Raghavan}, \& {Henry}}]{Mason(2018)}
{Mason}, B.~D., {Hartkopf}, W.~I., {Miles}, K.~N., {et~al.} 2018, \aj, 155, 215

\bibitem[{{McAlister} {et~al.}(1987){McAlister}, {Hartkopf}, {Hutter}, \&
  {Franz}}]{McAlister(1987c)}
{McAlister}, H.~A., {Hartkopf}, W.~I., {Hutter}, D.~J., \& {Franz}, O.~G. 1987,
  \aj, 93, 688

\bibitem[{{McLean} {et~al.}(2003){McLean}, {McGovern}, {Burgasser},
  {Kirkpatrick}, {Prato}, \& {Kim}}]{McLean(2003)}
{McLean}, I.~S., {McGovern}, M.~R., {Burgasser}, A.~J., {et~al.} 2003, \apj,
  596, 561

\bibitem[{{Meyer} {et~al.}(2006){Meyer}, {Horch}, {Ninkov}, {van Altena}, \&
  {Rothkopf}}]{Meyer(2006)}
{Meyer}, R.~D., {Horch}, E.~P., {Ninkov}, Z., {van Altena}, W.~F., \&
  {Rothkopf}, C.~A. 2006, \pasp, 118, 162

\bibitem[{{Monet} {et~al.}(2003){Monet}, {Levine}, {Canzian}, {Ables}, {Bird},
  {Dahn}, {Guetter}, {Harris}, {Henden}, {Leggett}, {Levison}, {Luginbuhl},
  {Martini}, {Monet}, {Munn}, {Pier}, {Rhodes}, {Riepe}, {Sell}, {Stone},
  {Vrba}, {Walker}, {Westerhout}, {Brucato}, {Reid}, {Schoening}, {Hartley},
  {Read}, \& {Tritton}}]{Monet(2003)}
{Monet}, D.~G., {Levine}, S.~E., {Canzian}, B., {et~al.} 2003, \aj, 125, 984

\bibitem[{{Montagnier} {et~al.}(2006){Montagnier}, {S{\'e}gransan}, {Beuzit},
  {Forveille}, {Delorme}, {Delfosse}, {Perrier}, {Udry}, {Mayor}, {Chauvin},
  {Lagrange}, {Mouillet}, {Fusco}, {Gigan}, \& {Stadler}}]{Montagnier(2006)}
{Montagnier}, G., {S{\'e}gransan}, D., {Beuzit}, J.-L., {et~al.} 2006, \aap,
  460, L19

\bibitem[{{Montet} {et~al.}(2014){Montet}, {Crepp}, {Johnson}, {Howard}, \&
  {Marcy}}]{Montet(2014)}
{Montet}, B.~T., {Crepp}, J.~R., {Johnson}, J.~A., {Howard}, A.~W., \& {Marcy},
  G.~W. 2014, \apj, 781, 28

\bibitem[{{Morgan}(1995)}]{Morgan(1995)}
{Morgan}, D.~H. 1995, in Astronomical Society of the Pacific Conference Series,
  Vol.~84, IAU Colloq. 148: The Future Utilisation of Schmidt Telescopes, ed.
  J.~{Chapman}, R.~{Cannon}, S.~{Harrison}, \& B.~{Hidayat}, 137

\bibitem[{{Nidever} {et~al.}(2002){Nidever}, {Marcy}, {Butler}, {Fischer}, \&
  {Vogt}}]{Nidever(2002)}
{Nidever}, D.~L., {Marcy}, G.~W., {Butler}, R.~P., {Fischer}, D.~A., \& {Vogt},
  S.~S. 2002, \apjs, 141, 503

\bibitem[{{Perryman} {et~al.}(1997){Perryman}, {Lindegren}, {Kovalevsky},
  {Hoeg}, {Bastian}, {Bernacca}, {Cr{\'e}z{\'e}}, {Donati}, {Grenon},
  {Grewing}, {van Leeuwen}, {van der Marel}, {Mignard}, {Murray}, {Le Poole},
  {Schrijver}, {Turon}, {Arenou}, {Froeschl{\'e}}, \&
  {Petersen}}]{Perryman(1997)}
{Perryman}, M.~A.~C., {Lindegren}, L., {Kovalevsky}, J., {et~al.} 1997, \aap,
  323, L49

\bibitem[{{Platais} {et~al.}(2003){Platais}, {Pourbaix}, {Jorissen}, {Makarov},
  {Berdnikov}, {Samus}, {Lloyd Evans}, {Lebzelter}, \&
  {Sperauskas}}]{Platais(2003)}
{Platais}, I., {Pourbaix}, D., {Jorissen}, A., {et~al.} 2003, \aap, 397, 997

\bibitem[{{Pokorny} {et~al.}(2004){Pokorny}, {Jones}, {Hambly}, \&
  {Pinfield}}]{Pokorny(2004)}
{Pokorny}, R.~S., {Jones}, H.~R.~A., {Hambly}, N.~C., \& {Pinfield}, D.~J.
  2004, \aap, 421, 763

\bibitem[{{Pourbaix} {et~al.}(2003){Pourbaix}, {Platais}, {Detournay},
  {Jorissen}, {Knapp}, \& {Makarov}}]{Pourbaix(2003)}
{Pourbaix}, D., {Platais}, I., {Detournay}, S., {et~al.} 2003, \aap, 399, 1167

\bibitem[{{Pourbaix} {et~al.}(2004){Pourbaix}, {Tokovinin}, {Batten}, {Fekel},
  {Hartkopf}, {Levato}, {Morrell}, {Torres}, \& {Udry}}]{Pourbaix(2004)}
{Pourbaix}, D., {Tokovinin}, A.~A., {Batten}, A.~H., {et~al.} 2004, \aap, 424,
  727

\bibitem[{{Poveda} {et~al.}(1994){Poveda}, {Herrera}, {Allen}, {Cordero}, \&
  {Lavalley}}]{Poveda(1994)}
{Poveda}, A., {Herrera}, M.~A., {Allen}, C., {Cordero}, G., \& {Lavalley}, C.
  1994, \rmxaa, 28, 43

\bibitem[{{Pravdo} \& {Shaklan}(2009)}]{Pravdo(2009)}
{Pravdo}, S.~H., \& {Shaklan}, S.~B. 2009, \apj, 700, 623

\bibitem[{{Pravdo} {et~al.}(2004){Pravdo}, {Shaklan}, {Henry}, \&
  {Benedict}}]{Pravdo(2004)}
{Pravdo}, S.~H., {Shaklan}, S.~B., {Henry}, T., \& {Benedict}, G.~F. 2004,
  \apj, 617, 1323

\bibitem[{{Pravdo} {et~al.}(2006){Pravdo}, {Shaklan}, {Wiktorowicz},
  {Kulkarni}, {Lloyd}, {Martinache}, {Tuthill}, \& {Ireland}}]{Pravdo(2006)}
{Pravdo}, S.~H., {Shaklan}, S.~B., {Wiktorowicz}, S.~J., {et~al.} 2006, \apj,
  649, 389

\bibitem[{{Raghavan} {et~al.}(2010){Raghavan}, {McAlister}, {Henry}, {Latham},
  {Marcy}, {Mason}, {Gies}, {White}, \& {ten Brummelaar}}]{Raghavan(2010)}
{Raghavan}, D., {McAlister}, H.~A., {Henry}, T.~J., {et~al.} 2010, \apjs, 190,
  1

\bibitem[{{Reid} {et~al.}(2001){Reid}, {Gizis}, {Kirkpatrick}, \&
  {Koerner}}]{Reid(2001a)}
{Reid}, I.~N., {Gizis}, J.~E., {Kirkpatrick}, J.~D., \& {Koerner}, D.~W. 2001,
  \aj, 121, 489

\bibitem[{{Reid} {et~al.}(1995){Reid}, {Hawley}, \& {Gizis}}]{Reid(1995)}
{Reid}, I.~N., {Hawley}, S.~L., \& {Gizis}, J.~E. 1995, \aj, 110, 1838

\bibitem[{{Reid} {et~al.}(2002){Reid}, {Kilkenny}, \& {Cruz}}]{Reid(2002)}
{Reid}, I.~N., {Kilkenny}, D., \& {Cruz}, K.~L. 2002, \aj, 123, 2822

\bibitem[{{Reid} {et~al.}(2003){Reid}, {Cruz}, {Laurie}, {Liebert}, {Dahn},
  {Harris}, {Guetter}, {Stone}, {Canzian}, {Luginbuhl}, {Levine}, {Monet}, \&
  {Monet}}]{Reid(2003a)}
{Reid}, I.~N., {Cruz}, K.~L., {Laurie}, S.~P., {et~al.} 2003, \aj, 125, 354

\bibitem[{{Reiners} \& {Basri}(2010)}]{Reiners(2010)}
{Reiners}, A., \& {Basri}, G. 2010, \apj, 710, 924

\bibitem[{{Reiners} {et~al.}(2012){Reiners}, {Joshi}, \&
  {Goldman}}]{Reiners(2012)}
{Reiners}, A., {Joshi}, N., \& {Goldman}, B. 2012, \aj, 143, 93

\bibitem[{{Riddle} {et~al.}(1971){Riddle}, {Priser}, \&
  {Strand}}]{Riddle(1971)}
{Riddle}, R.~K., {Priser}, J.~B., \& {Strand}, K.~A. 1971, Publications of the
  U.S.~Naval Observatory Second Series, 20, 1

\bibitem[{{Riedel} {et~al.}(2017){Riedel}, {Blunt}, {Lambrides}, {Rice},
  {Cruz}, \& {Faherty}}]{Riedel(2017)}
{Riedel}, A.~R., {Blunt}, S.~C., {Lambrides}, E.~L., {et~al.} 2017, \aj, 153,
  95

\bibitem[{{Riedel} {et~al.}(2011){Riedel}, {Murphy}, {Henry}, {Melis}, {Jao},
  \& {Subasavage}}]{Riedel(2011)}
{Riedel}, A.~R., {Murphy}, S.~J., {Henry}, T.~J., {et~al.} 2011, \aj, 142, 104

\bibitem[{{Riedel} {et~al.}(2018){Riedel}, {Silverstein}, {Henry}, {Jao},
  {Winters}, {Subasavage}, {Malo}, \& {Hambly}}]{Riedel(2018)}
{Riedel}, A.~R., {Silverstein}, M.~L., {Henry}, T.~J., {et~al.} 2018, ArXiv
  e-prints, arXiv:1804.08812

\bibitem[{{Riedel} {et~al.}(2010){Riedel}, {Subasavage}, {Finch}, {Jao},
  {Henry}, {Winters}, {Brown}, {Ianna}, {Costa}, \& {Mendez}}]{Riedel(2010)}
{Riedel}, A.~R., {Subasavage}, J.~P., {Finch}, C.~T., {et~al.} 2010, \aj, 140,
  897

\bibitem[{{Riedel} {et~al.}(2014){Riedel}, {Finch}, {Henry}, {Subasavage},
  {Jao}, {Malo}, {Rodriguez}, {White}, {Gies}, {Dieterich}, {Winters},
  {Davison}, {Nelan}, {Blunt}, {Cruz}, {Rice}, \& {Ianna}}]{Riedel(2014)}
{Riedel}, A.~R., {Finch}, C.~T., {Henry}, T.~J., {et~al.} 2014, \aj, 147, 85

\bibitem[{{Salim} \& {Gould}(2003)}]{Salim(2003)}
{Salim}, S., \& {Gould}, A. 2003, \apj, 582, 1011

\bibitem[{{Schilbach} {et~al.}(2009){Schilbach}, {R{\"o}ser}, \&
  {Scholz}}]{Schilbach(2009)}
{Schilbach}, E., {R{\"o}ser}, S., \& {Scholz}, R.-D. 2009, \aap, 493, L27

\bibitem[{{Schmidt} {et~al.}(2007){Schmidt}, {Cruz}, {Bongiorno}, {Liebert}, \&
  {Reid}}]{Schmidt(2007)}
{Schmidt}, S.~J., {Cruz}, K.~L., {Bongiorno}, B.~J., {Liebert}, J., \& {Reid},
  I.~N. 2007, \aj, 133, 2258

\bibitem[{{Schneider} {et~al.}(2011){Schneider}, {Melis}, {Song}, \&
  {Zuckerman}}]{Schneider(2011)}
{Schneider}, A., {Melis}, C., {Song}, I., \& {Zuckerman}, B. 2011, \apj, 743,
  109

\bibitem[{{Scholz}(2010)}]{Scholz(2010b)}
{Scholz}, R.-D. 2010, \aap, 515, A92

\bibitem[{{S{\'e}gransan} {et~al.}(2000){S{\'e}gransan}, {Delfosse},
  {Forveille}, {Beuzit}, {Udry}, {Perrier}, \& {Mayor}}]{Segransan(2000)}
{S{\'e}gransan}, D., {Delfosse}, X., {Forveille}, T., {et~al.} 2000, \aap, 364,
  665

\bibitem[{{Shakht}(1997)}]{Shakht(1997)}
{Shakht}, N.~A. 1997, Astronomical and Astrophysical Transactions, 13, 327

\bibitem[{{Shkolnik} {et~al.}(2012){Shkolnik}, {Anglada-Escud{\'e}}, {Liu},
  {Bowler}, {Weinberger}, {Boss}, {Reid}, \& {Tamura}}]{Shkolnik(2012)}
{Shkolnik}, E.~L., {Anglada-Escud{\'e}}, G., {Liu}, M.~C., {et~al.} 2012, \apj,
  758, 56

\bibitem[{{Shkolnik} {et~al.}(2010){Shkolnik}, {Hebb}, {Liu}, {Reid}, \&
  {Collier Cameron}}]{Shkolnik(2010)}
{Shkolnik}, E.~L., {Hebb}, L., {Liu}, M.~C., {Reid}, I.~N., \& {Collier
  Cameron}, A. 2010, \apj, 716, 1522

\bibitem[{{Siegler} {et~al.}(2005){Siegler}, {Close}, {Cruz}, {Mart{\'{\i}}n},
  \& {Reid}}]{Siegler(2005)}
{Siegler}, N., {Close}, L.~M., {Cruz}, K.~L., {Mart{\'{\i}}n}, E.~L., \&
  {Reid}, I.~N. 2005, \apj, 621, 1023

\bibitem[{{Simons} {et~al.}(1996){Simons}, {Henry}, \&
  {Kirkpatrick}}]{Simons(1996)}
{Simons}, D.~A., {Henry}, T.~J., \& {Kirkpatrick}, J.~D. 1996, \aj, 112, 2238

\bibitem[{{Skrutskie} {et~al.}(1989){Skrutskie}, {Forrest}, \&
  {Shure}}]{Skrutskie(1989)}
{Skrutskie}, M.~F., {Forrest}, W.~J., \& {Shure}, M. 1989, \aj, 98, 1409

\bibitem[{{Skrutskie} {et~al.}(2006){Skrutskie}, {Cutri}, {Stiening},
  {Weinberg}, {Schneider}, {Carpenter}, {Beichman}, {Capps}, {Chester},
  {Elias}, {Huchra}, {Liebert}, {Lonsdale}, {Monet}, {Price}, {Seitzer},
  {Jarrett}, {Kirkpatrick}, {Gizis}, {Howard}, {Evans}, {Fowler}, {Fullmer},
  {Hurt}, {Light}, {Kopan}, {Marsh}, {McCallon}, {Tam}, {Van Dyk}, \&
  {Wheelock}}]{Skrutskie(2006)}
{Skrutskie}, M.~F., {Cutri}, R.~M., {Stiening}, R., {et~al.} 2006, \aj, 131,
  1163

\bibitem[{{Smart} {et~al.}(2010){Smart}, {Ioannidis}, {Jones}, {Bucciarelli},
  \& {Lattanzi}}]{Smart(2010b)}
{Smart}, R.~L., {Ioannidis}, G., {Jones}, H.~R.~A., {Bucciarelli}, B., \&
  {Lattanzi}, M.~G. 2010, \aap, 514, A84

\bibitem[{{Smart} {et~al.}(2007){Smart}, {Lattanzi}, {Jahrei{\ss}},
  {Bucciarelli}, \& {Massone}}]{Smart(2007)}
{Smart}, R.~L., {Lattanzi}, M.~G., {Jahrei{\ss}}, H., {Bucciarelli}, B., \&
  {Massone}, G. 2007, \aap, 464, 787

\bibitem[{{Snodgrass} \& {Carry}(2013)}]{Snodgrass(2013)}
{Snodgrass}, C., \& {Carry}, B. 2013, The Messenger, 152, 14

\bibitem[{{S{\"o}derhjelm}(1999)}]{Soderhjelm(1999)}
{S{\"o}derhjelm}, S. 1999, \aap, 341, 121

\bibitem[{{Subasavage} {et~al.}(2005{\natexlab{a}}){Subasavage}, {Henry},
  {Hambly}, {Brown}, \& {Jao}}]{Subasavage(2005a)}
{Subasavage}, J.~P., {Henry}, T.~J., {Hambly}, N.~C., {Brown}, M.~A., \& {Jao},
  W.-C. 2005{\natexlab{a}}, \aj, 129, 413

\bibitem[{{Subasavage} {et~al.}(2005{\natexlab{b}}){Subasavage}, {Henry},
  {Hambly}, {Brown}, {Jao}, \& {Finch}}]{Subasavage(2005b)}
{Subasavage}, J.~P., {Henry}, T.~J., {Hambly}, N.~C., {et~al.}
  2005{\natexlab{b}}, \aj, 130, 1658

\bibitem[{{Subasavage} {et~al.}(2009){Subasavage}, {Jao}, {Henry}, {Bergeron},
  {Dufour}, {Ianna}, {Costa}, \& {M{\'e}ndez}}]{Subasavage(2009)}
{Subasavage}, J.~P., {Jao}, W.-C., {Henry}, T.~J., {et~al.} 2009, \aj, 137,
  4547

\bibitem[{{Subasavage}(2007)}]{Subasavage(2007)}
{Subasavage}, Jr., J.~P. 2007, PhD thesis, Georgia State University

\bibitem[{{Tamazian} {et~al.}(2006){Tamazian}, {Docobo}, {Melikian}, \&
  {Karapetian}}]{Tamazian(2006)}
{Tamazian}, V.~S., {Docobo}, J.~A., {Melikian}, N.~D., \& {Karapetian}, A.~A.
  2006, \pasp, 118, 814

\bibitem[{{Tanner} {et~al.}(2010){Tanner}, {Gelino}, \& {Law}}]{Tanner(2010)}
{Tanner}, A.~M., {Gelino}, C.~R., \& {Law}, N.~M. 2010, \pasp, 122, 1195

\bibitem[{{Teegarden} {et~al.}(2003){Teegarden}, {Pravdo}, {Hicks}, {Lawrence},
  {Shaklan}, {Covey}, {Fraser}, {Hawley}, {McGlynn}, \&
  {Reid}}]{Teegarden(2003)}
{Teegarden}, B.~J., {Pravdo}, S.~H., {Hicks}, M., {et~al.} 2003, \apjl, 589,
  L51

\bibitem[{{Teixeira} {et~al.}(2009){Teixeira}, {Ducourant}, {Chauvin},
  {Krone-Martins}, {Bonnefoy}, \& {Song}}]{Teixeira(2009)}
{Teixeira}, R., {Ducourant}, C., {Chauvin}, G., {et~al.} 2009, \aap, 503, 281

\bibitem[{{Tinney}(1996)}]{Tinney(1996)}
{Tinney}, C.~G. 1996, \mnras, 281, 644

\bibitem[{{Tinney} {et~al.}(1995){Tinney}, {Reid}, {Gizis}, \&
  {Mould}}]{Tinney(1995)}
{Tinney}, C.~G., {Reid}, I.~N., {Gizis}, J., \& {Mould}, J.~R. 1995, \aj, 110,
  3014

\bibitem[{{Tody}(1986)}]{Tody(1986)}
{Tody}, D. 1986, in \procspie, Vol. 627, Instrumentation in astronomy VI, ed.
  D.~L. {Crawford}, 733

\bibitem[{{Tody}(1993)}]{Tody(1993)}
{Tody}, D. 1993, in Astronomical Society of the Pacific Conference Series,
  Vol.~52, Astronomical Data Analysis Software and Systems II, ed. R.~J.
  {Hanisch}, R.~J.~V. {Brissenden}, \& J.~{Barnes}, 173

\bibitem[{{Tokovinin} \& {L{\'e}pine}(2012)}]{Tokovinin(2012c)}
{Tokovinin}, A., \& {L{\'e}pine}, S. 2012, \aj, 144, 102

\bibitem[{{Tokovinin}(1992)}]{Tokovinin(1992b)}
{Tokovinin}, A.~A. 1992, \aap, 256, 121

\bibitem[{{van Altena} {et~al.}(1995){van Altena}, {Lee}, \&
  {Hoffleit}}]{vanAltena(1995)}
{van Altena}, W.~F., {Lee}, J.~T., \& {Hoffleit}, D. 1995, VizieR Online Data
  Catalog, 1174, 0

\bibitem[{{van Biesbroeck}(1974)}]{vanBiesbroeck(1974)}
{van Biesbroeck}, G. 1974, \apjs, 28, 413

\bibitem[{{van de Kamp}(1975)}]{vandeKamp(1975)}
{van de Kamp}, P. 1975, \araa, 13, 295

\bibitem[{{van de Kamp} \& {Worth}(1972)}]{vandeKamp(1972)}
{van de Kamp}, P., \& {Worth}, M.~D. 1972, \aj, 77, 762

\bibitem[{{van Dessel} \& {Sinachopoulos}(1993)}]{vanDessel(1993)}
{van Dessel}, E., \& {Sinachopoulos}, D. 1993, \aaps, 100, 517

\bibitem[{{van Leeuwen}(2007)}]{vanLeeuwen(2007)}
{van Leeuwen}, F. 2007, \aap, 474, 653

\bibitem[{{von Braun} {et~al.}(2011){von Braun}, {Boyajian}, {Kane}, {van
  Belle}, {Ciardi}, {L{\'o}pez-Morales}, {McAlister}, {Henry}, {Jao}, {Riedel},
  {Subasavage}, {Schaefer}, {ten Brummelaar}, {Ridgway}, {Sturmann},
  {Sturmann}, {Mazingue}, {Turner}, {Farrington}, {Goldfinger}, \&
  {Boden}}]{vonBraun(2011)}
{von Braun}, K., {Boyajian}, T.~S., {Kane}, S.~R., {et~al.} 2011, \apjl, 729,
  L26

\bibitem[{{Wahhaj} {et~al.}(2011){Wahhaj}, {Liu}, {Biller}, {Clarke},
  {Nielsen}, {Close}, {Hayward}, {Mamajek}, {Cushing}, {Dupuy}, {Tecza},
  {Thatte}, {Chun}, {Ftaclas}, {Hartung}, {Reid}, {Shkolnik}, {Alencar},
  {Artymowicz}, {Boss}, {de Gouveia Dal Pino}, {Gregorio-Hetem}, {Ida},
  {Kuchner}, {Lin}, \& {Toomey}}]{Wahhaj(2011)}
{Wahhaj}, Z., {Liu}, M.~C., {Biller}, B.~A., {et~al.} 2011, \apj, 729, 139

\bibitem[{{Wang} {et~al.}(2014){Wang}, {Fischer}, {Xie}, \&
  {Ciardi}}]{Wang(2014)}
{Wang}, J., {Fischer}, D.~A., {Xie}, J.-W., \& {Ciardi}, D.~R. 2014, \apj, 791,
  111

\bibitem[{{Ward-Duong} {et~al.}(2015){Ward-Duong}, {Patience}, {De Rosa},
  {Bulger}, {Rajan}, {Goodwin}, {Parker}, {McCarthy}, \&
  {Kulesa}}]{Ward-Duong(2015)}
{Ward-Duong}, K., {Patience}, J., {De Rosa}, R.~J., {et~al.} 2015, \mnras, 449,
  2618

\bibitem[{{Weis}(1984)}]{Weis(1984)}
{Weis}, E.~W. 1984, \apjs, 55, 289

\bibitem[{{Weis}(1986)}]{Weis(1986)}
---. 1986, \aj, 91, 626

\bibitem[{{Weis}(1987)}]{Weis(1987)}
---. 1987, \aj, 93, 451

\bibitem[{{Weis}(1988)}]{Weis(1988)}
---. 1988, \apss, 142, 223

\bibitem[{{Weis}(1991{\natexlab{a}})}]{Weis(1991b)}
---. 1991{\natexlab{a}}, \aj, 102, 1795

\bibitem[{{Weis}(1991{\natexlab{b}})}]{Weis(1991a)}
---. 1991{\natexlab{b}}, \aj, 101, 1882

\bibitem[{{Weis}(1993)}]{Weis(1993)}
---. 1993, \aj, 105, 1962

\bibitem[{{Weis}(1994)}]{Weis(1994)}
---. 1994, \aj, 107, 1135

\bibitem[{{Weis}(1996)}]{Weis(1996)}
---. 1996, \aj, 112, 2300

\bibitem[{{Weis}(1999)}]{Weis(1999)}
---. 1999, \aj, 117, 3021

\bibitem[{{Winn} \& {Fabrycky}(2015)}]{Winn(2015)}
{Winn}, J.~N., \& {Fabrycky}, D.~C. 2015, \araa, 53, 409

\bibitem[{{Winters} {et~al.}(2011){Winters}, {Henry}, {Jao}, {Subasavage},
  {Finch}, \& {Hambly}}]{Winters(2011)}
{Winters}, J.~G., {Henry}, T.~J., {Jao}, W.-C., {et~al.} 2011, \aj, 141, 21

\bibitem[{{Winters} {et~al.}(2015){Winters}, {Henry}, {Lurie}, {Hambly}, {Jao},
  {Bartlett}, {Boyd}, {Dieterich}, {Finch}, {Hosey}, {Ianna}, {Riedel},
  {Slatten}, \& {Subasavage}}]{Winters(2015)}
{Winters}, J.~G., {Henry}, T.~J., {Lurie}, J.~C., {et~al.} 2015, \aj, 149, 5

\bibitem[{{Winters} {et~al.}(2017){Winters}, {Sevrinsky}, {Jao}, {Henry},
  {Riedel}, {Subasavage}, {Lurie}, {Ianna}, \& {Finch}}]{Winters(2017)}
{Winters}, J.~G., {Sevrinsky}, R.~A., {Jao}, W.-C., {et~al.} 2017, \aj, 153, 14

\bibitem[{{Winters} {et~al.}(2018){Winters}, {Irwin}, {Newton}, {Charbonneau},
  {Latham}, {Han}, {Muirhead}, {Berlind}, {Calkins}, \&
  {Esquerdo}}]{Winters(2018a)}
{Winters}, J.~G., {Irwin}, J., {Newton}, E.~R., {et~al.} 2018, \aj, 155, 125

\bibitem[{{Woitas} {et~al.}(2003){Woitas}, {Tamazian}, {Docobo}, \&
  {Leinert}}]{Woitas(2003)}
{Woitas}, J., {Tamazian}, V.~S., {Docobo}, J.~A., \& {Leinert}, C. 2003, \aap,
  406, 293

\bibitem[{{Worley}(1961)}]{Worley(1961)}
{Worley}, C.~E. 1961, \pasp, 73, 167

\bibitem[{{Worley}(1962)}]{Worley(1962)}
---. 1962, \aj, 67, 396

\bibitem[{{Worley} \& {Mason}(1998)}]{Worley(1998)}
{Worley}, C.~E., \& {Mason}, B.~D. 1998, \aj, 116, 917

\bibitem[{{Young} {et~al.}(1987){Young}, {Sadjadi}, \& {Harlan}}]{Young(1987)}
{Young}, A., {Sadjadi}, S., \& {Harlan}, E. 1987, \apj, 314, 272

\bibitem[{{Zechmeister} {et~al.}(2009){Zechmeister}, {K{\"u}rster}, \&
  {Endl}}]{Zechmeister(2009b)}
{Zechmeister}, M., {K{\"u}rster}, M., \& {Endl}, M. 2009, \aap, 505, 859

\end{thebibliography}

\clearpage




%
%
%
 
\end{document}